%% file: main.tex
\documentclass[manuscript, screen]{acmart}
\AtBeginDocument{%
  }

\setcopyright{acmlicensed}
\copyrightyear{2024}
\acmYear{2024}
\acmDOI{XXXXXXX.XXXXXXX}

\usepackage{multirow}
\usepackage{array}
\usepackage{tcolorbox}
\usepackage{tabularx} 
\usepackage{colortbl} 
\usepackage{xcolor}   
\usepackage{float}
\usepackage{algpseudocode}
\usepackage{amsmath}
\usepackage{algorithm}
\usepackage{booktabs}
\usepackage{graphicx}
\begin{document}

\title{Establishing a Foundation for Tetun Ad-hoc Text Retrieval: Stemming, Indexing, Retrieval, and Ranking}

\author{Gabriel de Jesus}
\email{gabriel.jesus@inesctec.pt}
\orcid{1234-5678-9012}
\affiliation{%
  \institution{Institute for Systems and Computer Engineering, Tech. and Science (INESC TEC)}
  \city{Porto}
  \country{Portugal}
}

\author{Sérgio Nunes}
\orcid{0000-0002-2693-988X}
\email{sergio.nunes@fe.up.pt}
\affiliation{%
  \institution{INESC TEC and Faculty of Engineering, University of Porto (FEUP)}
  \city{Porto}
  \country{Portugal}
}






\renewcommand{\shortauthors}{Gabriel de Jesus and Sérgio Nunes}

\begin{abstract}
Searching for information on the internet and digital platforms requires effective retrieval solutions. However, such solutions are not yet available for Tetun, making it difficult to find relevant documents for search queries in this language. To address this gap, we investigate Tetun text retrieval with a focus on the ad-hoc retrieval task. The study begins with the development of essential language resources---including a list of stopwords, a stemmer, and a test collection---that serve as a foundation for Tetun text retrieval. Various strategies are evaluated using document titles and content. The results show that retrieving document titles, after removing hyphens and apostrophes but without applying stemming, improves performance compared to the baseline. Efficiency increases by 31.37\%, while effectiveness achieves an average relative gains of +9.40\% in MAP@10 and +30.35\% in NDCG@10 with DFR BM25. Beyond the top-10 cutoff point, Hiemstra LM demonstrates strong performance across multiple retrieval strategies and evaluation metrics. The contributions of this work include the development of \textit{Labadain-Stopwords} (a list of 160 Tetun stopwords), \textit{Labadain-Stemmer} (a Tetun stemmer with three variants), and \textit{Labadain-Avaliadór} (a Tetun test collection comprising 59 topics, 33,550 documents, and 5,900 \textit{qrels}). These resources are publicly available to support future research in Tetun information retrieval.
\end{abstract}


\ccsdesc[500]{Information Retrieval~Text Retrieval}
\ccsdesc[300]{Low-Resource Languages~Tetun}

\keywords{Stopwords, Stemming, Test Collection, Ad-hoc Retrieval}


\maketitle

\section{Introduction}

Ad-hoc text retrieval is the task of retrieving documents from large text collections in response to user queries without prior knowledge of the topics that users are likely to search, highlighting the unpredictable nature and short duration of each search~\cite{voorhees-harman-1998-trec, voorhees-et-al-trec-2004}. Users typically express their information needs through natural language text queries and submit them to a search system. The retrieval system then retrieves, ranks, and returns documents relevant to the query, presenting the most relevant documents at the top of the list, with less relevant ones further down. 

Effective information retrieval (IR) systems are essential for accessing the extensive digital content available on the web and digital platforms. Evaluating the effectiveness of these IR systems relies on robust test collections. High-resource languages benefit from readily available test collections sourced from various publicly accessible repositories, such as the IR dataset catalog~\cite{macavaney-et-al-2021}\footnote{\url{https://ir-datasets.com}} and HuggingFace.\footnote{\url{https://huggingface.co/datasets}} However, this scenario differs for low-resource languages (LRLs), where data scarcity and linguistic complexities make accessing test collections challenging.

The classical approach for constructing test collections follows the Cranfield paradigm~\cite{cleverdon-1976}, which became widely recognized through the Text REtrieval Conference (TREC) series of large-scale evaluation campaigns~\cite{harman-et-al-1992}. In this TREC-style adaptation of the Cranfield approach, a test collection comprises three components: a document collection, a set of information needs (or topics), and corresponding relevance judgments. In ad-hoc text retrieval, a set of topics is formulated and then tested by searching large document collections to estimate the number of relevant documents returned for each topic~\cite{sanderson-2010}. These query-document pairs are then provided to assessors for relevance judgment. Traditionally, relevance judgments are made by human assessors, involving a process that is both time-intensive and costly. Due to financial constraints, relevance assessment tasks for constructing test collections for LRLs are often carried out by volunteer native language speakers, such as students~\cite{sahu-pal-2023, aleahmad-et-al-2009}.

To identify effective retrieval strategies, several approaches are explored and tested using a reliable test collection. The classical approach to configuring these strategies involves preprocessing documents and queries, primarily focusing on stopword removal and stemming. For stopwords removal, a readily available list of stopwords is necessary, and a proper stemmer is required to process the input text. However, these resources are often unavailable for most LRLs. 

These challenges are also faced in the development of resources for Tetun, a LRL spoken by over 923,000 people in Timor-Leste~\cite{de-jesus-2023}. Timor-Leste is a Southeast Asian island country characterized by its multilingualism, comprising two official languages (Tetun and Portuguese), two working languages (English and Indonesian)~\cite{vasconcelos-et-al-2011}, and over 30 dialects spoken across the territory~\cite{de-jesus-2023}. Tetun, which was a dialect, became one of Timor-Leste's official languages when the country restored its independence in 2002~\cite{vasconcelos-et-al-2011}. Despite its status as an official language, Tetun is characterized by data scarcity, with fewer than 45,000 documents available on the web as of 2023~\cite{kudugunta-et-al-2023, de-jesus-nunes-2024-labadain}. Moreover, Tetun is a less-studied and computerized language, lacking essential resources for effective text retrieval, including a stopword list, a stemmer, and a test collection for the ad-hoc retrieval task.

To tackle the aforementioned challenges, we investigated strategies for Tetun ad-hoc text retrieval, including evaluating the impact of stemming and stopwords, to identify the most effective retrieval solutions for Tetun. The research questions (RQs) we addressed in this study are the following:

\paragraph{\textbf{RQ1}}~\emph{How can text preprocessing techniques tailored to Tetun's linguistic characteristics improve retrieval effectiveness?}

\paragraph{\textbf{RQ2}}~\emph{What strategies provide the most effective solutions for Tetun text-based search?}

Given that Tetun words contain accented letters (á, é, í, ó, ú, ñ), apostrophes (`), and hyphens in monosemantic compound words, our objective is to investigate the impact of query and document preprocessing on the effectiveness of text retrieval in Tetun text-based search. In line with the research questions above, we hypothesize that applying language-specific preprocessing to queries and documents can improve retrieval effectiveness without the need for stemming, particularly when retrieving short texts such as document titles. This hypothesis is grounded in findings from our preliminary study on Tetun ad-hoc text retrieval, which reported a 3.1\% relative improvement in overall MAP when stemming was not applied~\cite{de-jesus-2022-ir}.

To test this hypothesis, we began by developing a list of Tetun stopwords (\textit{Labadain-Stopwords}), a Tetun stemmer (\textit{Labadain-Stemmer}), and a Tetun test collection (\textit{Labadain-Avaliadór}) using the Labadain-30k+ dataset~\cite{de-jesus-nunes-2024-labadain}. For \textit{Labadain-Stemmer}, three variants were developed: \textit{light}, \textit{moderate}, and \textit{heavy}. The \textit{Labadain-Stemmer} performance was evaluated both as standalone systems and for their impact within the retrieval system (intrinsic) and extrinsic assessments~\cite{molla-hutchinson-2003, jones-galliers-1996}. The \textit{Labadain-Avaliadór} was developed following TREC guidelines and assessed by native Tetun-speaking students. This collection was then used to evaluate the retrieval effectiveness for Tetun ad-hoc text retrieval. 
%
%
The contributions of this work include: (i) the development of \textit{Labadain-Stopwords}, (ii) the creation of \textit{Labadain-Stemmer} with three variants, (iii) the construction of \textit{Labadain-Avaliadór}, and (iv) the establishment of baselines for Tetun ad-hoc text retrieval.

The remainder of this paper is organized as follows. Section~\ref{sec:background-related-works} reviews background and related work. Section~\ref{sec:tetun} provides an overview of Tetun and its linguistic characteristics. The dataset used in this study is described in Section~\ref{sec:dataset}, and Section~\ref{sec:development-phases} outlines the methodology for establishing baselines in Tetun ad-hoc text retrieval. Section~\ref{sec:stopwords} presents the construction of \textit{Labadain-Stopwords}, while Section~\ref{sec:stemmer} describes the development of \textit{Labadain-Stemmer}. Section~\ref{sec:test-collection} details the creation of \textit{Labadain-Avaliadór}, and Section~\ref{sec:indexing-retrieva-ranking} reports the baselines for Tetun ad-hoc text retrieval. Finally, Section~\ref{sec:conclusions-future-work} concludes the paper and discusses directions for future work.

\section{Background and Related Work}\label{sec:background-related-works}

Text retrieval typically involves several preprocessing steps, such as stopword removal and stemming. The key topics relevant to this study are the development of stopword lists, stemming approaches, test collections for evaluation, and baselines for Tetun ad-hoc text retrieval. Background on each of these topics is provided in the following subsections.

\subsection{Stopwords}
Stopwords are function words---such as articles, prepositions, and conjunctions---that appear frequently in documents. Traditionally, stopword lists are created by selecting the most frequent terms in a corpus, often choosing the top-$n$ most common terms~\cite{croft-et-al-2009, manning-et-al-2009}. This approach typically relies on classical term weighting techniques such as term frequency (TF)~\cite{luhn-1957}, inverse document frequency (IDF)~\cite{sparck-jones-1972}, or term frequency-inverse document frequency (TF-IDF)~\cite{salton-et-al-1975}.

\citet{fox-1990} applied a term frequency approach to the Brown Corpus to generate a stopword list for English, a method that has since been widely adopted to develop stopwords for other languages, including French~\cite{savoy-1999}; Marathi, Czech, Hungarian, and five other languages~\cite{ghosh-bhattacharya-2017}; and Kinyarwanda and Kirundi~\cite{niyongabo-et-al-2020}. 

Furthermore, \citet{lo-et-al-2005} introduced the normalized inverse document frequency (NIDF) for stopword detection, evaluating it on four English TREC collections and showing that NIDF outperforms TF and IDF. Later, Ferilli~\cite{ferilli-2021} proposed the term-document frequency (TDF) metric and, after testing it on two Italian corpora, found that TDF surpasses TF, IDF, and NIDF, particularly in smaller datasets. More recently, \citet{ali-etal-2024-network} demonstrated that a network-based approach exploiting topological properties of co-occurrence networks---such as in-degree, out-degree, and degree---outperforms traditional term-weighting techniques, with in-degree yielding the most consistent results across both high- and low-resource languages, including Tetun.

In IR, stopwords generally contribute minimal value to retrieving relevant documents for a given query~\cite{baeza-yates-netos-2011, manning-et-al-2009}. Therefore, removing these words from both queries and documents can enhance retrieval efficiency and effectiveness~\cite{sahu-et-al-2023, sahu-pal-2023, baeza-yates-netos-2011, croft-et-al-2009}. Despite this, other studies have reported that the effectiveness of stopword removal varies between ranking models and languages, demonstrating that it is beneficial in some cases but not in others~\cite{ghosh-bhattacharya-2017, dolamic-savoy-2010, savoy-1999}. In a study on French, \citet{savoy-1999} found that retaining stopwords performed better than removing them when using BM25. Similarly, \citet{dolamic-savoy-2010} did not observe significant differences in retrieval effectiveness for Marathi and Bengali, while for Hindi, including stopwords improved retrieval effectiveness, with average gains of approximately 20\% in mean average precision (MAP). \citet{ghosh-bhattacharya-2017} further demonstrated variability in retrieval effectiveness across datasets within the same language, observing that stopword removal did not lead to noticeable differences in retrieval effectiveness when evaluated on datasets from FIRE, CLEF, and TREC collections across several languages.

\subsection{Stemming}
Stemming is an essential component of text processing that captures the relationship between different variations and forms of a word resulting from inflection (e.g., plurals, tenses, and gender) or derivation (e.g., converting a verb into a noun by adding suffixes) and reduces them to a common root~\cite{croft-et-al-2009}. A stem is the root form of a word that remains after the removal of its affixes~\cite{baeza-yates-netos-2011}. 
%
%
Stemming can be useful for improving retrieval effectiveness by minimizing index size and reducing the number of distinct terms.

The first stemming algorithm was proposed by \citet{lovins-1968}, based on the principles of iteration and longest match. The iteration principle assumes that affixes are attached to stems in a particular order from a predefined set of affixes. The algorithm removes affixes from either the beginning or the end of the word, depending on which affix class is detected. According to the longest match principle, if multiple endings within a class match, the longest one should be removed. Since then, various stemming techniques have been developed, including rule-based, dictionary-based, and automatic stemmers. One of the most notable examples is the Porter Stemmer~\cite{porter-1980}, a widely used rule-based suffix removal algorithm for English stemming due to its simplicity and performance~\cite{baeza-yates-netos-2011}. 

Several Asian languages have adopted suffix-stripping stemmers based on the Porter and Lovins approaches, including Sanskrit~\cite{sahu-pal-2023}, Sundanese~\cite{ardiyanti-et-al-2018}, Czech~\cite{dolamic-savoy-2009}, and Indonesian~\cite{adriani-et-al-2007}. An advanced version of the Porter Stemmer is the Snowball stemming algorithm, which supports multiple languages, including Portuguese, Spanish, and German~\cite{snowball-2002}. Snowball applies a set of predefined stemming rules tailored to the specific morphological structure of each language, primarily focusing on removing suffixes.

In an experiment on French, \citet{savoy-1999} found that stemming was particularly beneficial for retrieving short documents (scientific abstracts averaging 24.5 indexing terms), with improvements in average precision (AP) in various retrieval models. For longer documents (news articles averaging 182.2 terms per article) or cases where accents were ignored, stemming yielded only marginal benefits. Similarly, \citet{braschler-ripplinger-2004}, evaluating on German data from the CLEF 2000 and 2001 datasets, reported stemming gains in MAP of up to 23\% for short queries (title only) and up to 11\% for long queries (a combination of title, description, and narrative).

However, \citet{hollink-et-al-2004}, in a monolingual document retrieval experiment using the CLEF 2002 dataset across eight languages, reported inconsistent results. Specifically, stemming improved MAP for Finnish, French, German, and Swedish but had no positive effect on Dutch, English, Italian, or Spanish. Likewise, \citet{flores-moreira-2016}, in an experiment with four different languages from the CLEF 2005 and 2006 datasets, reported that stemming was generally beneficial in MAP for Portuguese, French, and Spanish but not for English when tested with different stemmers.

In studies on Asian LRLs, \citet{sahu-pal-2023} reported that stemming improved retrieval effectiveness for Sanskrit by 4.31\% in MAP across multiple ranking models. Likewise, \citet{sahu-et-al-2023} observed a 1.41\% MAP improvement for Urdu when testing several stemming approaches, while \citet{adriani-et-al-2007} found a 2.00\% MAP improvement in experiments with Indonesian.

\subsection{Test Collection}
A reliable test collection is essential for evaluating the effectiveness of retrieval systems. For high-resource languages, these collections are typically made available through large-scale campaigns such as the TREC,\footnote{\url{https://trec.nist.gov}} the Conference and Labs of the Evaluation Forum (CLEF),\footnote{\url{https://www.clef-initiative.eu}} the NII Testbeds and Community for Information Access Research project (NTCIR),\footnote{\url{http://research.nii.ac.jp/ntcir/index-en.html}} and the Forum for Information Retrieval Evaluation (FIRE).\footnote{\url{http://fire.irsi.res.in/}} Following the Cranfield paradigm implemented in the TREC, developing a test collection involves selecting various retrieval strategies to compare and produce top-ranked lists of documents (\textit{runs}). These \textit{runs} are then merged to create a pooled set of documents for each query. This pool is manually judged for relevance by human assessors, producing a list of relevant documents (\textit{qrels})~\cite{sanderson-2010}.

Relevance judgments typically fall into two categories: binary and graded relevance. Binary relevance categorizes each document as either relevant or non-relevant to the user's query, assigning a score of 1 for relevant and 0 for non-relevant documents. The graded relevance evaluates documents on multiple levels of relevance, with the most relevant documents awarding higher scores. Binary relevance is predominantly used for experimental research in the TREC collections. In the TREC-9 Web Track, three-level graded relevance was introduced: not relevant, relevant, and highly relevant~\cite{hawking-2000}. Later, Sormunen proposed a four-level relevance scale consisting of non-relevant, marginally relevant, relevant, and highly relevant~\cite{sormunen-2002}. \citet{kekalainen-2005} adopted a similar four-point scale but labeled the third level as ``fairly relevant'' rather than ``relevant''. This four-point scale was subsequently implemented across multiple TREC tracks. The ad-hoc retrieval task was a central focus of the TREC tracks held from 1992 to 1999 and was revisited on the robust track from 2003 to 2005~\cite{voorhees-2006}.

The TREC-style approach, derived from the Cranfield paradigm, is commonly used to develop test collections for LRLs. \citet{sahu-pal-2023} applied this method to create a Sanskrit test collection comprising 7,057 news articles and 50 topics, with queries and relevance judgments produced by two Ph.D. students. Similarly, \citet{chavula-suleman-2021} constructed a test collection for three Bantu languages---Chichewa, Citumbuka, and Cinyanja---using documents from newspapers, Wikipedia, and web pages. Their collection includes 13,627 documents and 387 topics, with queries and relevance assessments carried out by six recruited assessors. Furthermore, \citet{aleahmad-et-al-2009} developed the Hamshahri test collection for Persian, based on a news corpus of 166,774 documents and 65 queries, with queries and relevance assessments performed by 17 volunteer students.

\subsection{Summary}
In ad-hoc text retrieval, preprocessing steps, such as stopword removal and stemming, are often employed to enhance retrieval efficiency and effectiveness. However, studies show that the impact of these techniques on retrieval effectiveness varies across languages, proving beneficial in some instances but less so in others. Furthermore, \citet{ghosh-bhattacharya-2017} highlighted that the influence of stopwords can differ even within the same language across different collections, such as Bangla and Hindi in the FIRE datasets of 2010 and 2011.

The evaluation of retrieval system effectiveness relies on robust test collections, which are typically developed following TREC guidelines, with human assessors conducting relevance judgments. For LRLs, the same methodologies are adapted to create test collections. However, due to financial constraints, relevance assessments in these less-resourced contexts are often carried out by students who are native language speakers.

This study addresses a critical gap in Tetun text retrieval by introducing three essential resources: a stopword list (\textit{Labadain-Stopwords}), a language-specific stemmer (\textit{Labadain-Stemmer}), and a Tetun test collection (\textit{Labadain-Avaliadór}). Through a series of experiments, we evaluate various retrieval strategies to establish baselines and identify the most effective approach for ad-hoc text retrieval in Tetun. The subsequent sections provide a detailed overview of the development of each resource and its application in the experiments.

\section{Tetun}\label{sec:tetun}

This section presents an overview of Tetun, including its orthography, morphology, and Portuguese loanwords.

\subsection{Overview}

Tetun, alternatively written as Tetum (in English) or Tétum (in Portuguese), is an Austronesian language spoken in Timor-Leste, an island nation in Southeast Asia. It has two primary varieties: Tetun Dili, also known as Tetun \textit{Prasa} (commonly referred to simply as Tetun), and Tetun Terik~\cite{klinken-et-al-2002}. Tetun has two standardized forms: one developed by the \emph{Instituto Nacional de Linguística} (Tetun INL) and another by the Dili Institute of Technology (Tetun DIT). Tetun Terik, meanwhile, remains one of the dialects spoken in Timor-Leste.

Tetun is one of Timor-Leste's official languages alongside Portuguese~\cite{vasconcelos-et-al-2011}. The government recognized Tetun INL as the official Tetun, which is used in the education system, official publications, and media~\cite{standard-tetun-inl-2004}. Tetun DIT was developed by linguists at the Dili Institute of Technology with some standardized differences from Tetun INL in terms of writing conventions~\cite{klinken-et-al-2002}. For example, the words [~\textit{fó} (give), \textit{ne'ebé} (which/that) ] in Tetun INL correspond to [~\textit{foo}, \textit{neebe}~] in Tetun DIT.
According to the 2015 census report, Timor-Leste’s population was 1.18 million, with 78.78\% of the population being Tetun speakers~\cite{de-jesus-nunes-2024-labadain}. Among them, 30.50\% considered Tetun their home language, while 48.28\% spoke it as a second or third language. Census 2022 reported a population growth of 13.40\%, increasing from 1.18 million to 1.34 million~\cite{census-timor-leste-2022}, but it did not provide specific indicators for Tetun speakers.

\subsection{Orthography}

Tetun INL is based on the Latin alphabet, consisting of 5 vowels: \emph{a, e, i, o, u}, and 21 consonants: \emph{b, d, f, g, h, j, k, l, ll, m, n, ñ, p, r, rr, s, t, ', v, x, z}~\cite{inl-2004-tetum-orthography}. The letters \emph{c, q, w} and \emph{y} are not used in Tetun except for the proper names and international symbols. Accented vowels \emph{á, é, í, ó, ú} are also used, and the apostrophe (') denotes a glottal stop. Additionally, the hyphen is also used to indicate monosemantic compound words. Some basic phrases in Tetun are presented in \autoref{tab:basic-tetun}.

\input{tables/tetun-examples}

\subsection{Morphology}\label{tet:morphology}

Morphology is conventionally divided into inflection and word formation, with word formation further classified into derivation and compounding~\citep{aronoff-1983}. Inflection refers to the different syntactic variations of a word that do not alter its core meaning, while word formation involves the creation of new nouns, verbs, and adjectives. Derivation creates a new word from an existing one, whereas compounding combines two or more words to form a new word. 

Morphological processes such as circumfixes and reduplication also contribute to both the formation of new words and the modification of existing word structures. Circumfixes involve the simultaneous addition of a prefix and a suffix to a base word, while reduplication is a morphological process in which a part of a word is copied, either fully or partially, to form a new word that may have additional morphemes attached to it~\cite{greksakova-2018, suzanne-2017}.

Tetun does not have rich inflectional and derivational morphology, with only a few inflectional affixes~\cite{greksakova-2018, hull-correia-2005, klinken-et-al-2002}. Tetun affixes include both native Tetun elements and those derived from Portuguese. Prefixes are exclusive of native Tetun, whereas suffixes can derive from either native Tetun or Portuguese. In compounding, words are combined using hyphens, exclusively with native Tetun words. Additionally, Tetun uses circumfixes and reduplication within its native vocabulary and adopts zero derivation for Portuguese-derived words. Examples of Tetun inflection and derivation are provided in \autoref{tab:morphology}.

\input{tables/tetun-affix-examples}

In Tetun, both circumfixes and reduplication are not as widely used as in other languages. The circumfixes in Tetun are not productive~\cite{greksakova-2018} and are confined to simple verbs, typically consisting of one or two syllables derived from verbs~\cite{hull-correia-2005}. Reduplication is similarly limited, being applied only to nouns, adjectives, adverbs, and numerals, with only a few instances of its use for pluralization~\cite{greksakova-2018}.

\subsection{Portuguese Loanwords}

A significant portion of Tetun's verbs, nouns, and adjectives are derived from Portuguese, where this influence is particularly noticeable in the news media, such as newspapers~\cite{hajek-klinken-2019, greksakova-2018, klinken-hajek-2018, klinken-et-al-2002}. Klinken et al.~\cite{klinken-et-al-2002} highlighted that the prevalence of Portuguese loanwords can be traced back to Portuguese-educated political leaders who continued to use Portuguese in their homes after 1975. As these leaders frequently appeared in the news media, the incorporation of Portuguese loanwords into Tetun rapidly increased.

\input{tables/pt-loanword-examples}

Klinken and Hajek~\cite{klinken-hajek-2018} studied a selection of seven articles from different newspapers in 2009 and reported an average of 32\% of words are Portuguese loanwords. Similarly, \citet{greksakova-2018} highlighted 35\% of Portuguese loanwords in the analysis of 73,892 words from interview transcripts. Moreover, Hajek and Klinken~\cite{hajek-klinken-2019} described Tetun's influence from Portuguese in newspaper and technical writing, rising to over 40\%, with headlines often almost entirely in Portuguese. In a recent study, \citet{de-jesus-nunes-2024-labadain} reported 28.20\% of Portuguese loanwords in Tetun when analyzing approximately 10.69 million words extracted from the Labadain-30k+ dataset~\cite{labadain30k-dataset} for an interval time from 2017 to 2023. Additionally, they observed a 5.09 percentage point increase in Portuguese loanwords when comparing documents created before and after 2017 in the Labadain-30k+ dataset. Examples of Portuguese loanwords are presented in \autoref{tab:portuguese-loanwords}.

\section{Dataset}\label{sec:dataset}

In this work, we employed the Labadain-30k+ dataset~\cite{labadain30k-dataset}, comprising 33,550 Tetun documents acquired through web crawling.
%
%
The dataset was thoroughly audited by native Tetun speakers at the document level and comprised a diverse range of categories, including news articles, Wikipedia entries, legal and government documents, and research papers, among others~\cite{de-jesus-nunes-2024-labadain}. A detailed description of the dataset is provided in \autoref{tab:dataset-description}, with a summary grouped by category of documents presented in \autoref{tab:documents-per-category}. This dataset was employed to develop \textit{Labadain-Stopwords}, \textit{Labadain-Stemmer}, and \textit{Labadain-Avaliadór}, which were subsequently used to evaluate the retrieval effectiveness of Tetun ad-hoc text retrieval.

\input{tables/labadain-30k-dataset-desc}

\input{tables/labadain-30k-dataset-summary}

\section{Methodology}\label{sec:development-phases}
\begin{figure}[!ht]
  \centering
  \includegraphics[width=0.9\textwidth]{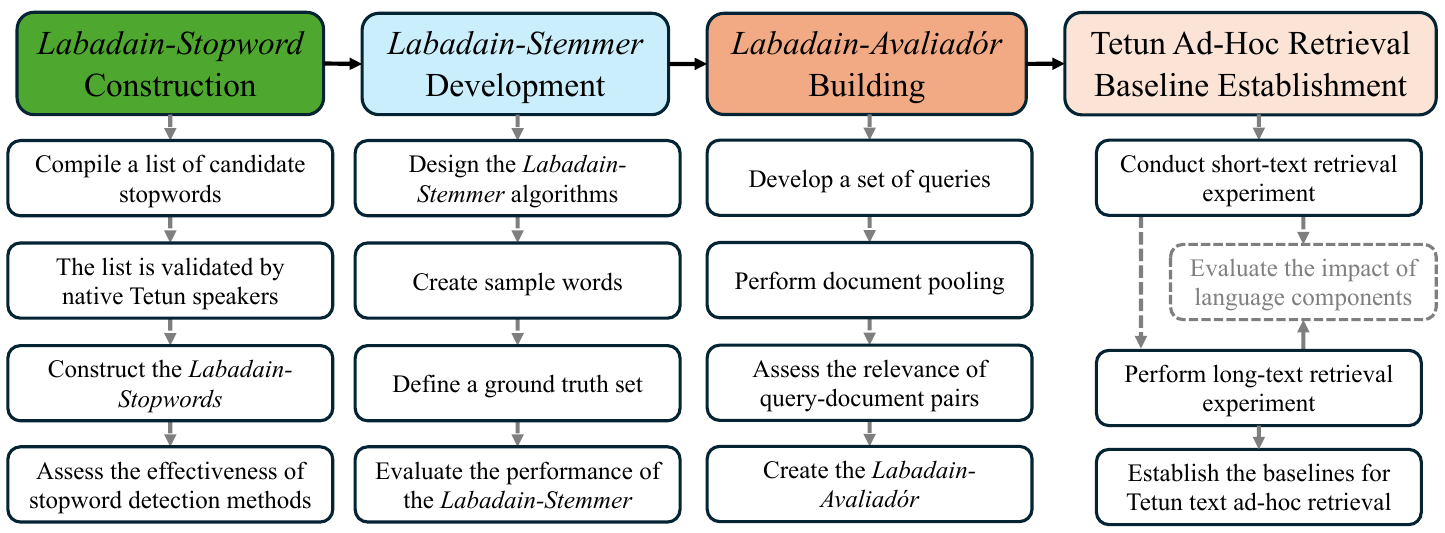}
  \caption{Methodology for Establishing Baselines in Tetun Ad-Hoc Text Retrieval.}
  \Description{Methodology for Establishing Baselines in Tetun Ad-Hoc Text Retrieval.}
  \label{fig:phases-development}
\end{figure}

To establish baselines for Tetun ad-hoc text retrieval, we employ the methodology illustrated in \autoref{fig:phases-development}. The process begins with the creation of a Tetun stopword list, continues with the development of a stemmer and a test collection, and concludes with experiments to establish the baselines. Each stage is described in the following subsections.

\subsection{\textit{Labadain-Stopwords} Construction}
This initial stage focuses on constructing a Tetun stopword list. Since manually creating stopword lists is both time-consuming and expensive, we adopted a corpus-based approach using the Labadain-30k+ dataset. Candidate stopwords were generated using frequency- and network-based detection methods, and the resulting lists were merged and validated by two native Tetun speakers to produce the final Tetun stopword list (called \textit{Labadain-Stopwords}).

Building on the findings of \citet{ali-etal-2024-network}, which demonstrated the superior effectiveness of network-based methods compared to traditional frequency-based techniques for stopword detection, we extended the evaluation to Tetun using the \textit{Labadain-Stopwords} as the ground-truth set. We then compared the results with those for Portuguese and English to gain further insight.

\subsection{\textit{Labadain-Stemmer} Development}
This stage focused on developing the stemmer algorithms for Tetun, called \textit{Labadain-Stemmer}. Since a substantial portion of Tetun verbs, nouns, and adjectives are Portuguese loanwords, and Tetun suffixes encompass Portuguese-derived words and native Tetun, we created three stemmer variants: \textit{light}, \textit{moderate}, and \textit{heavy}. The \textit{light} variant removes only the suffixes of Portuguese loanwords, the \textit{moderate} variant addresses both Portuguese loanwords and native Tetun suffixes, and the \textit{heavy} variant handles Portuguese loanword suffixes, as well as native Tetun prefixes and suffixes.

To evaluate the proposed stemmer, we conducted both intrinsic and extrinsic assessments. For the intrinsic evaluation, we systematically extracted a subset of vocabularies from the Labadain-30k+ dataset~\cite{labadain30k-dataset} and collaborated with native Tetun-speaking students to construct a sample of words. These students assessed the sample word list provided to establish a ground truth list, with each word paired with its corresponding lemma (root). The ground truth set was then used to evaluate the accuracy of each stemmer variant using the Paice metrics~\cite{paice-1994}. For extrinsic evaluation, we tested the effectiveness of Tetun stemmers in the ad-hoc text retrieval task.

\subsection{\textit{Labadain-Avaliadór} Building}
Since no test collection exists for Tetun ad-hoc text retrieval, this stage focused on creating one following TREC guidelines. Native Tetun-speaking students developed queries by examining real-world search logs and the document collection sourced from the Labadain-30k+ dataset~\cite{labadain30k-dataset}. The same students also assessed the relevance of query-document pairs using a user-friendly interface we developed, with the document pooling process automated to streamline the assessment workflow. The resulting Tetun test collection is called \textit{Labadain-Avaliadór}.

\subsection{Tetun Ad-Hoc Retrieval Baseline Establishment}\label{subsec:ad-hoc-retrieval}
This stage focused on investigating various retrieval strategies for Tetun ad-hoc text retrieval. Documents and queries were initially preprocessed by converting text to lowercase, normalizing apostrophes, removing punctuation and special characters, tokenizing into individual tokens, and then performing document indexing, retrieval, and ranking to establish the baselines. Additional preprocessing steps, such as handling accented letters, apostrophes, and hyphens, were applied individually to assess their impact on retrieval effectiveness relative to the baselines. The process also included stopword removal and stemming.

The features that demonstrated improvements over the baseline were selected and combined for subsequent experiments to create the baselines. This approach was applied to both document titles (short text) and content (long text), employing various retrieval and ranking models. The effectiveness of these preprocessing steps and models was then assessed using various evaluation metrics to identify the most effective retrieval strategy for Tetun ad-hoc text retrieval.

\section{\textit{Labadain-Stopwords} Construction}\label{sec:stopwords}
This section introduces the frequency- and network-based approaches used to create \textit{Labadain-Stopwords}, a Tetun stopword list. It describes the methodology for constructing the list and compares the effectiveness of network-based methods with traditional frequency-based techniques for stopword detection.

\subsection{Overview}
Frequency- and network-based approaches were applied in the development of the \textit{Labadain-Stopwords}. Frequency-based methods, such as TF, IDF, and TF-IDF, rely on term-weighting techniques to identify frequently occurring words. TF measures the frequency of a term within a document, IDF evaluates the importance of a term by assessing how many documents in the collection it contains, and TF-IDF is the product of these metrics, representing the importance of a term within a document relative to its occurrence across the entire collection.

Network-based methods exploit the topological properties of co-occurrence networks modeled as directed graphs, including in-degree, out-degree, and degree. The in-degree represents the number of incoming connections, indicating how often a word is preceded by others. The out-degree captures the number of outgoing connections, showing how frequently a word precedes subsequent terms. The degree is defined as the sum of the in-degree and out-degree.

\subsection{Approach}
The \textit{Labadain-Stopwords} was constructed using the Labadain-30k+ dataset~\cite{labadain30k-dataset}. The process began with preprocessing steps, including lowercase, normalizing apostrophes, removing punctuation, special characters, numbers, and extra spaces, followed by tokenization using the Tetun tokenizer~\cite{de-jesus-nunes-2024-labadain-crawler} and deduplication to create a vocabulary. The traditional term-weighting techniques (TF, IDF, and TF-IDF) were then applied to the vocabulary to assign weights to each word.

To analyze network properties, we constructed a vocabulary-level co-occurrence network as a directed graph from the preprocessed text, where each word corresponds to a node. For each node, we calculated in-degree (number of incoming links), out-degree (number of outgoing links), and degree (the sum of incoming and outgoing links), thereby quantifying the connectivity of words within the network.

Using these scores, the top 1,000 words from each method were selected in descending order based on their scores to create lists of potential stopwords. These lists were then merged, with duplicates and misspelled words excluded, to produce a candidate stopword list. Two native Tetun speakers---a Ph.D.\ student and an undergraduate student---reviewed and validated this list, resulting in \textit{Labadain-Stopwords}, containing 160 Tetun stopwords~\cite{de-jesus-nunes-2025-stopwords}. The complete list with English translations is provided in Appendix~\ref{app:list-of-stopwords}.

Some stopwords appeared in misspelled forms, such as for \textit{``ne'ebé''} (meaning ``which/that'' in English) was found in variations like \textit{``nebe''},~\textit{``neebe''}, and \textit{``neebé''}. These variations were compiled into a separate list of stopword variations, which was subsequently used to develop a stopword corrector for application during the preprocessing step.

\subsection{Experiment and Evaluation}

In the network-based approach for detecting stopwords proposed by \citet{ali-etal-2024-network}, Tetun stopwords were manually translated from the English stopwords in NLTK\footnote{\url{https://www.nltk.org}} to establish the ground-truth set. In this study, we used the Labadain-30k+ dataset~\cite{labadain30k-dataset} and evaluated the effectiveness of the approach with \textit{Labadain-Stopwords}. To further assess performance across both low- and high-resource languages, we also conducted experiments with English and Portuguese.

For Portuguese and English, we used documents extracted from the CC-100 dataset~\cite{wenzek-et-al-2020} and employed stopword lists from NLTK as the ground truth. The process of assigning weights to Portuguese and English words followed the same approach used for Tetun. A summary of the datasets used to create the stopword lists is provided in \autoref{tab:summary-dataset-for-stopwords}.

\input{tables/dataset-for-stopword-detection}

For evaluation, we used precision at $n$ (P@$n$) to measure the proportion of stopwords among the top-$n$ words. While \citet{ali-etal-2024-network} limited their analysis to P@200, we extended the P@$n$ cutoff to 1,000. For this purpose, we applied intervals of approximately 25 for cutoffs up to 100 and intervals of 250 for cutoffs between 100 and 1,000.

\subsection{Results}
The results of the experiment with Tetun are presented in \autoref{tab:stops-tetun}, demonstrating that network-based approaches generally outperform traditional term weighting methods in identifying stopwords. Specifically, in-degree consistently demonstrates superior performance across most cutoffs, except at P@75, where degree slightly surpasses it. At P@10, P@25, and P@1000, in-degree and degree achieve identical performance scores. Notably, at the P@10 cutoff, all techniques perform equally well, achieving perfect precision. Among traditional term weighting methods, the results are comparable, with IDF slightly outperforming TF and TF-IDF at P@500 and P@1000.

\input{tables/stopword-precision-tetun}

When evaluated on Portuguese, similar patterns were observed, as shown in \autoref{tab:stops-portuguese}, with network-based methods again demonstrating superior performance. The degree slightly surpassed the in-degree at P@50, P@250, P@500, and P@750. At P@10, P@25, P@100, and P@1000, both in-degree and out-degree achieved identical scores. In-degree outperformed degree at P@75. At P@10 and P@25, all techniques performed equally well, achieving perfect precision. At P@1000, all methods achieved identical scores. Traditional term-weighting approaches yielded identical results across all evaluated cutoffs, though slight variations appeared at certain cutoffs when the dataset size was reduced.

\input{tables/stopword-precision-tetun-pt}

Similarly, in English, network-based approaches maintained their advantage, as shown in \autoref{tab:stops-english}. The results mirrored those of Portuguese, with degree slightly outperforming in-degree at P@50, P@250, P@500, and P@750. In-degree and degree attained identical scores at P@100, while in-degree outperformed degree at P@75. As with the other languages, all methods achieved perfect precision at P@10 and P@25 and identical scores at P@1000. As in Portuguese, traditional term-weighting methods in English yielded identical results across all evaluated cutoffs.

\input{tables/stopword-precision-english}

\subsection{Discussion}

To examine the stopword detection approaches across different levels, we divide precision into lower cutoffs (up to P@25), mid-range cutoffs (P@50 to P@100), and higher cutoffs (P@250 to P@750). At lower cutoffs, all methods yielded similar results, with network-based approaches, such as in-degree and degree, slightly outperforming traditional term-weighting methods by a small margin of +0.04 points in Tetun and Portuguese at P@25. 

In mid-range cutoffs, network-based methods maintained their advantage, surpassing traditional methods by up to +0.06 points in Portuguese, +0.07 points in English, and +0.18 points in Tetun. Among network-based methods, degree consistently outperformed in-degree at these mid-range cutoffs.

In higher cutoffs, network-based methods still outperformed traditional term weighting approaches, with in-degree consistently delivering the best results for Tetun and Portuguese, showing improvements of up to +0.18 points. However, in English, the degree marginally surpassed the in-degree. These results indicate that network-based methods maintain a stronger advantage as the stopword list expands, except in English at P@1000, where all methods produced identical scores, likely due to characteristics of the dataset.

\begin{figure}[!ht]
  \centering
  \includegraphics[width=\textwidth]{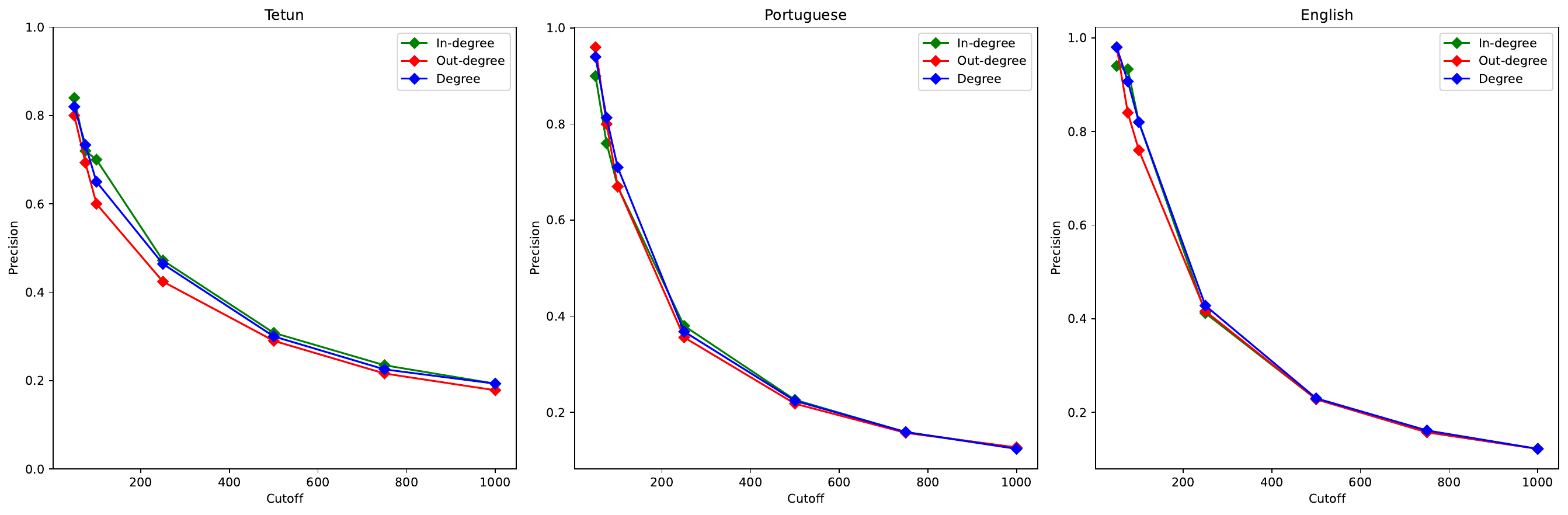}
  \caption{Comparison of the Network-Based Approach Performance at Mid-Range and Higher Cutoff Levels.}
  \Description{Comparison of the Network-Based Approach Performance at Mid-Range and Higher Cutoff Levels.}
  \label{fig:stopwords-performance}
\end{figure}

Overall, in Tetun, in-degree was slightly more effective than degree, while in English, degree marginally outperformed in-degree. In Portuguese, both methods performed similarly (see \autoref{fig:stopwords-performance}). Since the degree is defined as the sum of in-degree and out-degree and yields performance comparable to in-degree, the latter offers advantages in computational efficiency.

For traditional term-weighting approaches, IDF outperformed TF and TF-IDF at higher cutoffs in Tetun. At mid-range cutoffs, IDF and TF achieved identical scores, while at lower cutoffs, TF, IDF, and TF-IDF all produced identical scores. In contrast, for Portuguese and English, TF, IDF, and TF-IDF yielded identical results across all cutoffs. This consistency in Portuguese and English may be due to the more structured and mature linguistic resources available for these languages, such as well-established stopword lists and corpora, which minimize variations in term weighting effectiveness. In Tetun, the language's lesser-resourced nature likely results in greater sensitivity to different weighting methods, leading to performance differences at higher cutoffs.
Furthermore, the total number of stopwords we developed for Tetun is comparable to other LRLs such as Marathi (99 stopwords), Bengali (114 stopwords), and Hindi (165 stopwords)~\cite{sahu-et-al-2023}; and Kinyarwanda (80 stopwords) and Kirundi (59 stopwords)~\cite{niyongabo-et-al-2020}.

\subsection{Conclusion}

This study highlights the superiority of network-based approaches, particularly in-degree and degree, over traditional term weighting methods for stopword detection in both high- and low-resource languages, especially when dealing with larger stopword sets. Although traditional term weighting and network-based methods perform comparably at smaller cutoffs (up to 25 terms), network-based approaches demonstrate greater effectiveness as the number of evaluated terms increases. For smaller stopword lists, the differences between methods are less significant. However, when working with lists of 25 or more stopwords, network-based approaches are recommended for their superior performance at mid-range and higher cutoffs. The consistent in-degree performance observed in Tetun is aligned with the findings reported by \citet{ali-etal-2024-network}, further validating the effectiveness of network-based methods for stopword detection tasks, specifically in under-resourced scenarios.

\section{\textit{Labadain-Stemmer} Development}\label{sec:stemmer}

This section describes the development of \textit{Labadain-Stemmer}, a stemming algorithm specifically designed for Tetun. It covers the identification of Tetun affixes and the creation of \textit{Labadain-Stemmer} variants tailored to the language. Additionally, it details the process of generating a sample of words, which native Tetun speakers assessed to serve as the ground truth for evaluating the accuracy of \textit{Labadain-Stemmer}. Finally, the section presents the experiments conducted, their results, and corresponding discussion, concluding with a summary of limitations and key observations.

\subsection{Tetun Affixes}\label{subsec:tetun-affixes}

This study focuses on commonly used affixes in Tetun~\cite{klinken-et-al-2002, greksakova-2018, hull-correia-2005}, excluding circumfixes and reduplication due to their limited productivity and usage, as discussed in Subsection~\ref{tet:morphology}. The native Tetun prefixes are \textit{``ha''},~\textit{``nak''}, and \textit{``nam''}, while the native suffixes comprises \textit{``n''},~\textit{``-nain''},~\textit{``-teen''}, and \textit{``dór''}. Additionally, Portuguese-derived suffixes are adapted from the list of Portuguese suffixes used in the Portuguese stemmer in Snowball~\cite{snowball-pt-2005}, as presented in \autoref{tab:suffixes-description} in the Appendix~\ref{detail-tetun-stemmer}. Since Tetun has few inflectional affixes, stemming native Tetun words is a straightforward process that involves matching words with a predefined list of Tetun prefixes and suffixes at the beginning and end of each word.

\subsection{Stemmer Variants}
Tetun consists of both native words and a significant number of Portuguese loanwords, particularly verbs, nouns, and adjectives~\cite{hajek-klinken-2019, greksakova-2018, klinken-hajek-2018, klinken-et-al-2002}. To address this linguistic mix, the \textit{Labadain-Stemmer} is designed with three variants: \textit{light},~\textit{moderate}, and \textit{heavy}. Each variant is detailed in the following subsections.

\subsubsection{Light Stemmer}
The \textit{light} stemmer is designed to remove suffixes from Portuguese-derived words used in Tetun. This variant adapts the Portuguese stemmer from Snowball, incorporating a customized list of Portuguese suffixes. These suffixes were modified based on the loanword transformation rules defined by the INL~\cite{inl-2004-tetum-orthography}, as detailed in~\autoref{tab:portuguese-derived-transformation-rules}.

\input{tables/pt-tetun-transform-rules}

The Tetun \textit{light} stemmer is a simplified adaptation of the Portuguese stemmer, designed to handle loanwords while accounting for Tetun's unique morphological characteristics. It retains the linguistic regions utilized in the original Portuguese stemmer, which were adapted from the Spanish stemmer in Snowball~\cite{snowball-es-2005}. The definitions of these linguistic regions, as applied in the Tetun \textit{light} stemmer algorithm, are provided in~\autoref{tab:linguistic-regions} of Subsection~\ref{detail-tetun-stemmer}.

The Tetun \textit{light} stemmer processes words sequentially using a list of suffixes developed to account for the specific features of Portuguese loanwords in Tetun (see~\autoref{tab:suffixes-description}). The stemming procedure is summarized below, with the corresponding algorithm provided in Algorithm~\ref{alg:tetun-light-stemmer}:

\begin{enumerate}
    \item \textit{Word length validation}: After receiving an input word, the algorithm begins by validating its length. If the word contains fewer than four characters,  it is returned without stemming.
    \item \textit{Standard suffix removal:} For words longer than three characters, the algorithm searches for the longest matching suffix from the general suffix list. If a matching suffix is found in a specific region of the word, the suffix is deleted or replaced accordingly.
    \item \textit{Verb suffix removal:}  If no suffix is removed in step (2), the algorithm checks for verb-specific suffixes. It is removed if a matching suffix is found within the appropriate region of the word.
    \item \textit{Residual suffix removal:} If neither of the previous steps results in suffix removal, the algorithm looks at the remaining simple suffixes list and removes it as the final step.
    \item \textit{Return original word:} If none of the steps result in suffix removal, the input word is returned unchanged.
\end{enumerate}

\subsubsection{Moderate Stemmer}
The \textit{moderate} stemmer extends the functionality of the \textit{light} stemmer by handling suffixes from both Portuguese loanwords and native Tetun. It adheres to the same algorithm as the \textit{light} stemmer (outlined in Algorithm~\ref{alg:tetun-light-stemmer}), with the addition of a new step of 4.1, specifically designed to process native Tetun suffixes. This additional step is executed between steps 4 and 5 of the algorithm.

\subsubsection{Heavy Stemmer}
The \textit{heavy} stemmer builds on the functionality of the \textit{moderate} stemmer by introducing the removal of native Tetun prefixes. In this variant, the processing of native Tetun prefixes is integrated between steps 4.1 and 5 of the algorithm.

\subsection{Text Sample Construction for Evaluation}

This subsection outlines the creation of sample words that are used for intrinsic experimental and evaluation purposes. Selecting sample words to assess stemming performance poses challenges due to potential bias and limited generalization. To mitigate this, we designed a systematic methodology to construct this sample from a dataset containing a diverse collection of categories and sources, with human involvement in the loop. 

\begin{figure}[!ht]
  \centering
  \includegraphics[width=\textwidth]{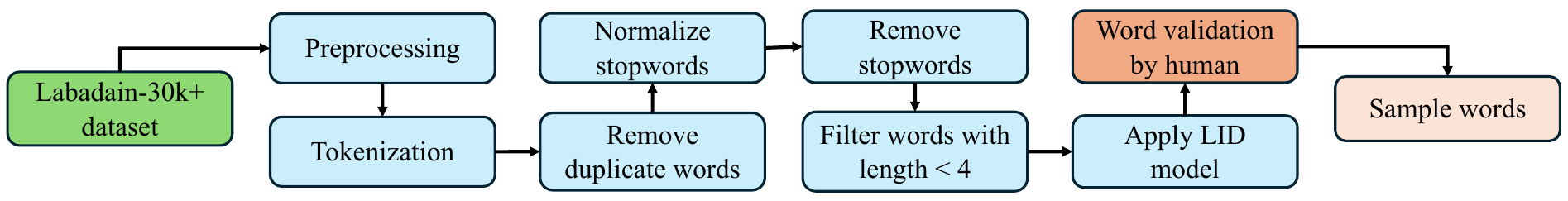}
  \caption{Process of Constructing a Text Sample for Evaluating Tetun Stemmer's Performance.}
  \Description{Process of Constructing a Collection for Evaluating Tetun Stemmer.}
  \label{fig:collection-for-stemming}
\end{figure}

The process of constructing sample words used for the experiment and evaluation is illustrated in \autoref{fig:collection-for-stemming}. First, we preprocessed the Labadain-30k+ dataset~\cite{labadain30k-dataset}, which involved lowercase, normalizing apostrophes, and removing punctuation, special characters, numbers, and extra spaces. After this, the text was tokenized into individual words using the Tetun tokenizer~\cite{de-jesus-nunes-2024-labadain-crawler}, and deduplication was performed to remove duplicate words.
This preprocessed text was then tokenized into individual words using the Tetun Word Tokenizer~\cite{dejesus-nunes-tetun-tokenizer-pypi}, and deduplication was performed to remove duplicate tokens. Stopwords were normalized and subsequently excluded, along with tokens shorter than four characters. To further refine the candidate sample words, the Tetun LID model~\cite{dejesus-nunes-tetun-lid-pypi} was applied with a threshold score of 0.95 to filter out words that did not meet this criterion. Finally, the refined candidate sample words were validated by native Tetun speakers to produce the final set of sample words.

The sample word verification with human involvement was carried out by six native Tetun speakers, consisting of one Ph.D.\ and five undergraduate students. Each student was tasked with verifying approximately 2,732 words, sorted in ascending alphabetical order. They checked the correctness of each word using both the INL dictionary~\cite{dictionary-tetun-inl-2005} and the Portuguese loanword dictionary~\cite{greksakova-2018} as reference materials. During the verification process, a considerable number of misspelled words were identified, such as the word \textit{``konsiderasaun''} (consideration, in English) appearing as [~\textit{konsidersaun},~\textit{konsiderasaunn},~\textit{konsideransaun} ]. Additionally, some words originating from Tetun Terik, Tetun DIT, or other variants not present in the reference dictionaries were excluded from the final sample. A summary of the resulting sample of words from this process is presented in~\autoref{tab:words-candidate-preprocessing-summary}.

\input{tables/summary-word-sample}

\subsection{Ground Truth Development}
The development of the ground truth set for evaluating the \textit{Labadain-Stemmer} performance was carried out by the same six Timorese students. To familiarize the assessors with the evaluation process, five example pairs from the sample of 1,839 words (see \autoref{tab:words-candidate-preprocessing-summary}) were provided during the training session. These pairs included the original words and their corresponding stemmed forms generated by each \textit{Labadain-Stemmer} variant. After this initial training, the complete list of input words and their stemmed results was distributed to two students per stemmer variant for evaluation. Their primary task was to determine whether each word correctly stemmed to its root form. When a word was incorrectly stemmed, the students provided the correct root form, using the suffixes detailed in \autoref{tab:suffixes-description} for Portuguese-derived words and the Tetun affixes described in Subsection~\ref{subsec:tetun-affixes} to guide their decisions.

Inter-annotator agreement was calculated to ensure consistency and reliability among the annotators. Discrepancies between annotators were analyzed and discussed, allowing them to reach a consensus for each stemmer variant. This procedure was followed by all annotators during the evaluation of the different stemmer variants. Inter-annotator agreement was measured using Cohen's kappa, as presented in \autoref{tab:cohen-k-creating-ground-truth}. In the final stage, the annotators pooled their evaluations and resolved any remaining discrepancies to finalize the correct stemmed forms. These consensus-based results were then used to compile the ground truth set, which is summarized in \autoref{tab:ground-truth-summary}.

\input{tables/cohens-k-score-ground-truth}

\input{tables/summary-ground-truth}

\subsection{Intrinsic Evaluation}

In intrinsic evaluation, the Paice metric~\cite{paice-1994} was used to evaluate the stemmer variant quality by measuring how effectively they reduce various word forms to a common root. This metric balances understemming and overstemming effects, both of which impact precision and recall in text-processing tasks. In IR, a high understemming lowers recall, resulting in relevant documents not being retrieved, while a high overstemming hurts precision by retrieving many irrelevant documents.

Paice introduced four intrinsic methods to assess stemming performance: understemming index (UI), overstemming index (OI), stemming weight (SW), and error rate relative to truncation (ERRT). UI measures how often the stemmer fails to reduce related words to the same root, while OI calculates the frequency of incorrectly merging unrelated words into the same root. SW is the ratio of OI/UI, representing the trade-off between overstemming and understemming. A lower value of SW indicates more understemming, whereas a higher value suggests a tendency toward overstemming. ERRT evaluates the stemmer’s ability to balance understemming and overstemming. This involves computing UI and OI values for various truncation lengths to establish a \textit{truncation line}, which serves as a baseline for stemmer performance. Any reasonable stemmer should have its (UI, OI) point located between the truncation line and the origin, with better performance indicated by a position further away from the truncation line or closer to the origin.

\subsection{Experimental Setting}

To compute the Paice metric, a list of words is first organized into conceptual groups based on semantic and morphological relationships. These groups serve as the target, and an ideal stemmer should conflate words according to these conceptual groupings. The stemmers were then applied to the word list, and their performance was evaluated by measuring how accurately they matched the predefined conceptual groups. Examples of these conceptual groupings are provided below, where the root word is shown on the left side and its corresponding conflated words are listed on the right side, separated by a colon delimiter.

\begin{verbatim}
    'ajente': ['ajénsia', 'ajénsias']
    'akompañ': ['akompaña', 'akompañadu', 'akompañamentu', 'akompañante']
    'akontes': ['akontese', 'akontesimentu', 'akontesimentus']
    'hatete': ['hatete', `hateten']
    'kbiit': ['kbiit-laek', 'kbiit-na'in', 'kbiit']
    'komunik': ['komunikadu', 'komunikadus', 'komunikadór', 'komunikasaun', 'komunikativa']
    'otél': ['otél']
\end{verbatim}

The application of the stemmer to conceptual groups resulted in understemming, overstemming, and the relative accuracy of the stemmers, represented by ERRT. To calculate ERRT, a baseline was established using length truncation, where the words in the list were truncated to their first $n$ letters, with $n$ set to 7, 8, and 9. The overstemming and understemming measures of these truncated lists define the truncation line.

\subsection{Evaluation and Results}

\begin{figure}[!ht]
  \centering
  \includegraphics[width=0.9\textwidth]{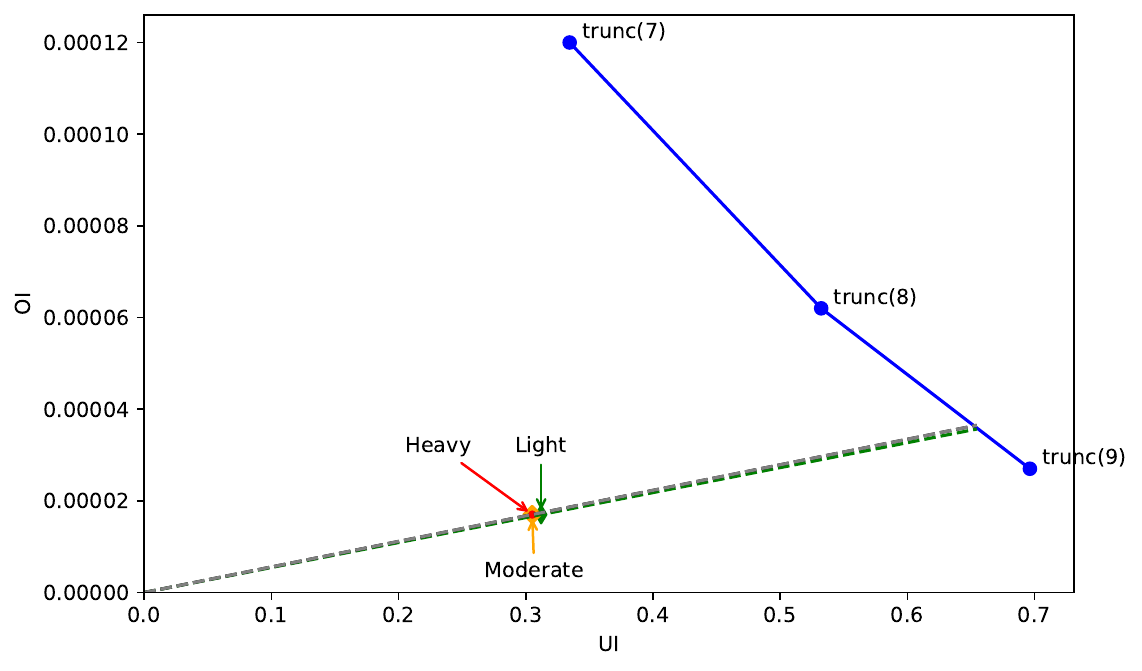}
  \caption{UI vs. OI Plot Showing ERRT Distances.}
  \Description{UI vs. OI Plot Showing ERRT Distances.}
  \label{fig:errt-distance}
\end{figure}

Using the Paice metric, we calculated the ERRT value for each stemmer variant by drawing a line from the origin through the point representing its understemming and overstemming indexes (UI, OI) and extending it to intersect the truncation line. The ERRT is calculated by dividing the distance from the origin to the (UI, OI) point by the distance from the origin to the truncation line intersection.  An ideal stemmer variant has low UI and OI values, indicating better performance when positioned closer to the origin or further away from the truncation line. Figure~\ref{fig:errt-distance} presents the UI and OI values for each stemmer with the truncation line, showing that the \textit{heavy} and \textit{moderate} stemmer variants slightly outperformed the \textit{light} variant.

The UI, OI, SW, and ERRT values are presented in Table~\ref{tab:stemming-performance-paice}. As expected, the \textit{light} variant exhibits the highest understemming value (by +0.007 points), while the \textit{moderate} and \textit{heavy} variants have identical lowest ERRT values (both lower by approximately -0.009 points), and all variants have the same overstemming values. Further investigation revealed that the difference in words stemmed from the \textit{heavy} variant compared to the \textit{moderate} variant was limited to only six words. This is due to the small proportion of native Tetun words, which make up only 18.21\% of the total (see \autoref{tab:ground-truth-summary}), and the limited number of Tetun prefixes (outlined in Subsection~\ref{subsec:tetun-affixes}).

\input{tables/stemming-performance-paice}

Regarding the overstemming value, since native Tetun has few inflectional forms in the word list, applying both the \textit{moderate} and \textit{heavy} variants had no impact on the overstemming value. The minimal difference in the ERRT values presented in Table~\ref{tab:stemming-performance-paice} indicates that the \textit{light},~\textit{moderate}, and \textit{heavy} stemmer variants perform quite similarly.

\subsection{Discussion}

Tetun, as a language with relatively few inflectional affixes~\cite{greksakova-2018,  hull-correia-2005, klinken-et-al-2002}, often includes short affixes, such as the prefix \textit{``ha''} and the suffix \textit{``n''}, which present challenges to stemming algorithms in correctly handling native Tetun affixes. Some verbs and nouns begin or end with these characters, though they are not affixes, as seen in words like ``\textit{\underline{ha}limar''} (play), ``\textit{\underline{ha}riis''} (bathe), ``\textit{ama\underline{n}''} (father), ``\textit{ina\underline{n}''} (mother), ``\textit{ibu\underline{n}''} (mouth), ``\textit{lima\underline{n}''} (hand), ``\textit{ulu\underline{n}''} (head), among others. Additionally, removing these characters from certain words changes their meaning. For example, removing \textit{``ha''} from ``\textit{\underline{ha}limar''} (play) results in \textit{``limar''} (rasp), and removing the suffix \textit{``n''} from \textit{``lima\underline{n}''} (hand) becomes \textit{``lima''} (five).

Given that three variants of the \textit{Labadain-Stemmer} show similar performance, it might be affected by the characteristics of the sample of words used for evaluation. Factors such as the proportion of native Tetun words, the presence of affixes, and the term distribution within the sample could influence the stemming algorithms' outcomes. Developing a more balanced sample of word composition could provide deeper insight into the observed results.

\subsection{Limitations}

The \textit{Instituto Nacional de Linguística} (INL) launched the \textit{Kursu Gramátika Tetun} (Tetun Grammar Course) in 2005, which served as a reference for teachers, translators, journalists, and students~\cite{hull-correia-2005}. It includes several Tetun affixes, such as prefixes [~\textit{hak, na, ma} ] and suffixes [~\textit{-laek, k} ]. However, more recent research by ~\citet{greksakova-2018} in 2018 reported that many of these affixes have been largely replaced by words such as \textit{``sai''} (meaning ``become'') and \textit{``laiha''} (meaning ``without''), making these affixes less productive in Tetun. As the INL has not updated its 2005 publication on Tetun grammar, the current state of Tetun morphology remains unclear.

Furthermore, the absence of linguistic experts in this study, due to the lack of funding to hire linguists, represents a limitation. Nevertheless, we have established a baseline that can serve as a foundation for future research in Tetun.

\subsection{Conclusion}
This study developed and assessed the effectiveness of the \textit{Labadain-Stemmer}, incorporating suffixes of Portuguese loanwords and the affixes of native Tetun words. The Tetun affixes used were based on those commonly reported by Klinken et al.~\cite{klinken-et-al-2002}, the INL~\cite{hull-correia-2005}, and \citet{greksakova-2018}. To evaluate stemmer performance, we systematically constructed sample words and established a baseline for \textit{Labadain-Stemmer}, testing three variants (light, moderate, and heavy). Results showed that integrating native Tetun affixes into the stemming process was marginally more effective than focusing solely on the suffixes of Portuguese loanwords.

However, one of the limitations in this study is the unbalanced representation of Portuguese loanwords and native Tetun words in the sample set. Future research should address this by using more balanced datasets that adequately represent both Portuguese loanwords and native Tetun words. Furthermore, the involvement of expert linguists specializing in Tetun will be crucial for enhancing the accuracy and overall effectiveness of the \textit{Labadain-Stemmer}. To enable reproducibility, the stemmer algorithms have been released under the MIT License~\cite{labadain-stemmer-algorithm}, to encourage further research and development in Tetun information retrieval.

\section{\textit{Labadain-Avaliadór} Building}\label{sec:test-collection}
This section provides an overview of the test collection and details the process of constructing \textit{Labadain-Avaliadór} (\textit{avaliadór}, a Tetun word meaning ``evaluator''), a Tetun test collection for evaluation. It covers the dataset used, query formulation, document pooling, and relevance judgments.

\subsection{Overview}
The effectiveness of information retrieval systems relies on the availability of reliable test collections for evaluation. The traditional approach to building such collections follows the Cranfield paradigm~\cite{cleverdon-1976}, which is widely adopted through the TREC evaluation campaigns~\cite{harman-et-al-1992}. A TREC-style test collection typically comprises three core components: a document collection, a set of information needs (or topics), and relevance judgments.

\subsection{Documents}
\begin{figure}[!ht]
  \centering
  \includegraphics[width=0.95\linewidth]{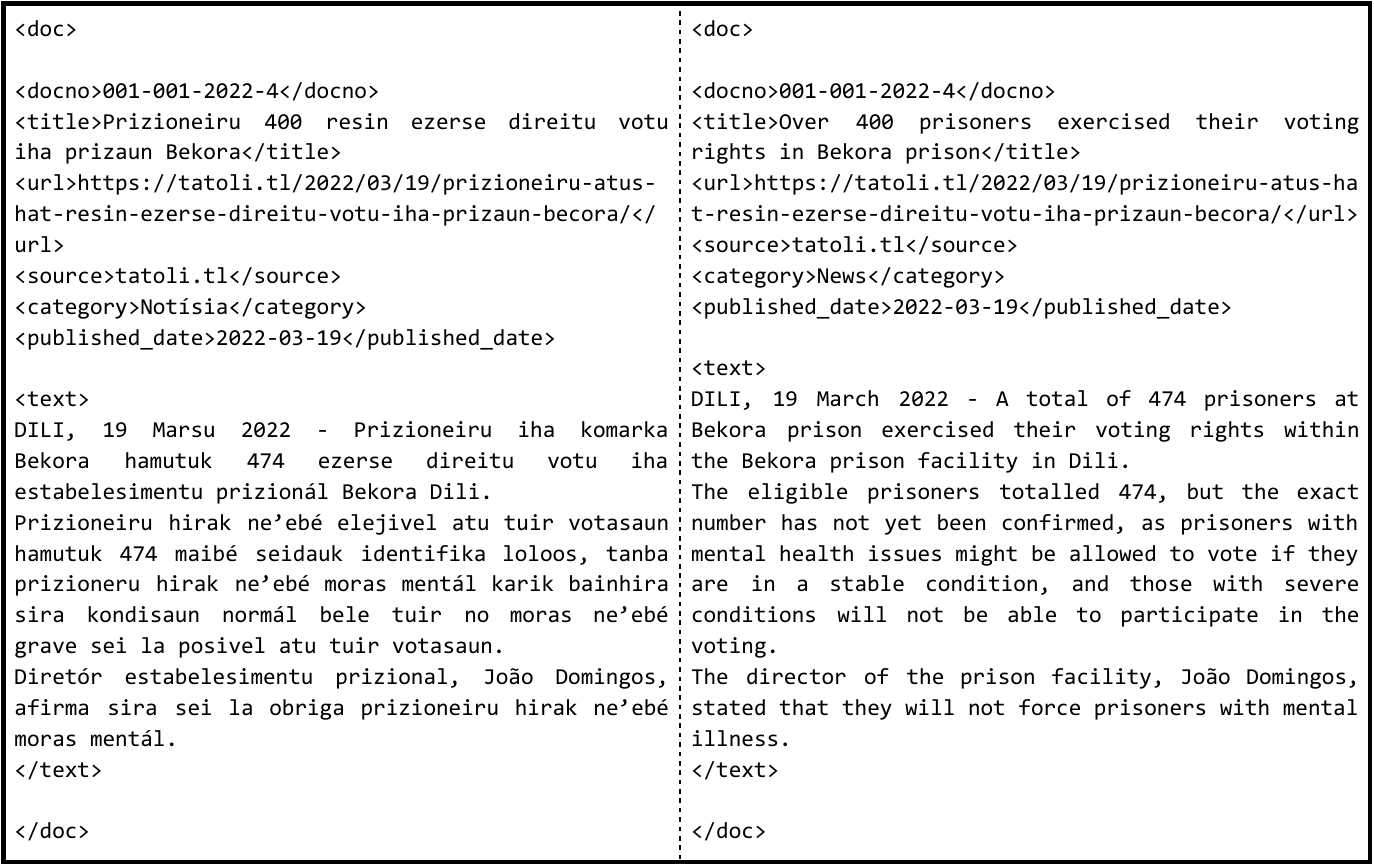}
  \caption{Sample of Document Formatted Following TREC Guidelines: Original (left) and English translation (right).}
  \Description{Sample of Document Formatted Following TREC Guidelines.}
  \label{fig:document}
\end{figure}

\begin{figure}[!ht]
  \centering
  \includegraphics[width=0.9\linewidth]{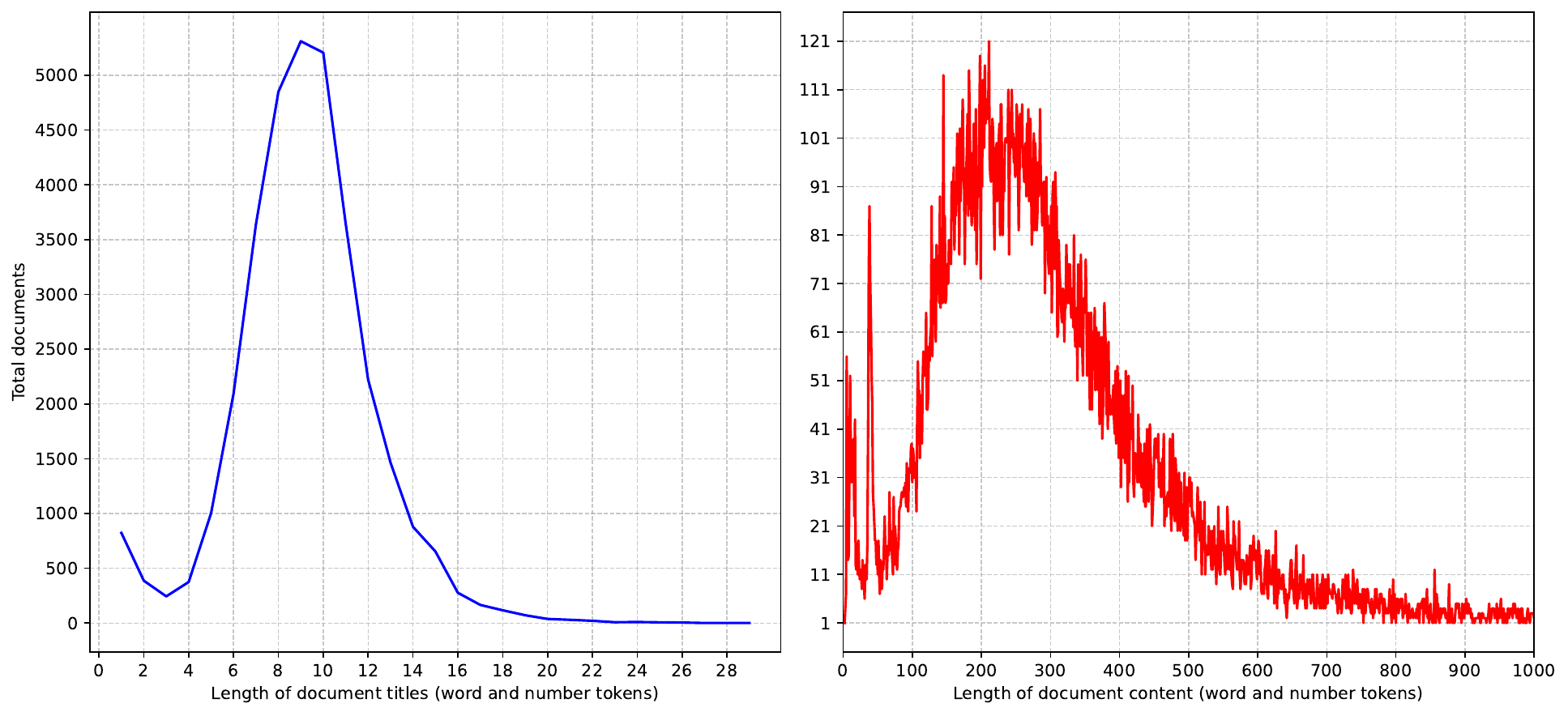}
  \caption{Length Distribution of Titles and Body Content in the Labadain-30k+ Dataset. In the right-hand figure, the x-axis of the document length distribution is limited to 1,000 tokens (words and numbers) for improved visualization.}
  \Description{Length Distribution of Titles and Body Content in the Labadain-30k+ Dataset.}
  \label{fig:document-distribution}
\end{figure}

The document collection comprises 33,550 Tetun documents sourced from the Labadain-30k+ dataset~\cite{labadain30k-dataset}, each enriched with metadata such as title, URL, source, publication date, and content. This dataset was collected from web crawling and covers a wide range of categories, including news articles, Wikipedia entries, legal and government documents, research papers, technical documents, blogs, forums, and more~\cite{de-jesus-nunes-2024-labadain}. The diversity of its sources and topics makes this dataset particularly suitable for constructing a test collection for Tetun. A sample of the documents, formatted according to TREC guidelines, is shown in \autoref{fig:document}. The collection is 84 MB in size, with approximately 12.3 million tokens and 162,466 unique tokens. A summary of the collection is provided in \autoref{tab:docs-summary}, and the length distribution of titles and content is illustrated in \autoref{fig:document-distribution}.

\input{tables/doc-collection-summary}

\subsection{Query Formulation}\label{subsec:query-formulation}
Queries were collected from two sources: Google Search Console\footnote{\url{https://search.google.com/search-console}} for Timor News and the user search logs from the Timor News platform.\footnote{\url{https://www.timornews.tl}}
The Google Search Console queries cover the period from November 1, 2021, to January 31, 2024, while the search logs from Timor News span from May 7, 2021, to January 31, 2024. Timor News is an online news agency based in Dili, Timor-Leste, founded in May 2019 and launched its news portal on May 7, 2019. The platform registered an average of 1,400 unique visitors per day and exclusively publishes news in Tetun.

The collected queries were compiled and distributed among five second-year undergraduate volunteers, all native Tetun speakers from Timor-Leste. The group comprised two students from Environmental Engineering, two from Information Systems, and one from Medicine. These students were tasked with developing queries following the established guidelines.
Initially, each student was assigned 250 queries, resulting in a total of 1,250 queries analyzed. To understand user information needs, the students reviewed the provided search logs and either retained, modified, or formulated new queries based on the contextual information interpreted from the logs (see examples in \autoref{tab:query-formulation}).

\input{tables/query-formulation}

Before beginning query development, the students attended a training session that provided practical examples of query formulation. Following this, three pilot testing sessions were conducted, during which each student created a query, defined the associated information need, and described the types of documents they would consider relevant.

The queries were then entered into a search prototype\footnote{\url{https://www.labadain.tl}} built on top of Apache Solr\footnote{\url{https://solr.apache.org}} using the BM25 ranking model. This prototype allowed the students to analyze the documents retrieved for each input query. For each query, each student selected four documents from the top 50 retrieved list, ensuring that one document represented each category: non-relevant, marginally relevant, relevant, or highly relevant. These sessions facilitated discussions about the results, highlighted challenges faced, and provided feedback to deepen their understanding of query development.

\begin{figure}[!ht]
  \centering
  \includegraphics[width=0.95\linewidth]{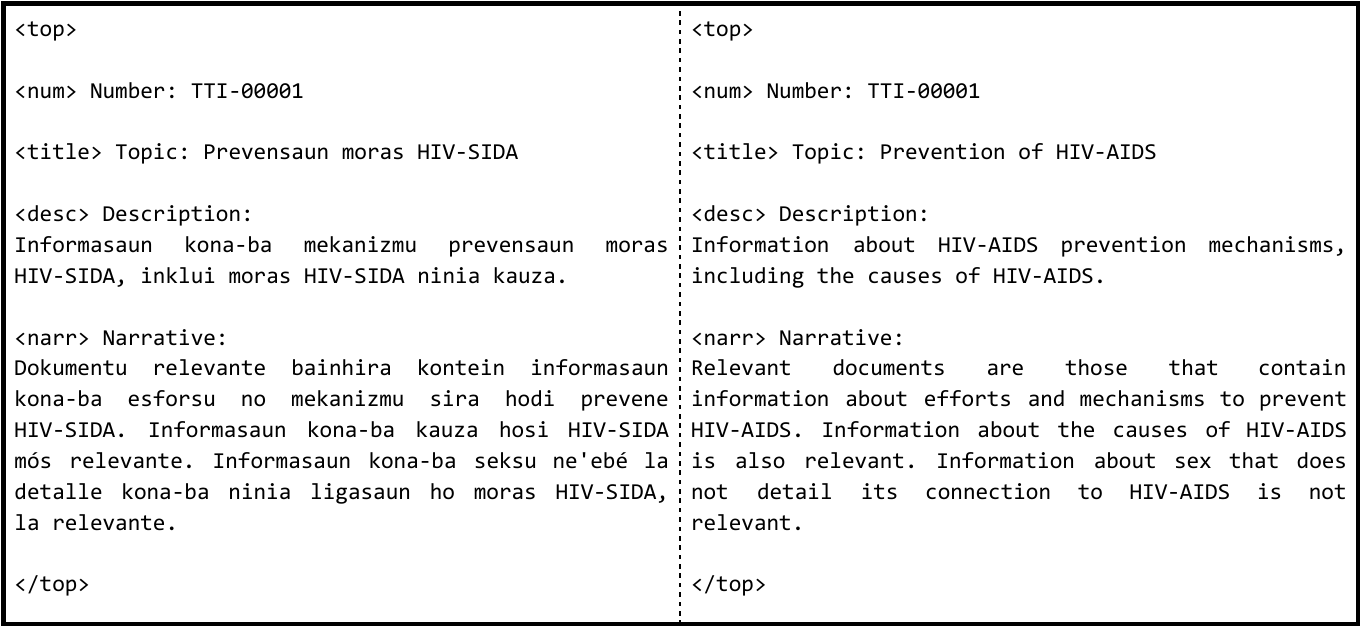}
  \caption{Sample of Topic Formatted According to TREC Guidelines: Original (left) and English translation (right).}
  \Description{Sample Topic Formatted According to TREC Guidelines: Original (left) and English Translation (right).}
  \label{fig:topic}
\end{figure}

\begin{figure}[!ht]
  \centering
  \includegraphics[width=0.75\textwidth]{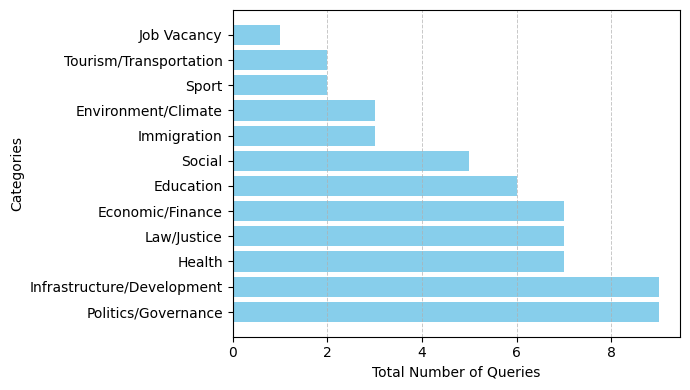}
  \caption{Distribution of Queries Over Categories.}
  \Description{Distribution of Queries Over Categories.}
  \label{fig:query-categories}
\end{figure}

Subsequently, students were tasked to develop short queries following TREC best practices~\cite{sanderson-2010}, ranging from three to five words, specifying information needs and describing the types of documents they expected the system to retrieve. Using the search prototype mentioned earlier, they input their queries and analyzed the retrieved documents. To finalize each query, they ensured that at least five relevant documents were identified. In total, 61 queries were developed. A sample query (or \textit{topic}), formatted according to TREC guidelines, is shown in \autoref{fig:topic}. A summary of the queries is provided in \autoref{tab:queries-summary}, and their distribution across categories is shown in \autoref{fig:query-categories}. For categorization, we adapted the query topic categorization frameworks of \citet{beitzel-et-al-2004} and \citet{rohatgi-et-al-2021}.

\input{tables/query-summary}

\subsection{Document Pooling}
Since Query-document relevance judgments are carried out by human assessors, it is not feasible to evaluate every document in a large collection. To address this challenge, \citet{sparck-jones-rijsbergen-1975} introduced the \textit{pooling} technique, in which a small subset containing a sufficiently representative sample of relevant documents is selected from the larger collection and provided to human assessors for relevance judgments~\cite{sanderson-2010}. 

Given the limited availability of retrieval models and techniques for LRLs, and to maximize the retrieval of relevant documents for constructing a robust test collection, we created the document pool using two retrieval models: BM25 and a language model (LM) with Dirichlet smoothing. BM25 is widely recognized for its effectiveness in ad-hoc retrieval~\cite{robertson-zaragoza-2009}, while the LM with Dirichlet smoothing has been shown to perform particularly well on short queries~\cite{zhai-lafferty-2001}. To balance the contributions of these models, we applied the balanced interleaving technique~\cite{radlinski-et-al-2008} to merge their results into a pool, which was then presented to assessors for evaluation.

\begin{figure}[!ht]
  \centering
  \includegraphics[width=0.9\textwidth]{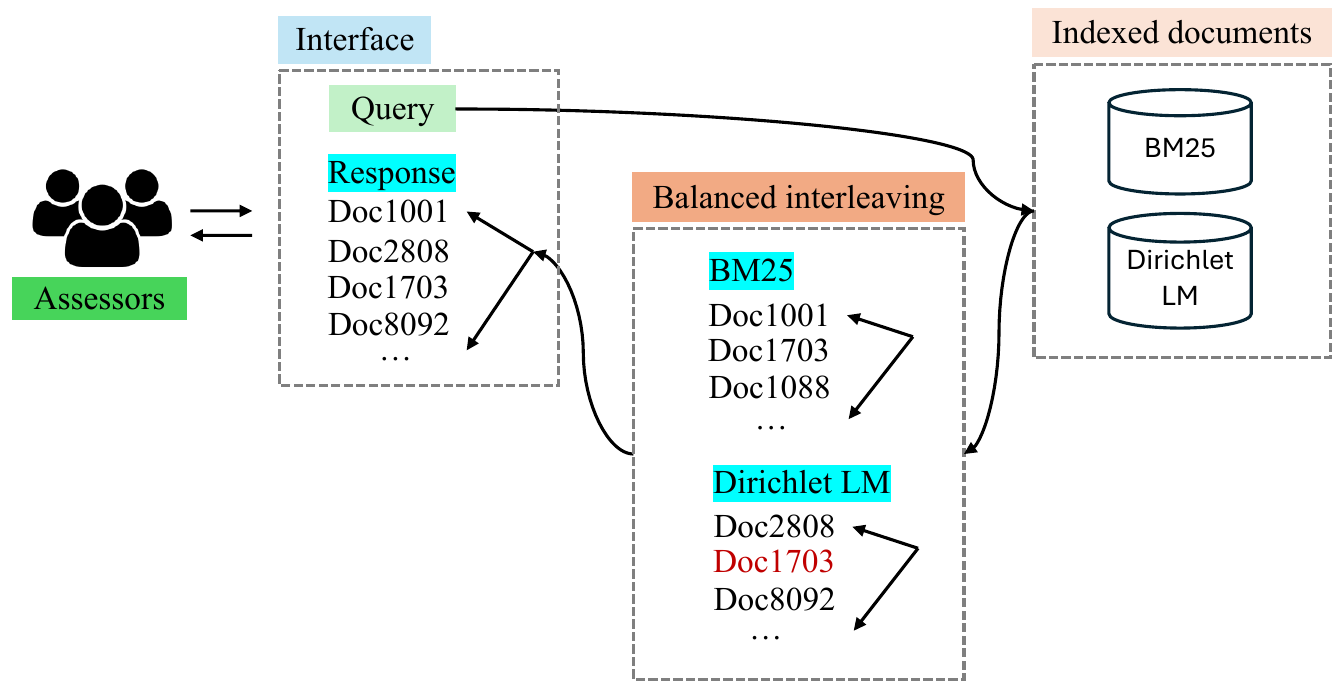}
  \caption{General Architecture of the Retrieval System used for Relevance Judgments. The document highlighted in \textit{red} indicates a duplicate that was excluded from the final list.}
  \Description{General Architecture of the Retrieval System used for Relevance Judgments. The document highlighted in \textit{red} indicates a duplicate that was excluded from the final list.}
  \label{fig:assessment-architecture}
\end{figure}

Documents were indexed in separate instances of Solr, each configured with either the BM25 or the Dirichlet LM ranking models. The document retrieval and pooling process was fully automated and integrated with a relevance assessment interface to streamline the workflow.
When a query was received, the system retrieved candidate documents ranked by each of the two models, merged the results into a pool, and presented the top 100 documents to the assessors for relevance judgments. The architecture of the retrieval system used for these assessments is shown in \autoref{fig:assessment-architecture}.

\subsection{Relevance Judgment}

\begin{figure}[!ht]
  \centering
  \includegraphics[width=\linewidth]{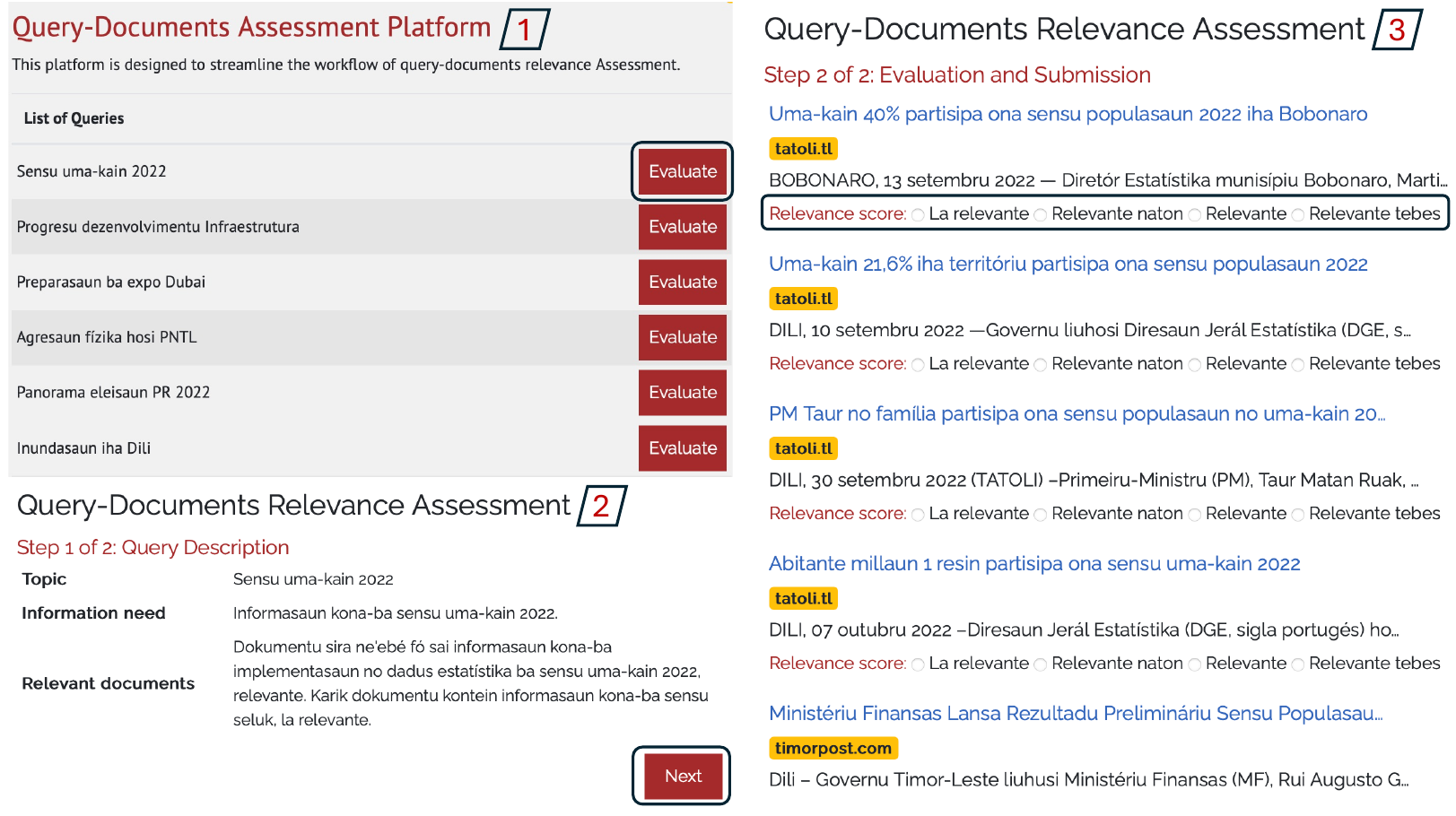}
  \caption{Web Interface Used by Human Assessors for Conducting Relevance Assessments.}
  \Description{Web Interface Used by Human Assessors for Conducting Relevance Assessments.}
  \label{fig:assessment-platform}
\end{figure}

Five native Tetun-speaking students who developed the queries conducted relevance judgments for the query-document pairs, categorizing them into four graded levels of topical relevance: non-relevant, marginally relevant, relevant, and highly relevant, as proposed by \citet{sormunen-2002}, following the guidelines outlined in Subsection~\ref{subsec:query-formulation}. A user-friendly web interface was created to streamline the process, allowing assessors to log in with individual accounts to conduct assessments. The interface used for the evaluation is illustrated in \autoref{fig:assessment-platform}.

Once logged in, the assessors began by selecting a query (label 1 in \autoref{fig:assessment-platform}) and reviewing the associated information needs and relevance criteria (label 2). They then evaluated each of the 100 documents, assigning a relevance score to each, and submitted their judgments (label 3). Upon submission, the system redirected the assessors to the homepage, displaying a list of queries. The option to reassess previously evaluated queries was automatically disabled.

\begin{figure}[!ht]
  \centering
  \includegraphics[width=0.9\linewidth]{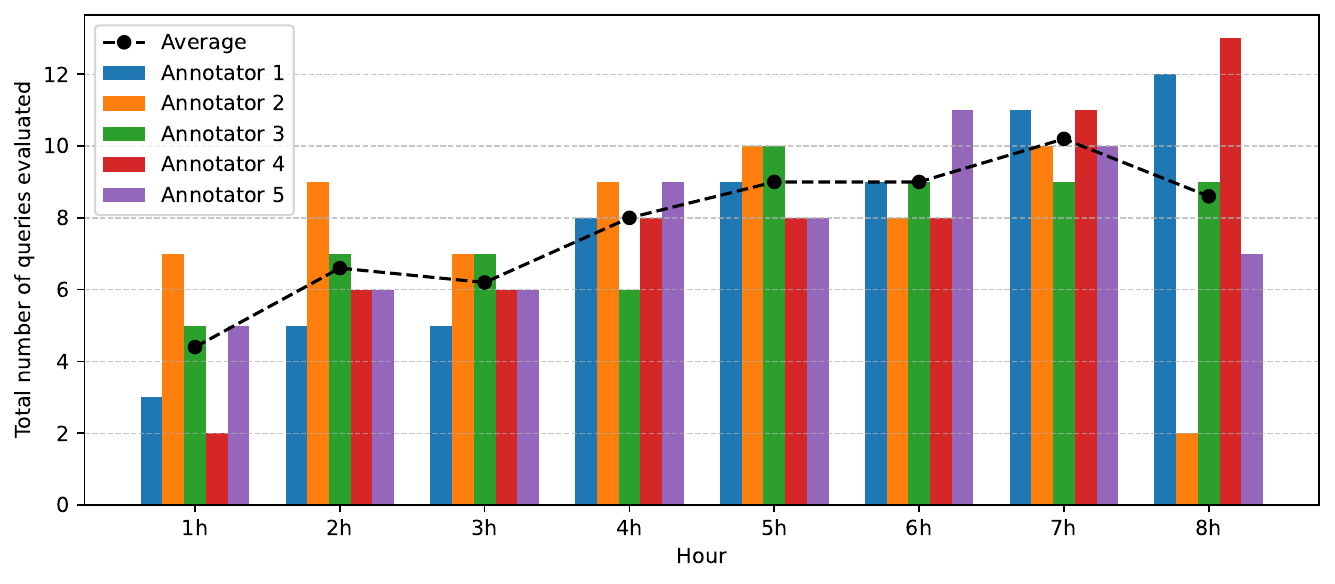}
  \caption{Hourly Statistics of the Total Number of Queries Evaluated by Each Annotator.}
  \Description{Hourly Statistics of Queries Evaluated per Annotator.}
  \label{fig:statistics-annotators-per-hour}
\end{figure}

Each assessor evaluated the same set of 61 queries, with 100 documents to assess per query, resulting in a total of 6,100 documents being assessed. Assessors were instructed to focus on topical relevance, check document details if the query was too long and not fully displayed on the interface, and disregard the retrieval order when determining the relevance of the document for the given query. The assessment process was completed within eight hours, and the number of queries assessed by each annotator per hour is illustrated in \autoref{fig:statistics-annotators-per-hour}. Documents not included in the judgment list were considered non-relevant. 

To assess inter-annotator reliability, we used Cohen's kappa measure~\cite{cohen_1960}, interpreting the strength of agreement according to the scale provided by \citet{landis-koch-1977}. The overall average score of inter-annotator agreement among the five annotators is 0.4236, indicating \textit{moderate} agreement, with detailed results presented in \autoref{tab:cohen-k-agreement-annotators}. 

\input{tables/cohens-k-assessors}

Since all assessors evaluated the same queries, a majority voting approach was applied to determine each document's relevance to its corresponding query, based on the scores assigned to the query-document pairs. According to the majority voting rule, the most frequently chosen label for each query-document pair must exceed 50\% of the total votes to qualify as the majority label~\cite{snow-et-al-2008}, meaning that at least three annotators must select the same label. The evaluation results revealed that, out of 6,100 documents assessed, 9.87\% did not meet this threshold or resulted in ties (e.g., two groups of annotators selected different labels, such as annotators 1 and 2 choosing score 3, annotators 3 and 4 choosing score 1, and the fifth annotator selecting score 0). Details of these ties are presented in \autoref{tab:tied-documents}.

\input{tables/tied-score-summary}

An approach to addressing tied scores is to use the tie-breaker strategy, which suggests that using a strong signal to break ties is more effective than a weak one~\cite{wu-fang-2013}. By applying this tie-breaker method to all instances of tied scores, we obtained the results shown in \autoref{tab:assessment-details} (referring to the 1st round). However, after conducting an in-depth analysis of the tied scores, we observed significant discrepancies in some cases, such as ties between scores of 0 and 2 or 1 and 3. To resolve these inconsistencies, we re-invited three of the five original assessors for a second round of evaluations on the tied documents. During this phase, assessors were presented with the two tied score options from the initial assessment. The reassessment was conducted using Microsoft Excel, with separate tabs for each query and its corresponding documents, and was completed in approximately one hour and 15 minutes.

\input{tables/human-judgment-results}

After completing the second round, we merged the evaluation results with those from the first round and applied a majority voting method, selecting the most frequent score for each query-document pair as the final relevance score. To ensure reliability, queries with 100 or more relevant documents or fewer than ten relevant documents were excluded~\cite{sanderson-2010, aleahmad-et-al-2009}, resulting in the exclusion of two queries with ten or fewer relevant documents in the second round.

\input{tables/final-test-collection-summary}

\begin{figure}[!ht]
  \centering
  \includegraphics[width=0.9\linewidth]{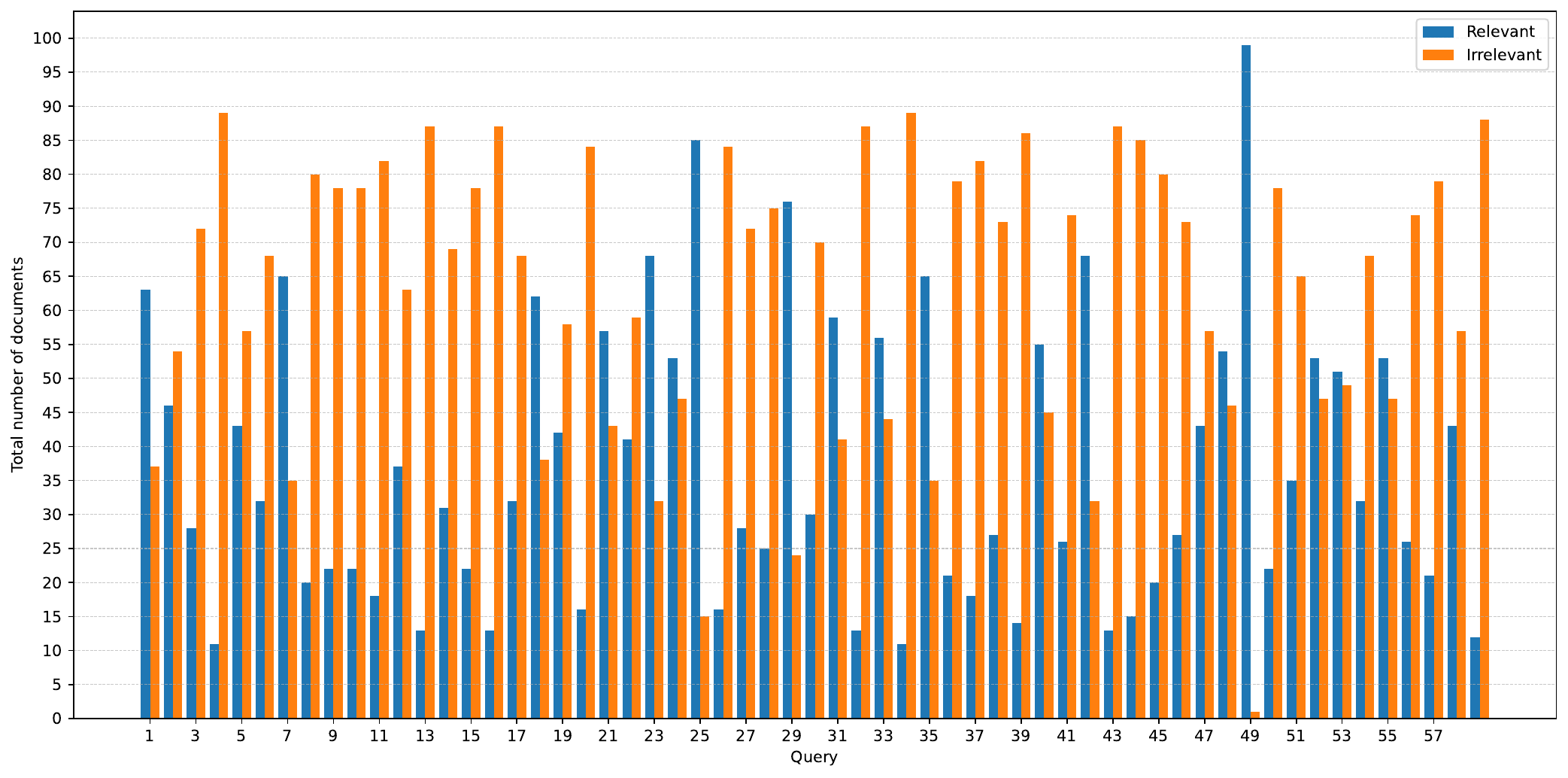}
  \caption{Total Number of Relevant and Non-relevant Documents per Query. Relevant documents consist of marginally relevant, relevant, and highly relevant.}
  \Description{Total Number of Relevant and Non-relevant Documents per Query. Relevant documents consist of marginally relevant, relevant, and highly relevant.}
  \label{fig:statistics-rel-docs-per-topic}
\end{figure}

The final test collection, called \textit{Labadain-Avaliadór}~\cite{de-jesus-nunes-2025-avaliador}, contains an average of 36.76 relevant documents per query, detailed in \autoref{tab:final-test-collection}---comprising 9.59\% highly relevant, 17.86\% relevant, 9.31\% marginally relevant, and 63.24\% non-relevant documents---as shown in the 2nd round column in \autoref{tab:assessment-details}. The distribution of document relevance per query is illustrated in \autoref{fig:statistics-rel-docs-per-topic}.

\subsection{Results and Discussion}

\autoref{tab:assessment-details} shows that, after reassessing the tied documents from the first round results using the tie-breaker strategy, the number of highly relevant documents decreased by 144 and relevant documents by 48, while marginally relevant documents increased by 73, and non-relevant documents decreased by 81. This shift suggests that some documents initially classified as highly relevant or relevant were reclassified as marginally relevant, and some marginally relevant documents were reclassified as non-relevant. Additionally, 200 documents from two excluded queries were removed, indicating a change in relevance interpretation after resolving the ties. This reclassification adjusted the distribution of documents across categories, enhancing the overall quality of the test collection.

When analyzing the changes in document relevance, \textit{annotator 2} exhibited a distinct pattern, as illustrated in~\autoref{fig:statistics-annotators-per-hour}. This annotator judged seven queries (700 documents) in the first hour at an average rate of 5.14 seconds per query-document pair, significantly faster than the average of approximately 8.18 seconds per pair observed among other annotators, who assessed an average of 4.4 queries (440 documents) in the first hour. This fast pace persisted until the seventh hour, leaving only two queries for the eighth hour. Additionally, \autoref{tab:cohen-k-agreement-annotators} shows that \textit{annotator 2} had only fair agreement with three other annotators (2, 3, and 4), which contributed to increased discrepancies. These observations suggest that the quality of annotations from \textit{annotator 2} was lower, probably due to the speed of the evaluation process, which affected the changes in the relevance of the document observed after the second round of evaluations.

\input{tables/compare-lab-avaliador-other-lrls}

Finally, \autoref{tab:comparison-test-collection-lrls} compares the \textit{Labadain-Avaliadór} to other LRL collections. The \textit{Labadain-Avaliadór} offers a balanced combination of scale and relevance density, making it a valuable resource for ad-hoc text retrieval in low-resource settings like Tetun. With 59 topics and 33,550 documents, it provides a moderately sized corpus that surpasses smaller collections like Sanskrit and Citumbuka but is smaller than larger collections like Hamshahri. Additionally, the diversity of its topics (see \autoref{fig:query-categories}) ensures its suitability for ad-hoc retrieval tasks.

The \textit{Labadain-Avaliadór} has high relevance density with an average of 36.76 relevant documents per topic, the highest among the collections analyzed. This level of relevance provides a robust foundation for retrieval experiments, supported by a substantial pool of annotated relevance judgments that enable precise and reliable evaluations. Compared to collections with size variations of up to 50,000 documents, such as Sanskrit and Chichewa, the \textit{Labadain-Avaliadór} stands out as a well-annotated collection, particularly suited for ad-hoc text retrieval in LRL contexts.

\subsection{Conclusion}

This study describes the development of \textit{Labadain-Avaliadór} following TREC guidelines. Five native Tetun-speaking students conducted both the query development and the query-document relevance assessment. The queries were derived from real-world search logs in Tetun, and the test collection was graded on a scale from zero to four. The results indicate that the assessors agreed on the relevance of more than 90\% of the query-document pairs assessments, with an average inter-annotator agreement of Cohen's kappa score of 0.4236, indicating moderate agreement.

Approximately 10\% of the 6,100 query-document pairs showed disagreement, resulting in tied scores. To resolve these discrepancies, three of the five assessors conducted a second evaluation with only two scoring options for each tied case. This process produced the \textit{Labadain-Avaliadór}, containing 5,900 \textit{qrels}, with 9.59\% highly relevant documents, 17.86\% relevant documents, 9.31\% marginally relevant documents, and 63.24\% non-relevant documents.

\section{Indexing, Retrieval, and Ranking}\label{sec:indexing-retrieva-ranking}
This section presents the experiments conducted on Tetun ad-hoc text retrieval, providing a detailed description of the retrieval and ranking strategies and the steps involved in text preprocessing, indexing, retrieval, and ranking.

\subsection{Overview}
The inverted index is one of the most widely used techniques in IR~\cite{baeza-yates-netos-2011}. It is a word-oriented mechanism that indexes all distinct words in the collection, pointing each word to a list of documents in which it appears. This full-text indexing allows direct access to each matching term and its position within the documents.

Studies in ad-hoc text retrieval have demonstrated the effectiveness of various retrieval and ranking models. TF-IDF serves as a foundational term-weighting scheme in IR~\cite{baeza-yates-netos-2011}, and BM25 is a widely recognized probabilistic model known for its effectiveness in classical IR~\cite{robertson-zaragoza-2009}. Similarly, the probabilistic language model (LM) with Dirichlet smoothing~\cite{mackay-peto-1995} has shown strong performance, particularly for short queries in ad-hoc retrieval tasks~\cite{zhai-lafferty-2001}. 

Moreover, the Divergence from Randomness (DFR) variant of BM25 (DFR BM25) has shown competitive performance in various retrieval settings and has been demonstrated in multiple TREC experiments~\cite{wolf-et-al-2009, plachouras-et-al-2004}. The Hiemstra LM~\cite{hiemstra-kraaij-1999, hiemstra-2001} has also been reported to perform well in ad-hoc text retrieval, especially in LRL scenarios~\cite{sahu-pal-2023}. These retrieval and ranking models are used in the experiments conducted in this study.

\subsection{Text Preprocessing}\label{subsection:preprocessing}
Given the language-specific characteristics of Tetun, text preprocessing was divided into several stages as follows:

\begin{enumerate}
    \item \textbf{Standard preprocessing:} This stage included converting text to lowercase, normalizing apostrophes, removing punctuation and special characters, tokenizing text into tokens (words and numbers), filtering out words longer than 60 characters, and removing extra spaces.
    \item \textbf{Language-specific preprocessing:} To address Tetun's unique linguistic features, additional text preprocessing techniques were applied, including the removal of apostrophes, accents, and hyphens (splitting hyphen-connected words). Each of these steps was independently implemented within the preprocessing workflow.
    \item \textbf{Stopwords Removal and Stemming:} Beyond character-based preprocessing, this stage included stopword removal and stemming, with \textit{light},~\textit{moderate}, and \textit{heavy} variants of the \textit{Labadain-Stemmer} applied for stemming.
\end{enumerate}

\subsection{Experimental Setting}
\begin{figure}[!ht]
  \centering
  \includegraphics[width=0.9\linewidth]{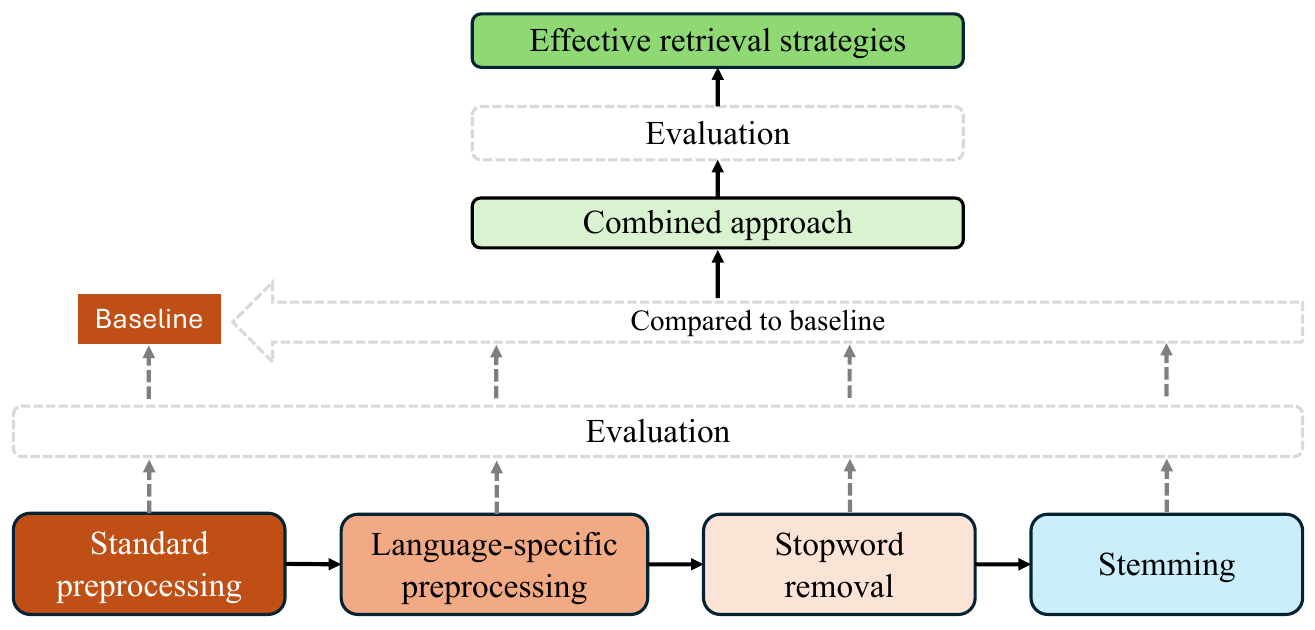}
  \caption{Overview of the Experiment Workflow.}
  \Description{Overview of the Experiment Workflow.}
  \label{fig:experimental-setting}
\end{figure}

We explored various retrieval strategies by applying multiple text preprocessing techniques to assess their impact on retrieval effectiveness. The experiment workflow is illustrated in \autoref{fig:experimental-setting}. First, we established the baseline by applying the standard preprocessing step, as detailed in Subsection~\ref{subsection:preprocessing}. Next, we tested each of the language-specific text preprocessing techniques, including stopword removal and stemming, to compare their results. Finally, we combined techniques that outperformed the baseline to further evaluate their effectiveness and determine effective strategies for Tetun ad-hoc text retrieval.

For stemming, although the \textit{moderate} and \textit{heavy} variants performed better than the \textit{light} variant in intrinsic assessment, the difference was minimal. Furthermore, \citet{flores-moreira-2016} noted that the most accurate stemmer does not always lead to the most effective retrieval, and therefore, we experimented with all stemmer variants.
We used PyTerrier~\cite{macdonald-tonellotto-2020}, a Python API for the Terrier IR platform~\cite{ounis-et-al-2005}\footnote{\url{http://terrier.org/}} for indexing, retrieval, and ranking, with the default settings maintained for each model. The same text preprocessing techniques were applied to both queries and documents, and experiments were conducted using document titles and content to evaluate retrieval effectiveness for each approach.

\subsection{Document Indexing}
Document titles and content were indexed independently, with the index created from standard preprocessing steps serving as the baseline. Additional indexes were generated for individual results obtained from applying language-specific preprocessing steps, stopword removal, and each of the stemmer variants. To evaluate the effects of these preprocessing methods on index size, index compression factors (ICF)~\cite{frakes-and-fox-2003} were calculated for each preprocessing configuration. A summary of the index compression results is presented in \autoref{tab:indexed-documents}.

The results show that all preprocessing methods generally reduce the index size compared to the baseline. Removing hyphens yielded the highest compression factor for title indexing, reducing the index size by up to 30.76\% compared to the baseline. For content indexing, the \textit{heavy} stemmer provided the highest compression, reducing the index size by up to 12.18\%. 

\input{tables/icf-comparison}

As expected, the \textit{moderate} stemmer variant compressed the index more efficiently than the \textit{light} variant, while the \textit{heavy} variant achieved the highest compression for both title and content indexing. Interestingly, removing apostrophes and accents also contributed to index size reduction, with accent removal achieving over a 7\% reduction for both titles and content. This suggests a substantial presence of identical words with and without accents in the documents.

\subsection{Short-Text Retrieval Results}
For short-text retrieval, document titles were indexed and used for retrieval. The impact of each preprocessing technique on retrieval effectiveness is presented~\autoref{tab:evaluation-preprocessing-technique-titles}. Scores highlighted in red indicate values lower than the baseline. As observed, removing accents and applying all stemmer variants did not improve retrieval effectiveness compared to the baseline.
Stopword removal yielded inconsistent results across models and metrics. It generally demonstrated improved performance only at the top-20 cutoffs (P@20, MAP@20, and NDCG@20) and overall MAP and NDCG, while exhibiting lower performance at top-5 and top-10 cutoffs across all metrics. Regarding the impact of stemming on retrieval effectiveness, the \textit{light} stemmer variant slightly outperformed the \textit{moderate} and \textit{heavy} variants across all cutoffs, with the \textit{moderate} and \textit{heavy} variants performing similarly.

\input{tables/short-text-retrieval-preliminary}

Based on the preliminary results in~\autoref{tab:evaluation-preprocessing-technique-titles}, all combinations of preprocessing techniques that outperformed the baseline, including stopword removal and the \textit{light} stemmer variant, were selected for further comparison. The best results from these combined preprocessing techniques are presented in~\autoref{tab:evaluation-titles}. The findings indicate that removing apostrophes and hyphens significantly enhances retrieval performance compared to the baseline.

Among the retrieval models, DFR BM25 consistently delivered the highest performance across all metrics up to top-10 cutoffs. Notable scores include P@5 (0.8881), P@10 (0.8390), MAP@5 (0.1589, removing stopwords), MAP@10 (0.2804), NDCG@5 (0.7512), and NDCG@10 (0.7356). DFR BM25 demonstrated relative improvements of up to 30.35\% over the baseline and up to 8.19\% over individual preprocessing techniques on multiple metrics, along with modest relative gains of up to 5.54\% in MAP@5 over other retrieval models within the same settings. While DFR BM25 achieved the same P@5 as BM25 and MAP@5 as TF-IDF, it demonstrated slightly higher scores in other cutoffs.

\input{tables/short-text-retrieval-final}

The removal of hyphens and stopwords also proved beneficial, particularly with Hiemstra LM, which demonstrated the highest scores for P@20 (0.7305), MAP@20 (0.4372), NDCG@20 (0.7152), and overall MAP (0.7040) and NDCG (0.8289, with apostrophe removal). This combination enhanced retrieval effectiveness overall, with Hiemstra LM consistently showing relative improvements at top-20 cutoffs and in global MAP and NDCG. It demonstrated relative improvements of up to 15.60\% over the baseline and up to 8.33\% over individual text preprocessing techniques in multiple metrics, with marginal relative gains of up to 3.94\% over other retrieval models within the same setting.

Stemming, however, had minimal impact on retrieval effectiveness, even when combined with other text preprocessing techniques that outperformed the baseline. The best result involving the \textit{light} stemmer variant was achieved when paired with hyphen and stopword removal. While this combination produced competitive results, it did not surpass the performance of strategies that excluded stemming.

\subsection{Long-Text Retrieval Results}
Document content was indexed and used for long-text retrieval. The impact of each preprocessing technique on retrieval effectiveness is summarized in~\autoref{tab:evaluation-preprocessing-technique-contents}. 
The results differ from those of short-text retrieval: apostrophe and accent removal did not provide any measurable benefit for BM25-based models in terms of Precision or MAP relative to the baseline. Likewise, accent removal offered no improvement in P@10, P@20, or MAP with the Dirichlet LM model.

However, there was evidence of improvement with Hiemstra LM in MAP and NDCG at various cutoffs and in overall NDCG. In contrast, hyphen removal, stopword removal, and stemming demonstrated clear advantages for long-text retrieval. Among the stemming variants, the \textit{light} stemmer consistently outperformed the \textit{moderate} and \textit{heavy} variants, though the margin of improvement over the \textit{moderate} variant was minimal.

\input{tables/long-text-retrieval-preliminary}

The results presented in~\autoref{tab:evaluation-preprocessing-technique-contents} were further analyzed by combining text preprocessing techniques to better understand their collective impact on long-text retrieval effectiveness. These combined preprocessing techniques yielded varied impacts across evaluation metrics and models. Hiemstra LM consistently outperformed other models in most combinations and cutoffs, except in P@5 and NDCG@5, where TF-IDF exhibited a slight performance advantage. Notable results for Hiemstra LM include the combination of apostrophe and hyphen removal, which achieved the highest scores at P@10 (0.5576), P@20 (0.4907), MAP@20 (0.2523), and NDCG@20 (0.4790).

Additional configurations, such as incorporating \textit{moderate} stemming, yielded the best results at MAP@5 (0.1645), while including stopword removal improved NDCG@10 (0.4760) and overall MAP (0.4358). Furthermore, Hiemstra LM showed notable gains in effectiveness when combining hyphen and accent removal, particularly at MAP@10 (0.1645) and NDCG (0.6855). Hyphen removal alone continued to demonstrate its effectiveness in long-text retrieval, achieving the same highest score as its combination with apostrophe removal at P@20 (0.4907).

\input{tables/long-text-retrieval-final}

In contrast, individual preprocessing techniques, such as accent and apostrophe removal, generally underperformed when applied in isolation, contributing minimally to retrieval effectiveness across different metrics (see~\autoref{tab:evaluation-preprocessing-technique-contents}). However, combining these techniques with hyphen and stopword removal, along with \textit{moderate} stemming, resulted in a strong performance, particularly with Hiemstra LM across most metrics and cutoffs. While the \textit{heavy} stemming variant slightly outperformed the \textit{moderate} variant when applied individually, the \textit{moderate} variant showed a slight advantage when combined with other preprocessing techniques.

Hiemstra LM consistently delivered the highest scores with the combined preprocessing approach across all metrics, except at P@5, where TF-IDF marginally outperformed it. Hiemstra LM demonstrated notable performance improvements over the baseline, achieving gains of up to 13.75\% for Precision cutoffs, 18.01\% for MAP cutoffs, 14.48\% for NDCG cutoffs, 19.23\% for overall MAP, and 14.44\% for overall NDCG. Additionally, Hiemstra LM achieved an average relative improvement of up to 16.40\% over the other retrieval models.

Conversely, BM25 and DFR BM25 did not produce the best results with any preprocessing strategies for long-text retrieval, suggesting that BM25-based models may be less effective in handling long-document contexts. These findings underscore the different impacts of text preprocessing techniques and retrieval models on short- and long-text retrieval tasks in Tetun.

\subsection{Discussion}
In Tetun ad-hoc text retrieval, text preprocessing techniques that involve the removal of apostrophes and hyphens generally lead to performance gains of up to 8.19\% with DFR BM25 in short-text retrieval up to the top-10 cutoffs. 
When stopword removal is combined with apostrophe and hyphen removal, further gains are observed, particularly at MAP@5 with both DFR BM25 and TF-IDF models and at NDCG with Hiemstra LM. 
When apostrophes are retained and only hyphens and stopwords are removed, Hiemstra LM shows improvements of up to 8.33\% compared to other models at P@20, MAP@20, NDCG@20, and overall MAP. These results suggest that in short-text retrieval, targeting hyphen removal can enhance retrieval effectiveness. 

Different strategies for handling stopwords and apostrophes yield varied benefits across retrieval models. Retaining stopwords while removing apostrophes proves more effective with DFR BM25, whereas retaining apostrophes while removing stopwords shows better results with Hiemstra LM at higher cutoffs and overall. Accent removal, however, does not enhance retrieval effectiveness in short-text retrieval, whether applied individually or in combination with other preprocessing techniques.

Removing stopwords does not lead to noticeable improvements in retrieval effectiveness. For instance, at P@20, the relative gain is limited to 0.58\% compared to Hiemstra LM with apostrophe and hyphen removal. Similarly, MAP@20 improves by 0.76\%, NDCG@20 by 0.07\%, and overall MAP by 1.22\% relative to the baselines. These findings align with Savoy's~\cite{savoy-1999} experiments with French, which showed that retaining stopwords yielded better results with the BM25 model. Similar conclusions were reported by \citet{ghosh-bhattacharya-2017}, who found variability in retrieval effectiveness within the same language and no significant improvement from stopword removal. Additionally, \citet{dolamic-savoy-2010} observed that retaining stopwords improved retrieval effectiveness in Hindi, with minimal differences for Marathi and Bengali.

Regarding stemming, it does not improve retrieval effectiveness in Tetun, whether applied independently or in combination with other preprocessing techniques. This result is consistent with findings from preliminary experiments on Tetun text retrieval~\cite{de-jesus-2022-ir} and Flores and Moreira's~\cite{flores-moreira-2016} experiments with Dutch, Italian, Spanish, and English. However, studies on other LRLs, including those by \citet{sahu-pal-2023} (Sanskrit), \citet{sahu-et-al-2023} (Urdu), and \citet{adriani-et-al-2007} (Indonesian), reported positive impacts of stemming on retrieval effectiveness. This discrepancy likely arises from Tetun's language-specific characteristics, which may respond differently to stemming than other languages. 

While intrinsic evaluations suggested that the \textit{moderate} and \textit{heavy} stemmers were slightly more accurate than the \textit{light} stemmer, these differences did not consistently translate into improved retrieval performance. These findings align with Flores and Moreira's~\cite{flores-moreira-2016} results that higher stemmer accuracy does not always improve retrieval effectiveness.

For long-text retrieval, Hiemstra LM consistently outperformed other models across most metrics and cutoffs. Hyphen removal delivered performance gains across various metrics and models, particularly when combined with apostrophe removal, achieving high scores at P@10, P@20, MAP@20, and NDCG@20. Adding \textit{moderate} stemming to this combination yielded the highest MAP@5 score, though the improvement was minimal compared to strategies without stemming.

Apostrophe removal showed inconsistent results as an individual preprocessing technique but became more effective when combined with hyphen removal. Similarly, accent removal had varying impacts across models and metrics when applied alone, but combined with other techniques, it achieved a relative improvement of 16.40\% in MAP@10 over other models. This outcome aligns with Savoy's~\cite{savoy-1999} findings on French, where accent removal slightly improved retrieval effectiveness in long-text retrieval but negatively affected short-text retrieval for several strategies.

\subsection{Conclusion}
The most effective retrieval strategy for Tetun ad-hoc text retrieval involves short-text retrieval (i.e., using document titles) combined with targeted preprocessing techniques, specifically splitting compound words by removing hyphens and eliminating apostrophes from queries and documents. Together, these techniques enhance both retrieval efficiency and effectiveness. The removal of stopwords shows minimal impact, while preserving accents proves crucial for effective text retrieval in Tetun, likely due to the language’s reliance on diacritics to distinguish word meanings.

Stemming does not improve retrieval effectiveness in short-text retrieval. In long-text retrieval, the \textit{moderate} stemming variant, when combined with other preprocessing techniques, shows slight improvements at MAP@5; however, the gains remain marginal and are still lower than the best-performing score achieved in short-text retrieval, representing -39.25\% relative performance drop. This minimal impact suggests that morphological normalization through stemming is not essential for effective text retrieval in Tetun, likely due to the language’s relatively simple morphology and minimal use of inflectional affixes.

The experimental results show that text preprocessing techniques, particularly the removal of hyphens and apostrophes, are crucial in enhancing the effectiveness of Tetun ad-hoc text retrieval. Among the retrieval models, DFR BM25 performs best for cutoffs up to the top-10 but exhibits greater sensitivity to stopword removal. In contrast, Hiemstra LM provides the most substantial improvements at top-20 cutoffs and beyond, particularly in overall MAP and NDCG.

Overall, this study demonstrates the importance of thoroughly investigating language-specific preprocessing strategies by segmenting them into distinct stages and systematically integrating them to uncover the linguistic components that impact retrieval effectiveness (see \autoref{fig:experimental-setting}). It emphasizes the benefit of developing approaches tailored to the unique characteristics of each language, rather than relying solely on established techniques in the literature. The study introduces a detailed and adaptable methodology for establishing ad-hoc text retrieval baselines, particularly for LRLs where such benchmarks have not yet been established. By accounting for the unique morphological and syntactic features of each language, researchers can design more effective and linguistically appropriate retrieval strategies.

\section{Conclusions and Future Work}\label{sec:conclusions-future-work}

This study presents the development of Tetun text retrieval and establishes the first baselines for the ad-hoc retrieval task. As part of this effort, we created three essential resources: \textit{Labadain-Stopwords}~\cite{de-jesus-nunes-2025-stopwords}, \textit{Labadain-Stemmer}~\cite{labadain-stemmer-algorithm}, and \textit{Labadain-Avaliadór}~\cite{de-jesus-nunes-2025-avaliador}. These resources are publicly available to the IR and NLP research community under the Creative Commons Attribution-ShareAlike license.

\textit{Labadain-Stopwords} contains 160 Tetun stopwords, a total comparable to those reported for other low-resource languages (LRLs) in the literature. \textit{Labadain-Stemmer} is available in three variants---\textit{light},~\textit{moderate}, and \textit{heavy}---with the \textit{light} variant specifically targeting Portuguese loanwords commonly used in Tetun. Due to the prominence of these loanwords, the Portuguese stemmer from Snowball was adapted for Tetun by adapting it to the linguistic characteristics of Portuguese-derived words in Tetun. This method can be extended to other LRLs with a similar linguistic nature.

\textit{Labadain-Avaliadór} consists of 59 topics, 33,550 documents, and 5,900 relevance judgments \textit{(qrels)}, with an average of 36.76 relevant documents per query. This average reflects a balanced representation of the relevant documents in the collection. The balance arises from the fact that the queries were derived from real-world search logs sourced from two platforms: Google Search Console logs for Timor News and logs from searches performed using the Timor News search functionality. This ensures that queries and documents reflect real-world scenarios where the documents exist in the Labadain-30k+ dataset~\cite{labadain30k-dataset}, which is used as the document collection.

Our investigation involved experimenting with different retrieval strategies, with a focus on the impact of various text preprocessing techniques tailored based on the linguistic characteristics of Tetun. We initially hypothesized that text preprocessing techniques would improve the effectiveness of Tetun text retrieval, and our findings confirmed this hypothesis. In response to our research questions, we conclude that removing hyphens to split compound words into individual words, combined with the removal of apostrophes, enhances retrieval effectiveness overall in both short- and long-text retrieval, with short-text retrieval being the most effective approach for Tetun.

For the retrieval and ranking models, DFR BM25 performs effectively with short-text retrieval up to the top-10 cutoff but shows slightly lower effectiveness when stopwords are removed. Meanwhile, Hiemstra LM consistently demonstrates effective performance across various metrics, particularly beyond the top-10 cutoffs and in overall MAP and NDCG. These findings suggest that Hiemstra LM is more effective for Tetun text retrieval when more than ten documents are prioritized for retrieval.
The effectiveness of Hiemstra LM for Tetun text retrieval is consistent with the findings of \citet{sahu-pal-2023} in their study on Sanskrit, a script-based language spoken in India. Although Tetun uses a Latin script, both languages demonstrate comparable retrieval effectiveness with the Hiemstra LM model in short-text retrieval, measured by MAP. This suggests that Hiemstra LM may adapt effectively to LRLs, regardless of their linguistic or script characteristics.

Future work will investigate semantic search, which captures the contextual meaning of queries and documents rather than relying solely on exact term matches. Integrating large language models (LLMs) into retrieval tasks may open new avenues for enhancing retrieval effectiveness and better aligning retrieval systems with user information needs. Additionally, investigating user search behavior influenced by LLM advancements may reveal evolving patterns in user search intent and information needs, particularly in LRLs like Tetun. Insights from these trends could guide the design of retrieval systems that effectively adapt to changing search behaviors.

\section{Acknowledgments}
This work is financed by National Funds through the Portuguese funding agency, FCT - Funda\c{c}\~{a}o para a Ci\^{e}ncia e a Tecnologia, under the Ph.D. studentship grant number SFRH/BD/151437/2021 (DOI: 10.54499/SFRH/BD/151437/2021).

We would like to extend our deepest gratitude to Fanézio Pinto, Altedio Araújo, Roohy Freitas, Alton Freitas, Rita Belo, and Paulo Barreto for their invaluable contributions to the development of \textit{Labadain-Stopwords}, \textit{Labadain-Stemmer}, and \textit{Labadain-Avaliadór}, which serve as key components of this study.

\newpage

\section{Appendices}
Subsection~\ref{app:list-of-stopwords} presents the \textit{Labadain-Stopwords}, and algorithm of the \textit{Labadain-Stemmer} is shown in Subsection~\ref{detail-tetun-stemmer}.

\subsection{The Labadain-Stopwords} \label{app:list-of-stopwords} The \textit{Labadain-Stopwords} list~\cite{de-jesus-nunes-2025-stopwords} with their English translations are presented in \autoref{tab:list-of-stopwords}.

\input{tables/tetun-stopword-list}
\newpage

\subsection{Details of the \textit{Labadain-Stemmer} Algorithms}\label{detail-tetun-stemmer}
The linguistic regions used in the \textit{Labadain-Stemmer} are shown in \autoref{tab:linguistic-regions}, while Table~\ref{tab:suffixes-description} provides a list of Portuguese-derived suffixes. The algorithm of the Tetun \textit{light} stemmer variant is detailed in Algorithm~\ref{alg:tetun-light-stemmer}.

\input{tables/linguistic-regions}

\input{tables/pt-derived-suffixes}

\input{algorithms/light_stemmer}
\newpage

\bibliographystyle{ACM-Reference-Format}
\bibliography{main}

\appendix

\end{document}

%% file: tables/tetun-examples.tex
\begin{table}[!ht]
  \centering
  \caption{Examples of Basic Tetun Phrase.}
  \label{tab:basic-tetun}
  \begin{tabular}{llll}
    \toprule
    Tetun & English & Tetun & English\\
    \midrule
    Dadeer di'ak! & Good morning! & Di'ak ka lae? & How are you?\\
    Ita-nia naran saida? & What is your name? & Ita-boot hela iha ne'ebé? & Where do you live?\\
  \bottomrule
\end{tabular}
\end{table}

%% file: tables/tetun-affix-examples.tex
\begin{table}[!ht]
\centering
\caption{Examples of Tetun Inflection and Derivation.~$^*$Suffix \textit{dór} is used in both native Tetun and Portuguese loanwords.}
\label{tab:morphology}
\begin{tabular}{lll}
\toprule
\textbf{Prefixes}              & \multicolumn{2}{c}{\textbf{Suffixes}}\\\cmidrule(l){2-3}
Native Tetun          & Native Tetun              & Portuguese Loanwords \\\midrule
\underline{ha}dame (reconcile)    & susu\underline{n} (breast)           & selebra\underline{saun} (celebration) \\
\underline{nak}fera (break)       & sala\underline{-na'in} (sinner)       & ezata\underline{mente} (exactly) \\
\underline{nam}kari (scatter)     & nakar\underline{-teen} (naughty)      & doador\underline{es} (donors)\\
\underline{hak}besik (get closer) & hemu\underline{dór} (drinker)$^*$ & toka\underline{dór} (musician)\\
\bottomrule
\end{tabular}
\end{table}

%% file: tables/pt-loanword-examples.tex
\begin{table}[!ht]
\centering
  \caption{Example of Portuguese Loanwords.}
  \label{tab:portuguese-loanwords}
  \begin{tabular}{llll}
    \toprule
    Vebs & Nouns & Adjectives\\
    \midrule
    estuda (\textit{estudar}, study)  & serveja (\textit{cerveja}, beer) &  baratu (\textit{barato}, cheap)\\
kanta (\textit{cantar}, sing)  & estudante (\textit{estudante}, student) &  forte (\textit{forte}, strong)\\
organiza (\textit{organizar}, organize)  & eskola (\textit{escola}, school) &  rápidu (\textit{rápido}, fast)\\
  \bottomrule
\end{tabular}
\end{table}

%% file: tables/labadain-30k-dataset-desc.tex
\begin{table}[!ht]
\centering
\caption{Description of the Labadain-30k+ Dataset. *Tokens consist of words and numbers.}
\label{tab:dataset-description}
\begin{tabular}{lr}
      \toprule
        Total documents in the dataset & 33,550\\
        Total paragraphs in the content & 334,875\\
        Total sentences in the content & 414,370\\
        Total tokens in the corpus$^*$ & 12,300,237\\
        Vocabulary in the corpus & 162,466\\
      \bottomrule
\end{tabular}
\end{table}

%% file: tables/labadain-30k-dataset-summary.tex
\begin{table}[!ht]
\centering
\caption{Summary of the Labadain-30k+ Dataset.}
\label{tab:documents-per-category}
\begin{tabular}{lrr}
  \toprule
  \textbf{Category} & \textbf{\#docs} & \textbf{Proportion}\\
  \midrule
    News articles & 30,150 & 89.87\%\\
    Wikipedia documents & 1,455 & 4.34\%\\
    Legal/government documents & 1,223 & 3.65\%\\
    Technical documents & 211 & 0.63\%\\
    Blogs and Forums & 145 & 0.43\%\\
    Advertisements/announcements & 124 & 0.37\%\\
    Research papers & 83 & 0.25\%\\
    Personal pages & 74 & 0.22\%\\
    Institutional information & 53 & 0.16\%\\
    Correspondence letters & 32 & 0.10\%\\
  \bottomrule
\end{tabular}
\end{table}

%% file: tables/dataset-for-stopword-detection.tex
\begin{table}[!ht]
\centering
\caption{Summary of Tetun, Portuguese and English Datasets Used for Stopword Detection.}
\label{tab:summary-dataset-for-stopwords}
\begin{tabular}{lrrr}
      \toprule
      Description & Tetun & Portuguese & English\\
      \midrule
        Total number of documents & 33,550 & 3,153 & 624\\
        Total number of words & 11,928,821 & 613,736 & 667,584 \\
        Total vocabulary size & 146,783 & 45,860 & 31,390 \\
      \bottomrule
\end{tabular}
\end{table}

%% file: tables/stopword-precision-tetun.tex
\begin{table}[!ht]
\centering
\caption{Stopword Precision for Tetun.}
\label{tab:stops-tetun}
\begin{tabular}{lrrrrrrrrrr}
\hline
Approach   & P@10            & P@25            & P@50            & P@75            & P@100           & P@250           & P@500           & P@750           & P@1000          \\ \hline
In-degree  & \textbf{1.0000} & \textbf{0.9600} & \textbf{0.8400} & 0.7200          & \textbf{0.7000} & \textbf{0.4720} & \textbf{0.3080} & \textbf{0.2347} & \textbf{0.1930} \\
Out-degree & \textbf{1.0000} & 0.8800          & 0.8000          & 0.6933          & 0.6000          & 0.4240          & 0.2900          & 0.2160          & 0.1780          \\
Degree     & \textbf{1.0000} & \textbf{0.9600} & 0.8200          & \textbf{0.7333} & 0.6500          & 0.4640          & 0.3000          & 0.2253          & \textbf{0.1930} \\
TF         & \textbf{1.0000} & 0.9200          & 0.6800          & 0.6000          & 0.5200          & 0.3600          & 0.2480          & 0.1933          & 0.1610          \\
IDF        & \textbf{1.0000} & 0.9200          & 0.6800          & 0.6000          & 0.5200          & 0.3600          & 0.2540          & 0.1973          & 0.1640          \\
TF-IDF     & \textbf{1.0000} & 0.9200          & 0.6800          & 0.5867          & 0.5100          & 0.3560          & 0.2500          & 0.1947          & 0.1620          \\ \hline
\end{tabular}
\end{table}

%% file: tables/stopword-precision-tetun-pt.tex
\begin{table}[!ht]
\centering
\caption{Stopword Precision for Portuguese.}
\label{tab:stops-portuguese}
\begin{tabular}{lrrrrrrrrrr}
\hline
Approach   & P@10            & P@25            & P@50            & P@75            & P@100           & P@250           & P@500           & P@750           & P@1000          \\ \hline
In-degree  & \textbf{1.0000} & \textbf{1.0000} & 0.9000 & 0.7600          & 0.6700 & \textbf{0.3800} & \textbf{0.2260} & \textbf{0.1587} & 0.1260 \\
Out-degree & \textbf{1.0000} & 0.9600          & \textbf{0.9600} & 0.8000          & 0.6700          & 0.3560          & 0.2180          & 0.1573          & \textbf{0.1270} \\
Degree     & \textbf{1.0000} & \textbf{1.0000} & 0.9400          & \textbf{0.8133} & \textbf{0.7100} & 0.3680          & 0.2240          & \textbf{0.1587} & 0.1240 \\
TF         & \textbf{1.0000} & 0.9600          & 0.9000          & 0.7600          & 0.6500          & 0.3320          & 0.2000          & 0.1480          & 0.1200          \\
IDF        & \textbf{1.0000} & 0.9600          & 0.9000          & 0.7600          & 0.6500          & 0.3320          & 0.2000          & 0.1480          & 0.1200          \\
TF-IDF     & \textbf{1.0000} & 0.9600          & 0.9000          & 0.7600          & 0.6500          & 0.3320          & 0.2000          & 0.1480          & 0.1200          \\ \hline
\end{tabular}
\end{table}

%% file: tables/stopword-precision-english.tex
\begin{table}[!ht]
\centering
\caption{Stopword Precision for English.}
\label{tab:stops-english}
\begin{tabular}{lrrrrrrrrrr}
\hline
Approach   & P@10            & P@25            & P@50            & P@75            & P@100           & P@250           & P@500           & P@750           & P@1000          \\ \hline
In-degree  & \textbf{1.0000} & \textbf{1.0000} & 0.9400 & \textbf{0.9333} & \textbf{0.8200} & 0.4120 & 0.2280 & 0.1573 & \textbf{0.1220} \\
Out-degree & \textbf{1.0000} & \textbf{1.0000} & \textbf{0.9800} & 0.8400 & 0.7600 & 0.4160          & 0.2280          & 0.1573          & \textbf{0.1220} \\
Degree     & \textbf{1.0000} & \textbf{1.0000} & \textbf{0.9800} & 0.9067 & \textbf{0.8200} & \textbf{0.4280} & \textbf{0.2300} & \textbf{0.1613} & \textbf{0.1220} \\
TF         & \textbf{1.0000} & \textbf{1.0000} & 0.9600 & 0.8933          & 0.7500          & 0.4240 & 0.2280 & 0.1573 & \textbf{0.1220} \\
IDF        & \textbf{1.0000} & \textbf{1.0000} & 0.9600          & 0.8933          & 0.7500          & 0.4240          & 0.2280          & 0.1573          & \textbf{0.1220} \\
TF-IDF     & \textbf{1.0000} & \textbf{1.0000} & 0.9600          & 0.8933         & 0.7500          & 0.4240          & 0.2280          & 0.1573          & \textbf{0.1220} \\ \hline
\end{tabular}
\end{table}

%% file: tables/pt-tetun-transform-rules.tex
\begin{table}[!ht]
\centering
\caption{Rules for Transforming Portuguese-Derived Words into Tetun Which Applied in Suffix Transformations.}
\label{tab:portuguese-derived-transformation-rules}
    \begin{tabular}{lll}
          \toprule
          Rule (Portuguese $\rightarrow$ Tetun) & Effect on Suffix & Example (Portuguese $\rightarrow$ Tetun)\\
          \midrule
            ão $\rightarrow$ aun & as\textbf{aun} & comemoraç\textbf{ão} $\rightarrow$ komemoras\textbf{aun} (celebration) \\
            ss, c (before e, i), ç (before a, o, u) $\rightarrow$ s & \textbf{s}aun & discu\textbf{ss}ão $\rightarrow$ disku\textbf{s}aun (discussion)\\
            qu, c (before a, o, u) $\rightarrow$ k & i\textbf{k} + amente & automati\textbf{c}amente $\rightarrow$ automati\textbf{k}amente (automatically)\\
            g (before e, i) $\rightarrow$ j & lo\textbf{j}ia & tecnolo\textbf{g}ia $\rightarrow$ teknolo\textbf{j}ia (technology)\\
            s (between vowels) $\rightarrow$ z & o\textbf{z}a & podero\textbf{s}a $\rightarrow$ podero\textbf{z}a (powerful)\\
            ê $\rightarrow$ é & \textbf{é}nsia & compet\textbf{ê}ncia $\rightarrow$ kompet\textbf{é}nsia (competence)\\
            â $\rightarrow$ á & \textbf{á}nsia & ignor\textbf{â}ncia $\rightarrow$ ignor\textbf{á}nsia (ignorance)\\
            o $\rightarrow$ u & \textbf{u} & infermeir\textbf{o} $\rightarrow$ infermeir\textbf{u} (nurse)\\
          \bottomrule
    \end{tabular}
\end{table}

%% file: tables/summary-word-sample.tex
\begin{table}[!ht]
\centering
  \caption{Summary of Words Produced at Each Stage of the Sample Word Generation Process.~$^*$Approximately 88.4\% reduction in the number of words was observed after applying the LID model, attributed to the model's average score per word of 0.3943, with a threshold set to 0.95.}
  \label{tab:words-candidate-preprocessing-summary}
  \begin{tabular}{lr}
    \toprule
    Description & Total of Words\\
    \midrule
    Initial total number of words & 146,387\\
    Remaining words after removing stopwords & 146,204\\
    Remaining words after excluding words with fewer than four characters & 141,487\\
    Remaining words after applying LID model & 16,391$^*$\\
    Remaining words after manual verification by human assessors & 1,839\\ 
  \bottomrule
\end{tabular}
\end{table}

%% file: tables/cohens-k-score-ground-truth.tex
\begin{table}[!ht]
\centering
  \caption{Cohen's Kappa Score for Inter-Annotator Agreement in the Construction of the Ground Truth Set.}
  \label{tab:cohen-k-creating-ground-truth}
  \begin{tabular}{lr}
    \toprule
    Algorithm &  \textit{k}-Score\\
    \midrule
    Light stemmer & 0.7006\\
    Moderate stemmer & 0.6990\\
    Heavy stemmer & 0.7683\\
  \bottomrule
\end{tabular}
\end{table}

%% file: tables/summary-ground-truth.tex
\begin{table}[!ht]
\centering
  \caption{Summary of the Ground Truth Set and Word Length.}
  \label{tab:ground-truth-summary}
  \begin{tabular}{lrrr}
    \toprule
    Description & All Words & Portuguese Loanwords & Native Tetun Words\\
    \midrule
    Total number of words & 1,839 & 81.79\% & 18.21\% \\
    Minimum character count per word & 4.00 & 4.00 & 4.00\\
    Maximum character count per word & 20.00 & 20.00 & 15.00\\
    Average character count per word & 9.50 & 10.21 & 5.97\\
    \bottomrule
  \end{tabular}
\end{table}

%% file: tables/stemming-performance-paice.tex
\begin{table}[!ht]
\centering
  \caption{Analysis of the Stemming Algorithms' Performance Using Paice Metrics.}
  \label{tab:stemming-performance-paice}
  \begin{tabular}{lrrrr}
    \toprule
     &  UI & OI & SW & ERRT\\\midrule
    Light & 0.312062  & 0.000017 & 0.000056 & 0.481367\\
    Moderate & 0.305049 & 0.000017 & 0.000057 & \textbf{0.472802}\\
    Heavy &  0.305049 & 0.000017 & 0.000057 & \textbf{0.472802}\\
  \bottomrule
\end{tabular}
\end{table}

%% file: tables/doc-collection-summary.tex
\begin{table}[!ht]
\centering
  \caption{Summary of Document Collection.~$^*$Tokens comprise words and numbers, excluding punctuation and special characters.}
  \label{tab:docs-summary}
  \begin{tabular}{lrrrrr}
    \toprule
    Description & Total & Min & Max & Avg & Std \\
    \midrule
    Number of tokens$^*$ (titles) & 306,840 & 1 & 29 & 9.15 & 3.05\\
    Number of tokens (content) & 11,997,420 & 2 & 27,166 & 357.48 & 473.99\\
  \bottomrule
\end{tabular}
\end{table}

%% file: tables/query-formulation.tex
\begin{table}[!ht]
\centering
\caption{Examples of original query logs and their reformulations. Words highlighted in \textit{green} background indicate newly added terms, while word in             \textit{orange} background indicates orthography correction.}
  \label{tab:query-formulation}
      \begin{tabular}{ll}
        \toprule
        Original query log & Reformulated query \\
        \midrule
        Problema lixu (Waste problem) & Problema lixu \colorbox{green!25}{iha Dili} (Waste problem in Dili) \\
        Soe bebe (Baby abandonment) & \colorbox{green!25}{Kazu} soe \colorbox{orange!25}{bebé} (Case of baby abandonment) \\
        Konsumu tabaku (Tobacco consumption) & \colorbox{green!25}{Dadus} konsumu tabaku (Tobacco consumption data) \\
      \bottomrule
    \end{tabular}
\end{table}

%% file: tables/query-summary.tex
\begin{table}[!ht]
\centering
  \caption{Summary of Queries.}
  \label{tab:queries-summary}
  \begin{tabular}{lr}
    \toprule
    Description & Value\\
    \midrule
    Total of queries & 61\\
    Total number of three-word queries & 37\\
    Total number of four-word queries & 22 \\ 
    Total number of five-word queries & 2\\
    Average numbers of words per query & 3.43\\
  \bottomrule
\end{tabular}
\end{table}

%% file: tables/cohens-k-assessors.tex
\begin{table}[!ht]
\centering
  \caption{Cohen's Kappa scores for Inter-Annotator Agreement Among the Five Assessors.}
  \label{tab:cohen-k-agreement-annotators}
  \begin{tabular}{lrrrrrr}
    \toprule
    & HA$_{1}$ & HA$_{2}$ & HA$_{3}$ & HA$_{4}$ & HA$_{5}$\\
    \midrule
    Human annotator 1 (HA$_{1}$) & --- & 0.4344 & 0.4999 & 0.4434 & 0.4380\\ 
    Human annotator 2 (HA$_{2}$) & 0.4344 & --- & 0.3745 & 0.3310 & 0.3500\\ 
    Human annotator 3 (HA$_{3}$) &  0.4999 & 0.3745 & --- & 0.4646 & 0.4199\\ 
    Human annotator 4 (HA$_{4}$) & 0.4434 & 0.3310 & 0.4646 & --- & 0.4063\\ 
    Human annotator 5 (HA$_{5}$) &  0.4380 & 0.3500 & 0.4199 & 0.4063 & ---\\\midrule
    \textbf{Average kappa score} & \multicolumn{5}{c}{\textbf{0.4236}}\\
  \bottomrule
\end{tabular}
\end{table}

%% file: tables/tied-score-summary.tex
\begin{table}[!ht]
\centering
  \caption{Summary of Documents with Tied Scores.}
  \label{tab:tied-documents}
  \begin{tabular}{lr}
    \toprule
    Description & Value\\
    \midrule
    Total number of tied documents & 602\\
    Proportion of tied documents & 9.87\%\\
    Minimum number of tied documents per query & 1\\
    Maximum number of tied documents per query & 38\\
    Average number of tied documents per query & 10.20\\
    Standard deviation in the number of tied documents per query & 7.17\\
  \bottomrule
\end{tabular}
\end{table}

%% file: tables/human-judgment-results.tex
\begin{table}[!ht]
\centering
\caption{Details of the Human Judgment Results.}
\label{tab:assessment-details}
\begin{tabular}{@{}lrrrr@{}}
\toprule
\multirow{2}{*}{Relevance Level} & \multicolumn{2}{l}{1st Round} & \multicolumn{2}{l}{2nd Round} \\ \cmidrule(l){2-5} 
                                 & \#            & \%            & \#            & \%            \\ \midrule
3 - Highly relevant               & 710           & 11.64         & 566           & 9.59          \\
2 - Relevant                     & 1,102         & 18.07        & 1,054         & 17.86         \\
1 - Marginally relevant            & 476           & 7.80          & 549           & 9.31          \\
0 - Irrelevant                   & 3,812         & 62.49         & 3,731         & 63.24        \\ \bottomrule
\end{tabular}
\end{table}

%% file: tables/final-test-collection-summary.tex
\begin{table}[!ht]
\centering
  \caption{Summary of the Final Test Collection.~\textsuperscript{*}Relevant documents consist of marginally relevant, relevant, and highly relevant.}
  \label{tab:final-test-collection}
  \begin{tabular}{lr}
    \toprule
    Description & Value\\
    \midrule
    Total number of topics & 59\\
    Total number of \textit{qrels} & 5,900\\
    Minimum number of relevant documents per query$^*$  & 11 \\ 
    Maximum number of relevant documents per query  & 99 \\ 
    Average number of relevant documents per query  & 36.76 \\ 
    Standard deviation of relevant documents per query  & 20.89 \\ 
    \bottomrule
  \end{tabular}
\end{table}

%% file: tables/compare-lab-avaliador-other-lrls.tex
\begin{table}[!ht]
\centering
  \caption{Comparison of the \textit{Labadain-Avaliadór} With Other LRLs.}
  \label{tab:comparison-test-collection-lrls}
  \begin{tabular}{lrrrr}
    \toprule
    Language & \#Docs & \#Topics & Avg. Relevant Docs\\
    \midrule
    Tetun & 33,550 & 59 & 36.76 \\
    Sanskrit~\cite{sahu-pal-2023} & 7,057 & 50 & 8.54\\
    Chichewa~\cite{chavula-suleman-2021} & 9,380 & 129 & 19\\
    Citumbuka~\cite{chavula-suleman-2021} & 2,258 & 129 & 17\\
    Cinyanja~\cite{chavula-suleman-2021} & 173 & 129 & 15\\
    Hamshahri~\cite{aleahmad-et-al-2009} & 166,774 & 65 & 36.18 \\
    \bottomrule
  \end{tabular}
\end{table}

%% file: tables/icf-comparison.tex
\begin{table}[!ht]
\centering
\caption{Index Compression Factor (\%) for Titles and Contents Compared to the Baseline.}
\label{tab:indexed-documents}
\resizebox{\textwidth}{!}{%
\begin{tabular}{lrrrrrrrrrrrrrrr}
\toprule
\multirow{2}{*}{Description} & \multirow{2}{*}{Baseline} & \multicolumn{2}{c}{No Apostrophes} & \multicolumn{2}{c}{No Accents} & \multicolumn{2}{c}{No Hyphens} & \multicolumn{2}{c}{No Stopwords} & \multicolumn{2}{c}{Light} & \multicolumn{2}{c}{Moderate} & \multicolumn{2}{c}{Heavy} \\
\cmidrule(lr){3-4} \cmidrule(lr){5-6} \cmidrule(lr){7-8} \cmidrule(lr){9-10} \cmidrule(lr){11-12} \cmidrule(lr){13-14} \cmidrule(lr){15-16} 
& & \multicolumn{1}{c}{\#} & \multicolumn{1}{c}{\%} & \multicolumn{1}{c}{\#} & \multicolumn{1}{c}{\%} & \multicolumn{1}{c}{\#} & \multicolumn{1}{c}{\%} & \multicolumn{1}{c}{\#} & \multicolumn{1}{c}{\%} & \multicolumn{1}{c}{\#} & \multicolumn{1}{c}{\%} & \multicolumn{1}{c}{\#} & \multicolumn{1}{c}{\%} & \multicolumn{1}{c}{\#} & \multicolumn{1}{c}{\%} \\
\midrule
Title & 25,412 & 25,258 & 0.61 & 23,568 & 7.25 & 17,596 & 30.76 & 25,256 & 0.61 & 23,416 & 7.86 & 23,377 & 8.01 & 23,283 & 8.38 \\
Content & 163,203 & 162,012 & 0.73 & 150,596 & 7.72 & 146,657 & 10.14 & 163,148 & 0.33 & 144,240 & 11.62 & 143,698 & 11.95 & 143,329 & 12.18 \\
\bottomrule
\end{tabular}%
}
\end{table}

%% file: tables/short-text-retrieval-preliminary.tex
\begin{table}[!ht]
\centering
\caption{Effectiveness of Text Preprocessing Techniques in Short-Text Retrieval. Red values indicate scores lower than the baseline.}
\label{tab:evaluation-preprocessing-technique-titles}
\resizebox{\textwidth}{!}{%
      \begin{tabular}{p{2.5cm}llrrrrrrrrrrr}
      \hline
     &  & \multicolumn{3}{c}{Precision at Cutoff} & \multicolumn{3}{c}{MAP at Cutoff} & \multicolumn{3}{c}{NDCG at Cutoff} & \multicolumn{1}{l}{} & \multicolumn{1}{l}{} \\ \cline{3-11}
     \multirow{-2}{*}{Retrieval Strategies} & \multirow{-2}{*}{Model} & \multicolumn{1}{c}{@5} & \multicolumn{1}{c}{@10} & \multicolumn{1}{c}{@20} & \multicolumn{1}{c}{@5} & \multicolumn{1}{c}{@10} & \multicolumn{1}{c}{@20} & \multicolumn{1}{c}{@5} & \multicolumn{1}{c}{@10} & \multicolumn{1}{c}{@20} & \multicolumn{1}{l}{\multirow{-2}{*}{MAP}} & \multicolumn{1}{l}{\multirow{-2}{*}{NDCG}} \\ \hline
     & BM25 & 0.8169 & 0.7763 & 0.6602 & 0.1444 & 0.2568 & 0.3903 & 0.6801 & 0.6668 & 0.6454 & 0.5925 & 0.7408 \\
     & DFR BM25 & 0.8169 & 0.7763 & 0.6619 & 0.1440 & 0.2563 & 0.3901 & 0.6811 & 0.6666 & 0.6468 & 0.5926 & 0.7407 \\
     & TF-IDF & 0.8136 & 0.7746 & 0.6458 & 0.1432 & 0.2546 & 0.3825 & 0.6739 & 0.6640 & 0.6380 & 0.5802 & 0.7364 \\
     & \cellcolor[HTML]{FFFFFF}{\color[HTML]{1F1F1F} Dirichlet LM} & 0.7898 & 0.7525 & 0.6398 & 0.1299 & 0.2361 & 0.3671 & 0.6359 & 0.6356 & 0.6174 & 0.5780 & 0.7208 \\
    \multirow{-5}{*}{Baseline} & \cellcolor[HTML]{FFFFFF}{\color[HTML]{1F1F1F} Hiemstra LM} & 0.8136 & 0.7695 & 0.6669 & 0.1428 & 0.2521 & 0.3928 & 0.6670 & 0.6588 & 0.6465 & 0.6090 & 0.7435 \\ \hline
     & BM25 & 0.8237 & 0.7763 & 0.6644 & 0.1453 & 0.2572 & 0.3930 & 0.6866 & 0.6685 & 0.6499 & 0.5938 & 0.7419 \\
     & DFR BM25 & 0.8237 & 0.7763 & 0.6661 & 0.1450 & 0.2568 & 0.3929 & 0.6878 & 0.6684 & 0.6515 & 0.5942 & 0.7420 \\
     & TF-IDF & 0.8203 & 0.7746 & 0.6500 & 0.1443 & 0.2552 & 0.3854 & 0.6808 & 0.6660 & 0.6428 & 0.5818 & 0.7377 \\
     & \cellcolor[HTML]{FFFFFF}{\color[HTML]{1F1F1F} Dirichlet LM} & 0.7898 & 0.7542 & 0.6432 & 0.1301 & 0.2365 & 0.3686 & 0.6380 & 0.6376 & 0.6206 & 0.5794 & 0.7219 \\
    \multirow{-5}{*}{Remove apostrophes} & \cellcolor[HTML]{FFFFFF}{\color[HTML]{1F1F1F} Hiemstra LM} & 0.8169 & 0.7712 & 0.6712 & 0.1429 & 0.2529 & 0.3953 & 0.6725 & 0.6609 & 0.6507 & 0.6102 & 0.7443 \\ \hline
     & BM25 & 0.8271 & 0.7881 & 0.6856 & 0.1459 & 0.2616 & 0.4069 & 0.7143 & 0.7014 & 0.6871 & 0.6498 & 0.8130 \\
     & DFR BM25 & 0.8271 & 0.7881 & 0.6856 & 0.1459 & 0.2616 & 0.4070 & 0.7138 & 0.7016 & 0.6873 & 0.6506 & 0.8135 \\
     & TF-IDF & 0.8271 & 0.7814 & 0.6805 & 0.1457 & 0.2573 & 0.4028 & 0.7118 & 0.6979 & 0.6845 & 0.6402 & 0.8077 \\
     & \cellcolor[HTML]{FFFFFF}{\color[HTML]{1F1F1F} Dirichlet LM} & 0.7898 & 0.7576 & 0.6797 & 0.1322 & 0.2420 & 0.3860 & 0.6578 & 0.6615 & 0.6652 & 0.6679 & 0.8039 \\
    \multirow{-5}{*}{Remove hyphens} & \cellcolor[HTML]{FFFFFF}{\color[HTML]{1F1F1F} Hiemstra LM} & 0.8339 & 0.7881 & 0.6898 & 0.1472 & 0.2635 & 0.4143 & 0.7142 & 0.6980 & 0.6914 & 0.6841 & 0.8239 \\ \hline
     & BM25 & {\color[HTML]{CC0000} 0.7831} & {\color[HTML]{CC0000} 0.7542} & {\color[HTML]{CC0000} 0.6424} & {\color[HTML]{CC0000} 0.1370} & {\color[HTML]{CC0000} 0.2445} & {\color[HTML]{CC0000} 0.3735} & {\color[HTML]{CC0000} 0.6564} & {\color[HTML]{CC0000} 0.6512} & {\color[HTML]{CC0000} 0.6308} & {\color[HTML]{CC0000} 0.5744} & {\color[HTML]{CC0000} 0.7329} \\
     & DFR BM25 & {\color[HTML]{CC0000} 0.7831} & {\color[HTML]{CC0000} 0.7542} & {\color[HTML]{CC0000} 0.6441} & {\color[HTML]{CC0000} 0.1366} & {\color[HTML]{CC0000} 0.2440} & {\color[HTML]{CC0000} 0.3734} & {\color[HTML]{CC0000} 0.6566} & {\color[HTML]{CC0000} 0.6509} & {\color[HTML]{CC0000} 0.6317} & {\color[HTML]{CC0000} 0.5747} & {\color[HTML]{CC0000} 0.7328} \\
     & TF-IDF & {\color[HTML]{CC0000} 0.7864} & {\color[HTML]{CC0000} 0.7508} & {\color[HTML]{CC0000} 0.6297} & {\color[HTML]{CC0000} 0.1364} & {\color[HTML]{CC0000} 0.2417} & {\color[HTML]{CC0000} 0.3666} & {\color[HTML]{CC0000} 0.6555} & {\color[HTML]{CC0000} 0.6477} & {\color[HTML]{CC0000} 0.6237} & {\color[HTML]{CC0000} 0.5631} & {\color[HTML]{CC0000} 0.7287} \\
     & \cellcolor[HTML]{FFFFFF}{\color[HTML]{1F1F1F} Dirichlet LM} & {\color[HTML]{CC0000} 0.7390} & {\color[HTML]{CC0000} 0.7237} & {\color[HTML]{CC0000} 0.6331} & {\color[HTML]{CC0000} 0.1159} & {\color[HTML]{CC0000} 0.2173} & {\color[HTML]{CC0000} 0.3503} & {\color[HTML]{CC0000} 0.5945} & {\color[HTML]{CC0000} 0.6081} & {\color[HTML]{CC0000} 0.6036} & {\color[HTML]{CC0000} 0.5588} & {\color[HTML]{CC0000} 0.7104} \\
    \multirow{-5}{*}{Remove accents} & \cellcolor[HTML]{FFFFFF}{\color[HTML]{1F1F1F} Hiemstra LM} & {\color[HTML]{CC0000} 0.7763} & {\color[HTML]{CC0000} 0.7441} & {\color[HTML]{CC0000} 0.6458} & {\color[HTML]{CC0000} 0.1341} & {\color[HTML]{CC0000} 0.2388} & {\color[HTML]{CC0000} 0.3732} & {\color[HTML]{CC0000} 0.6446} & {\color[HTML]{CC0000} 0.6406} & {\color[HTML]{CC0000} 0.6305} & {\color[HTML]{CC0000} 0.5880} & {\color[HTML]{CC0000} 0.7352} \\ \hline
    & BM25 & {\color[HTML]{CC0000} 0.8102} & {\color[HTML]{CC0000} 0.7729} & 0.6695 & {\color[HTML]{CC0000} 0.1438} & {\color[HTML]{CC0000} 0.2547} & 0.3976 & {\color[HTML]{CC0000} 0.6693} & {\color[HTML]{CC0000} 0.6600} & 0.6488 & 0.6030 & 0.7443 \\
     & DFR BM25 & {\color[HTML]{CC0000} 0.8102} & {\color[HTML]{CC0000} 0.7729} & 0.6712 & {\color[HTML]{CC0000} 0.1439} & {\color[HTML]{CC0000} 0.2549} & 0.3984 & {\color[HTML]{CC0000} 0.6695} & {\color[HTML]{CC0000} 0.6602} & 0.6503 & 0.6049 & 0.7451 \\
     & TF-IDF & {\color[HTML]{CC0000} 0.8102} & {\color[HTML]{CC0000} 0.7729} & 0.6686 & 0.1439 & 0.2549 & 0.3975 & {\color[HTML]{CC0000} 0.6691} & {\color[HTML]{CC0000} 0.6600} & 0.6484 & 0.6018 & 0.7438 \\
     & \cellcolor[HTML]{FFFFFF}{\color[HTML]{1F1F1F} Dirichlet LM} & 0.8034 & 0.7593 & 0.6653 & 0.1317 & 0.2379 & 0.3803 & {\color[HTML]{CC0000} 0.6329} & {\color[HTML]{CC0000} 0.6315} & 0.6299 & 0.5936 & 0.7255 \\
    \multirow{-5}{*}{Remove stopwords} & \cellcolor[HTML]{FFFFFF}{\color[HTML]{1F1F1F} Hiemstra LM} & 0.8203 & {\color[HTML]{CC0000} 0.7678} & 0.6864 & 0.1437 & 0.2521 & 0.4036 & 0.6702 & {\color[HTML]{CC0000} 0.6587} & 0.6577 & 0.6189 & 0.7483 \\ \hline
     & BM25 & {\color[HTML]{CC0000} 0.8000} & {\color[HTML]{CC0000} 0.7678} & {\color[HTML]{CC0000} 0.6500} & {\color[HTML]{CC0000} 0.1381} & {\color[HTML]{CC0000} 0.2464} & {\color[HTML]{CC0000} 0.3758} & {\color[HTML]{CC0000} 0.6693} & {\color[HTML]{CC0000} 0.6605} & {\color[HTML]{CC0000} 0.6355} & {\color[HTML]{CC0000} 0.5826} & {\color[HTML]{CC0000} 0.7364} \\
     & DFR BM25 & {\color[HTML]{CC0000} 0.8000} & {\color[HTML]{CC0000} 0.7661} & {\color[HTML]{CC0000} 0.6492} & {\color[HTML]{CC0000} 0.1376} & {\color[HTML]{CC0000} 0.2456} & {\color[HTML]{CC0000} 0.3748} & {\color[HTML]{CC0000} 0.6679} & {\color[HTML]{CC0000} 0.6588} & {\color[HTML]{CC0000} 0.6345} & {\color[HTML]{CC0000} 0.5824} & {\color[HTML]{CC0000} 0.7358} \\
     & TF-IDF & {\color[HTML]{CC0000} 0.7966} & {\color[HTML]{CC0000} 0.7610} & {\color[HTML]{CC0000} 0.6407} & {\color[HTML]{CC0000} 0.1371} & {\color[HTML]{CC0000} 0.2421} & {\color[HTML]{CC0000} 0.3700} & {\color[HTML]{CC0000} 0.6648} & {\color[HTML]{CC0000} 0.6551} & {\color[HTML]{CC0000} 0.6304} & {\color[HTML]{CC0000} 0.5711} & {\color[HTML]{CC0000} 0.7331} \\
     & \cellcolor[HTML]{FFFFFF}{\color[HTML]{1F1F1F} Dirichlet LM} & {\color[HTML]{CC0000} 0.7661} & {\color[HTML]{CC0000} 0.7203} & {\color[HTML]{CC0000} 0.6305} & {\color[HTML]{CC0000} 0.1225} & {\color[HTML]{CC0000} 0.2198} & {\color[HTML]{CC0000} 0.3518} & {\color[HTML]{CC0000} 0.6286} & {\color[HTML]{CC0000} 0.6192} & {\color[HTML]{CC0000} 0.6076} & {\color[HTML]{CC0000} 0.5613} & {\color[HTML]{CC0000} 0.7174} \\
    \multirow{-5}{*}{Light stemming} & \cellcolor[HTML]{FFFFFF}{\color[HTML]{1F1F1F} Hiemstra LM} & {\color[HTML]{CC0000} 0.7898} & {\color[HTML]{CC0000} 0.7492} & {\color[HTML]{CC0000} 0.6576} & {\color[HTML]{CC0000} 0.1315} & {\color[HTML]{CC0000} 0.2365} & {\color[HTML]{CC0000} 0.3760} & {\color[HTML]{CC0000} 0.6456} & {\color[HTML]{CC0000} 0.6410} & {\color[HTML]{CC0000} 0.6326} & {\color[HTML]{CC0000} 0.5907} & {\color[HTML]{CC0000} 0.7344}\\\hline   
     & BM25 & {\color[HTML]{CC0000} 0.7797} & {\color[HTML]{CC0000} 0.7610} & {\color[HTML]{CC0000} 0.6466} & {\color[HTML]{CC0000} 0.1336} & {\color[HTML]{CC0000} 0.2423} & {\color[HTML]{CC0000} 0.3714} & {\color[HTML]{CC0000} 0.6594} & {\color[HTML]{CC0000} 0.6553} & {\color[HTML]{CC0000} 0.6319} & {\color[HTML]{CC0000} 0.5790} & {\color[HTML]{CC0000} 0.7346} \\
     & DFR BM25 & {\color[HTML]{CC0000} 0.7797} & {\color[HTML]{CC0000} 0.7593} & {\color[HTML]{CC0000} 0.6458} & {\color[HTML]{CC0000} 0.1331} & {\color[HTML]{CC0000} 0.2415} & {\color[HTML]{CC0000} 0.3704} & {\color[HTML]{CC0000} 0.6581} & {\color[HTML]{CC0000} 0.6535} & {\color[HTML]{CC0000} 0.6309} & {\color[HTML]{CC0000} 0.5789} & {\color[HTML]{CC0000} 0.7341} \\
     & TF-IDF & {\color[HTML]{CC0000} 0.7763} & {\color[HTML]{CC0000} 0.7542} & {\color[HTML]{CC0000} 0.6373} & {\color[HTML]{CC0000} 0.1326} & {\color[HTML]{CC0000} 0.2380} & {\color[HTML]{CC0000} 0.3656} & {\color[HTML]{CC0000} 0.6550} & {\color[HTML]{CC0000} 0.6499} & {\color[HTML]{CC0000} 0.6268} & {\color[HTML]{CC0000} 0.5676} & {\color[HTML]{CC0000} 0.7313} \\
     & \cellcolor[HTML]{FFFFFF}{\color[HTML]{1F1F1F} Dirichlet LM} & {\color[HTML]{CC0000} 0.7593} & {\color[HTML]{CC0000} 0.7186} & {\color[HTML]{CC0000} 0.6254} & {\color[HTML]{CC0000} 0.1197} & {\color[HTML]{CC0000} 0.2175} & {\color[HTML]{CC0000} 0.3484} & {\color[HTML]{CC0000} 0.6225} & {\color[HTML]{CC0000} 0.6149} & {\color[HTML]{CC0000} 0.6025} & {\color[HTML]{CC0000} 0.5577} & {\color[HTML]{CC0000} 0.7138} \\
    \multirow{-5}{*}{Moderate stemming} & \cellcolor[HTML]{FFFFFF}{\color[HTML]{1F1F1F} Hiemstra LM} & {\color[HTML]{CC0000} 0.7729} & {\color[HTML]{CC0000} 0.7390} & {\color[HTML]{CC0000} 0.6525} & {\color[HTML]{CC0000} 0.1275} & {\color[HTML]{CC0000} 0.2319} & {\color[HTML]{CC0000} 0.3710} & {\color[HTML]{CC0000} 0.6365} & {\color[HTML]{CC0000} 0.6337} & {\color[HTML]{CC0000} 0.6277} & {\color[HTML]{CC0000} 0.5868} & {\color[HTML]{CC0000} 0.7322} \\ \hline
     & BM25 & {\color[HTML]{CC0000} 0.7797} & {\color[HTML]{CC0000} 0.7610} & {\color[HTML]{CC0000} 0.6466} & {\color[HTML]{CC0000} 0.1336} & {\color[HTML]{CC0000} 0.2423} & {\color[HTML]{CC0000} 0.3714} & {\color[HTML]{CC0000} 0.6594} & {\color[HTML]{CC0000} 0.6553} & {\color[HTML]{CC0000} 0.6319} & {\color[HTML]{CC0000} 0.5788} & {\color[HTML]{CC0000} 0.7346} \\ 
     & DFR BM25 & {\color[HTML]{CC0000} 0.7797} & {\color[HTML]{CC0000} 0.7593} & {\color[HTML]{CC0000} 0.6458} & {\color[HTML]{CC0000} 0.1333} & {\color[HTML]{CC0000} 0.2416} & {\color[HTML]{CC0000} 0.3705} & {\color[HTML]{CC0000} 0.6583} & {\color[HTML]{CC0000} 0.6537} & {\color[HTML]{CC0000} 0.6311} & {\color[HTML]{CC0000} 0.5788} & {\color[HTML]{CC0000} 0.7341} \\
     & TF-IDF & {\color[HTML]{CC0000} 0.7763} & {\color[HTML]{CC0000} 0.7542} & {\color[HTML]{CC0000} 0.6373} & {\color[HTML]{CC0000} 0.1326} & {\color[HTML]{CC0000} 0.2380} & {\color[HTML]{CC0000} 0.3656} & {\color[HTML]{CC0000} 0.6550} & {\color[HTML]{CC0000} 0.6499} & {\color[HTML]{CC0000} 0.6268} & {\color[HTML]{CC0000} 0.5674} & {\color[HTML]{CC0000} 0.7312} \\
     & \cellcolor[HTML]{FFFFFF}{\color[HTML]{1F1F1F} Dirichlet LM} & {\color[HTML]{CC0000} 0.7559} & {\color[HTML]{CC0000} 0.7186} & {\color[HTML]{CC0000} 0.6254} & {\color[HTML]{CC0000} 0.1189} & {\color[HTML]{CC0000} 0.2170} & {\color[HTML]{CC0000} 0.3479} & {\color[HTML]{CC0000} 0.6200} & {\color[HTML]{CC0000} 0.6145} & {\color[HTML]{CC0000} 0.6022} & {\color[HTML]{CC0000} 0.5571} & {\color[HTML]{CC0000} 0.7135} \\
    \multirow{-5}{*}{Heavy stemming} & \cellcolor[HTML]{FFFFFF}{\color[HTML]{1F1F1F} Hiemstra LM} & {\color[HTML]{CC0000} 0.7729} & {\color[HTML]{CC0000} 0.7390} & {\color[HTML]{CC0000} 0.6534} & {\color[HTML]{CC0000} 0.1275} & {\color[HTML]{CC0000} 0.2319} & {\color[HTML]{CC0000} 0.3715} & {\color[HTML]{CC0000} 0.6365} & {\color[HTML]{CC0000} 0.6337} & {\color[HTML]{CC0000} 0.6282} & {\color[HTML]{CC0000} 0.5867} & {\color[HTML]{CC0000} 0.7322} \\ \hline
    \end{tabular}%
}
\end{table}

%% file: tables/short-text-retrieval-final.tex
\begin{table}[!ht]
\centering
\caption{Effectiveness of Combined Text Preprocessing Techniques in Short-Text Retrieval. Values highlighted with a \textit{green} background indicate the best score at the respective metric and cutoff.}
\label{tab:evaluation-titles}
\resizebox{\textwidth}{!}{%
      \begin{tabular}{p{2.5cm}llrrrrrrrrrrr}
      \hline
     &  & \multicolumn{3}{c}{Precision at Cutoff} & \multicolumn{3}{c}{MAP at Cutoff} & \multicolumn{3}{c}{NDCG at Cutoff} & \multicolumn{1}{l}{} & \multicolumn{1}{l}{} \\ \cline{3-11}
    \multirow{-2}{*}{Retrieval Strategies} & \multirow{-2}{*}{Model} & \multicolumn{1}{c}{@5} & \multicolumn{1}{c}{@10} & \multicolumn{1}{c}{@20} & \multicolumn{1}{c}{@5} & \multicolumn{1}{c}{@10} & \multicolumn{1}{c}{@20} & \multicolumn{1}{c}{@5} & \multicolumn{1}{c}{@10} & \multicolumn{1}{c}{@20} & \multicolumn{1}{l}{\multirow{-2}{*}{MAP}} & \multicolumn{1}{l}{\multirow{-2}{*}{NDCG}} \\ \hline
     & BM25 & 0.8169 & 0.7763 & 0.6602 & 0.1444 & 0.2568 & 0.3903 & 0.6801 & 0.6668 & 0.6454 & 0.5925 & 0.7408 \\
     & DFR BM25 & 0.8169 & 0.7763 & 0.6619 & 0.1440 & 0.2563 & 0.3901 & 0.6811 & 0.6666 & 0.6468 & 0.5926 & 0.7407 \\
     & TF-IDF & 0.8136 & 0.7746 & 0.6458 & 0.1432 & 0.2546 & 0.3825 & 0.6739 & 0.6640 & 0.6380 & 0.5802 & 0.7364 \\
     & Dirichlet LM & 0.7898 & 0.7525 & 0.6398 & 0.1299 & 0.2361 & 0.3671 & 0.6359 & 0.6356 & 0.6174 & 0.5780 & 0.7208 \\
    \multirow{-5}{*}{Baseline} & Hiemstra LM & 0.8136 & 0.7695 & 0.6669 & 0.1428 & 0.2521 & 0.3928 & 0.6670 & 0.6588 & 0.6465 & 0.6090 & 0.7435 \\ \hline
    & BM25 & 0.8237 & 0.7763 & 0.6644 & 0.1453 & 0.2572 & 0.3930 & 0.6866 & 0.6685 & 0.6499 & 0.5938 & 0.7419 \\
     & DFR BM25 & 0.8237 & 0.7763 & 0.6661 & 0.1450 & 0.2568 & 0.3929 & 0.6878 & 0.6684 & 0.6515 & 0.5942 & 0.7420 \\
     & TF-IDF & 0.8203 & 0.7746 & 0.6500 & 0.1443 & 0.2552 & 0.3854 & 0.6808 & 0.6660 & 0.6428 & 0.5818 & 0.7377 \\
     & \cellcolor[HTML]{FFFFFF}{\color[HTML]{1F1F1F} Dirichlet LM} & 0.7898 & 0.7542 & 0.6432 & 0.1301 & 0.2365 & 0.3686 & 0.6380 & 0.6376 & 0.6206 & 0.5794 & 0.7219 \\
    \multirow{-5}{*}{Remove apostrophes} & \cellcolor[HTML]{FFFFFF}{\color[HTML]{1F1F1F} Hiemstra LM} & 0.8169 & 0.7712 & 0.6712 & 0.1429 & 0.2529 & 0.3953 & 0.6725 & 0.6609 & 0.6507 & 0.6102 & 0.7443 \\ \hline
     & BM25 & 0.8271 & 0.7881 & 0.6856 & 0.1459 & 0.2616 & 0.4069 & 0.7143 & 0.7014 & 0.6871 & 0.6498 & 0.8130 \\
     & DFR BM25 & 0.8271 & 0.7881 & 0.6856 & 0.1459 & 0.2616 & 0.4070 & 0.7138 & 0.7016 & 0.6873 & 0.6506 & 0.8135 \\
     & TF-IDF & 0.8271 & 0.7814 & 0.6805 & 0.1457 & 0.2573 & 0.4028 & 0.7118 & 0.6979 & 0.6845 & 0.6402 & 0.8077 \\
     & Dirichlet LM & 0.7898 & 0.7576 & 0.6797 & 0.1322 & 0.2420 & 0.3860 & 0.6578 & 0.6615 & 0.6652 & 0.6679 & 0.8039 \\
    \multirow{-5}{*}{Remove hyphens} & Hiemstra LM & 0.8339 & 0.7881 & 0.6898 & 0.1472 & 0.2635 & 0.4143 & 0.7142 & 0.6980 & 0.6914 & 0.6841 & 0.8239 \\ \hline

    & BM25 & 0.8102 & 0.7729 & 0.6695 & 0.1438 & 0.2547 & 0.3976 & 0.6693 & 0.6600 & 0.6488 & 0.6030 & 0.7443 \\
     & DFR BM25 & 0.8102 & 0.7729 & 0.6712 & 0.1439 & 0.2549 & 0.3984 & 0.6695 & 0.6602 & 0.6503 & 0.6049 & 0.7451 \\
     & TF-IDF & 0.8102 & 0.7729 & 0.6686 & 0.1439 & 0.2549 & 0.3975 & 0.6691 & 0.6600 & 0.6484 & 0.6018 & 0.7438 \\
     & Dirichlet LM & 0.8034 & 0.7593 & 0.6653 & 0.1317 & 0.2379 & 0.3803 & 0.6329 & 0.6315 & 0.6299 & 0.5936 & 0.7255 \\
    \multirow{-5}{*}{Remove stopwords} & Hiemstra LM & 0.8203 & 0.7678 & 0.6864 & 0.1437 & 0.2521 & 0.4036 & 0.6702 & 0.6587 & 0.6577 & 0.6189 & 0.7483 \\ \hline
     & BM25 & \cellcolor[HTML]{B7E1CD}0.8881 & 0.8373 & 0.7153 & 0.1553 & 0.2796 & 0.4304 & 0.7500 & 0.7347 & 0.7133 & 0.6648 & 0.8213 \\
     & DFR BM25 & \cellcolor[HTML]{B7E1CD}0.8881 & \cellcolor[HTML]{B7E1CD}0.8390 & 0.7169 & 0.1553 & \cellcolor[HTML]{B7E1CD}0.2804 & 0.4313 & \cellcolor[HTML]{B7E1CD}0.7512 & \cellcolor[HTML]{B7E1CD}0.7356 & 0.7149 & 0.6664 & 0.8219 \\
     & TF-IDF & 0.8780 & 0.8322 & 0.7119 & 0.1543 & 0.2759 & 0.4273 & 0.7401 & 0.7288 & 0.7086 & 0.6553 & 0.8149 \\
     & Dirichlet LM & 0.8407 & 0.8034 & 0.7068 & 0.1390 & 0.2561 & 0.4099 & 0.6834 & 0.6920 & 0.6829 & 0.6713 & 0.8018 \\
    \multirow{-5}{*}{\parbox{2.6cm}{Remove apostrophes and hyphens}} & Hiemstra LM & 0.8780 & 0.8305 & 0.7263 & 0.1524 & 0.2743 & 0.4339 & 0.7379 & 0.7245 & 0.7147 & 0.6955 & 0.8282 \\ \hline
     & BM25 & 0.8814 & 0.8237 & 0.7237 & 0.1576 & 0.2720 & 0.4356 & 0.7394 & 0.7221 & 0.7130 & 0.6752 & 0.8220 \\
     & DFR BM25 & 0.8847 & 0.8254 & 0.7237 & 0.1585 & 0.2729 & 0.4366 & 0.7416 & 0.7228 & 0.7139 & 0.6764 & 0.8224 \\
     & TF-IDF & 0.8847 & 0.8237 & 0.7229 & 0.1585 & 0.2722 & 0.4355 & 0.7416 & 0.7220 & 0.7126 & 0.6715 & 0.8202 \\
     & Dirichlet LM & 0.8508 & 0.8102 & 0.7220 & 0.1409 & 0.2549 & 0.4171 & 0.6933 & 0.6925 & 0.6918 & 0.6863 & 0.8065 \\
    \multirow{-5}{*}{\parbox{2.3cm}{Remove hyphens and stopwords}} & Hiemstra LM & 0.8746 & 0.8305 & \cellcolor[HTML]{B7E1CD}0.7305 & 0.1524 & 0.2720 & \cellcolor[HTML]{B7E1CD}0.4372 & 0.7294 & 0.7211 & \cellcolor[HTML]{B7E1CD}0.7152 & \cellcolor[HTML]{B7E1CD}0.7040 & 0.8288 \\ \hline
     & BM25 & 0.8814 & 0.8220 & 0.7212 & 0.1580 & 0.2725 & 0.4342 & 0.7395 & 0.7211 & 0.7123 & 0.6743 & 0.8223 \\
     & DFR BM25 & 0.8847 & 0.8237 & 0.7212 & \cellcolor[HTML]{B7E1CD}0.1589 & 0.2735 & 0.4353 & 0.7418 & 0.7218 & 0.7131 & 0.6756 & 0.8228 \\
     & TF-IDF & 0.8847 & 0.8220 & 0.7203 & \cellcolor[HTML]{B7E1CD}0.1589 & 0.2727 & 0.4341 & 0.7417 & 0.7210 & 0.7118 & 0.6705 & 0.8205 \\
     & Dirichlet LM & 0.8508 & 0.8068 & 0.7144 & 0.1418 & 0.2546 & 0.4142 & 0.6949 & 0.6913 & 0.6872 & 0.6831 & 0.8053 \\
    \multirow{-5}{*}{\parbox{2.3cm}{Remove hyphens, apostrophes, and stopwords}} & Hiemstra LM & 0.8746 & 0.8254 & 0.7263 & 0.1528 & 0.2711 & 0.4353 & 0.7287 & 0.7190 & 0.7134 & 0.7029 & \cellcolor[HTML]{B7E1CD}0.8289 \\ \hline
     & BM25 & 0.8678 & 0.8169 & 0.7110 & 0.1540 & 0.2698 & 0.4267 & 0.7367 & 0.7203 & 0.7053 & 0.6602 & 0.8154 \\
     & DFR BM25 & 0.8712 & 0.8153 & 0.7110 & 0.1549 & 0.2695 & 0.4268 & 0.7389 & 0.7200 & 0.7058 & 0.6608 & 0.8157 \\
     & TF-IDF & 0.8712 & 0.8169 & 0.7102 & 0.1549 & 0.2701 & 0.4267 & 0.7389 & 0.7201 & 0.7048 & 0.6584 & 0.8145 \\
     & Dirichlet LM & 0.8339 & 0.7932 & 0.7034 & 0.1390 & 0.2503 & 0.4056 & 0.6936 & 0.6860 & 0.6800 & 0.6622 & 0.8023 \\
    \multirow{-5}{*}{\parbox{2.3cm}{Remove hyphens and stopwords, and apply light stemmer}} & Hiemstra LM & 0.8576 & 0.8203 & 0.7229 & 0.1496 & 0.2678 & 0.4289 & 0.7256 & 0.7171 & 0.7083 & 0.6831 & 0.8208 \\ \hline
    \multicolumn{2}{l}{\parbox{4.5cm}{Average performance gains of the best model compared to the baseline}} & \textit{8.72\%} & \textit{8.08\%} & \textit{9.54\%} & \textit{10.66\%} & \textit{9.40\%} & \textit{11.30\%} & \textit{10.29\%} & \textit{30.35\%} & \textit{10.63\%} &\textit{15.60\%} & \textit{11.49\%} \\ \hline
 
    \multicolumn{2}{l}{\parbox{4.5cm}{Average performance gains of the best model compared to individual text preprocessing techniques}} & \textit{7.60\%} & \textit{7.39\%} & \textit{6.16\%} & \textit{9.75\%} & \textit{8.19\%} & \textit{6.93\%} & \textit{7.23\%} & \textit{7.45\%} & \textit{6.09\%} &\textit{8.33\%} & \textit{7.58\%} \\ \hline
    
    \multicolumn{2}{l}{\parbox{4.5cm}{Average performance gains of the best model compared to others within the same retrieval strategy}} & \textbf{2.65\%} & \textbf{1.62\%} & \textbf{1.03\%} & \textbf{5.54\%} & \textbf{3.41\%} & \textbf{1.43\%} & \textbf{3.35\%} & \textbf{2.22\%} & \textbf{1.06\%} &\textbf{3.94\%} & \textbf{1.38\%} \\ \hline
    \end{tabular}%
}
\end{table}

%% file: tables/long-text-retrieval-preliminary.tex
\begin{table}[!ht]
\centering
\caption{Effectiveness of Text Preprocessing Techniques in Long-Text Retrieval. Red values indicate scores lower than the baseline.}
\label{tab:evaluation-preprocessing-technique-contents}
\resizebox{\textwidth}{!}{%
      \begin{tabular}{p{2.5cm}llrrrrrrrrrrr}
      \hline
     &  & \multicolumn{3}{c}{Precision at Cutoff} & \multicolumn{3}{c}{MAP at Cutoff} & \multicolumn{3}{c}{NDCG at Cutoff} & \multicolumn{1}{l}{} & \multicolumn{1}{l}{} \\ \cline{3-11}
     \multirow{-2}{*}{Retrieval Strategies} & \multirow{-2}{*}{Model} & \multicolumn{1}{c}{@5} & \multicolumn{1}{c}{@10} & \multicolumn{1}{c}{@20} & \multicolumn{1}{c}{@5} & \multicolumn{1}{c}{@10} & \multicolumn{1}{c}{@20} & \multicolumn{1}{c}{@5} & \multicolumn{1}{c}{@10} & \multicolumn{1}{c}{@20} & \multicolumn{1}{l}{\multirow{-2}{*}{MAP}} & \multicolumn{1}{l}{\multirow{-2}{*}{NDCG}} \\ \hline  
    & BM25 & 0.4847 & 0.4525 & 0.3839 & 0.0769 & 0.1281 & 0.1931 & 0.3883 & 0.3800 & 0.3765 & 0.3429 & 0.5764 \\
    & DFR BM25 & 0.4746 & 0.4475 & 0.3839 & 0.0758 & 0.1259 & 0.1925 & 0.3826 & 0.3763 & 0.3753 & 0.3416 & 0.5754 \\
    & TF-IDF & 0.5288 & 0.4746 & 0.4110 & 0.0855 & 0.1382 & 0.2068 & 0.4275 & 0.4086 & 0.4045 & 0.3564 & 0.5927 \\
    & Dirichlet LM & 0.4576 & 0.4186 & 0.3669 & 0.0696 & 0.1111 & 0.1701 & 0.3577 & 0.3544 & 0.3545 & 0.3110 & 0.5558 \\
    \multirow{-5}{*}{Baseline} & Hiemstra LM & 0.5390 & 0.4915 & 0.4314 & 0.0856 & 0.1402 & 0.2138 & 0.4289 & 0.4158 & 0.4189 & 0.3655 & 0.5990 \\ \hline
    & BM25 & 0.4881 & {\color[HTML]{CC0000} 0.4458} & 0.3847 & 0.0775 & {\color[HTML]{CC0000} 0.1277} & 0.1942 & 0.3902 & {\color[HTML]{CC0000} 0.3787} & 0.3783 & 0.3441 & 0.5774 \\
    & DFR BM25 & 0.4780 & {\color[HTML]{CC0000} 0.4407} & 0.3839 & 0.0764 & {\color[HTML]{CC0000} 0.1256} & 0.1934 & 0.3843 & {\color[HTML]{CC0000} 0.3749} & 0.3765 & 0.3428 & 0.5763 \\
    & TF-IDF & 0.5288 & {\color[HTML]{CC0000} 0.4695} & 0.4110 & 0.0860 & 0.1388 & 0.2079 & 0.4278 & 0.4088 & 0.4061 & 0.3578 & 0.5938 \\
    & \cellcolor[HTML]{FFFFFF}Dirichlet LM & 0.4576 & 0.4220 & {\color[HTML]{CC0000} 0.3661} & {\color[HTML]{CC0000} 0.0691} & 0.1121 & 0.1707 & 0.3585 & 0.3571 & 0.3549 & 0.3120 & 0.5567 \\
    \multirow{-5}{*}{Remove apostrophes} & \cellcolor[HTML]{FFFFFF}Hiemstra LM & 0.5424 & 0.4915 & {\color[HTML]{CC0000} 0.4297} & 0.0862 & 0.1408 & 0.2149 & 0.4328 & 0.4183 & 0.4200 & 0.3671 & 0.6001 \\ \hline
    & BM25 & 0.5322 & 0.4966 & 0.4407 & 0.0869 & 0.1437 & 0.2265 & 0.4315 & 0.4270 & 0.4324 & 0.4042 & 0.6570 \\
    & DFR BM25 & 0.5288 & 0.4949 & 0.4407 & 0.0860 & 0.1425 & 0.2250 & 0.4259 & 0.4247 & 0.4302 & 0.4026 & 0.6547 \\
    & TF-IDF & 0.5966 & 0.5390 & 0.4686 & 0.0942 & 0.1580 & 0.2425 & 0.4812 & 0.4677 & 0.4649 & 0.4224 & 0.6783 \\
    & \cellcolor[HTML]{FFFFFF}Dirichlet LM & 0.5051 & 0.4729 & 0.4153 & 0.0813 & 0.1338 & 0.2034 & 0.3961 & 0.3967 & 0.4014 & 0.3729 & 0.6364 \\
    \multirow{-5}{*}{Remove hyphens} & \cellcolor[HTML]{FFFFFF}Hiemstra LM & 0.5797 & 0.5542 & 0.4907 & 0.0946 & 0.1620 & 0.2503 & 0.4688 & 0.4726 & 0.4771 & 0.4338 & 0.6827 \\ \hline
    & BM25 & 0.4915 & {\color[HTML]{CC0000} 0.4441} & {\color[HTML]{CC0000} 0.3754} & 0.0798 & 0.1283 & {\color[HTML]{CC0000} 0.1900} & 0.3977 & 0.3815 & 0.3773 & {\color[HTML]{CC0000} 0.3413} & 0.5798 \\
    & DFR BM25 & 0.4780 & {\color[HTML]{CC0000} 0.4373} & {\color[HTML]{CC0000} 0.3746} & 0.0784 & 0.1261 & {\color[HTML]{CC0000} 0.1889} & 0.3899 & 0.3778 & 0.3754 & {\color[HTML]{CC0000} 0.3396} & 0.5785 \\
    & TF-IDF & {\color[HTML]{CC0000} 0.5153} & {\color[HTML]{CC0000} 0.4627} & {\color[HTML]{CC0000} 0.3966} & 0.0870 & {\color[HTML]{CC0000} 0.1369} & {\color[HTML]{CC0000} 0.2013} & 0.4280 & {\color[HTML]{CC0000} 0.4074} & {\color[HTML]{CC0000} 0.4011} & {\color[HTML]{CC0000} 0.3540} & 0.5948 \\
    & \cellcolor[HTML]{FFFFFF}Dirichlet LM & {\color[HTML]{CC0000} 0.4542} & {\color[HTML]{CC0000} 0.4102} & {\color[HTML]{CC0000} 0.3534} & {\color[HTML]{CC0000} 0.0657} & {\color[HTML]{CC0000} 0.1055} & {\color[HTML]{CC0000} 0.1592} & {\color[HTML]{CC0000} 0.3517} & {\color[HTML]{CC0000} 0.3457} & {\color[HTML]{CC0000} 0.3421} & {\color[HTML]{CC0000} 0.3037} & {\color[HTML]{CC0000} 0.5497} \\
    \multirow{-5}{*}{Remove accents} & \cellcolor[HTML]{FFFFFF}Hiemstra LM & 0.5424 & {\color[HTML]{CC0000} 0.4881} & {\color[HTML]{CC0000} 0.4186} & 0.0861 & 0.1405 & {\color[HTML]{CC0000} 0.2108} & 0.4382 & 0.4198 & {\color[HTML]{CC0000} 0.4153} & {\color[HTML]{CC0000} 0.3634} & 0.6001 \\ \hline
    & BM25 & 0.5390 & 0.4847 & 0.4212 & 0.0891 & 0.1465 & 0.2182 & 0.4286 & 0.4157 & 0.4156 & 0.3716 & 0.6098 \\
    & DFR BM25 & 0.5356 & 0.4864 & 0.4195 & 0.0886 & 0.1463 & 0.2169 & 0.4267 & 0.4155 & 0.4138 & 0.3706 & 0.6088 \\
    & TF-IDF & 0.5559 & 0.4831 & 0.4203 & 0.0924 & 0.1470 & 0.2193 & 0.4423 & 0.4185 & 0.4174 & 0.3716 & 0.6123 \\
    & \cellcolor[HTML]{FFFFFF}Dirichlet LM & 0.4712 & 0.4271 & 0.3703 & 0.0723 & 0.1176 & 0.1788 & 0.3640 & 0.3553 & 0.3585 & 0.3184 & 0.5656 \\
    \multirow{-5}{*}{Remove stopwords} & Hiemstra LM & 0.5458 & 0.5102 & 0.4347 & 0.0895 & 0.1505 & 0.2243 & 0.4341 & 0.4283 & 0.4263 & 0.3796 & 0.6164 \\ \hline
    & BM25 & 0.4881 & 0.4695 & 0.3992 & 0.0819 & 0.1409 & 0.2096 & 0.3944 & 0.3970 & 0.3949 & 0.3650 & 0.6060 \\
    & DFR BM25 & 0.4881 & 0.4712 & 0.3983 & 0.0818 & 0.1410 & 0.2088 & 0.3941 & 0.3979 & 0.3939 & 0.3643 & 0.6056 \\
    & TF-IDF & 0.5525 & 0.4949 & 0.4339 & 0.0925 & 0.1500 & 0.2252 & 0.4449 & 0.4270 & 0.4298 & 0.3810 & 0.6244 \\
    & Dirichlet LM & 0.4712 & 0.4441 & 0.3805 & 0.0701 & 0.1197 & 0.1810 & {\color[HTML]{CC0000} 0.3563} & 0.3616 & 0.3631 & 0.3290 & 0.5762 \\ 
    \multirow{-5}{*}{Heavy stemming} & Hiemstra LM & 0.5458 & 0.5153 & 0.4492 & 0.0868 & 0.1479 & 0.2275 & 0.4312 & 0.4321 & 0.4366 & 0.3871 & 0.6261 \\ \hline
    & BM25 & 0.4915 & 0.4712 & 0.3992 & 0.0822 & 0.1413 & 0.2099 & 0.3970 & 0.3981 & 0.3953 & 0.3656 & 0.6065 \\
    & DFR BM25 & 0.4915 & 0.4712 & 0.3983 & 0.0821 & 0.1411 & 0.2090 & 0.3967 & 0.3982 & 0.3942 & 0.3648 & 0.6060 \\
    & TF-IDF & 0.5525 & 0.4949 & 0.4339 & 0.0925 & 0.1500 & 0.2252 & 0.4449 & 0.4270 & 0.4298 & 0.3814 & 0.6246 \\
    & Dirichlet LM & 0.4712 & 0.4441 & 0.3805 & 0.0701 & 0.1197 & 0.1810 & {\color[HTML]{CC0000} 0.3563} & 0.3616 & 0.3631 & 0.3293 & 0.5764 \\
    \multirow{-5}{*}{Moderate stemming} & \cellcolor[HTML]{FFFFFF}Hiemstra LM & 0.5458 & 0.5153 & 0.4492 & 0.0868 & 0.1479 & 0.2276 & 0.4312 & 0.4321 & 0.4366 & 0.3874 & 0.6263 \\ \hline
    & BM25 & 0.4949 & 0.4763 & 0.4008 & 0.0826 & 0.1421 & 0.2106 & 0.3996 & 0.4015 & 0.3963 & 0.3675 & 0.6079 \\
    & DFR BM25 & 0.4949 & 0.4763 & 0.4008 & 0.0825 & 0.1420 & 0.2102 & 0.3993 & 0.4016 & 0.3958 & 0.3668 & 0.6075 \\
    & TF-IDF & 0.5559 & 0.5000 & 0.4381 & 0.0929 & 0.1508 & 0.2266 & 0.4475 & 0.4305 & 0.4323 & 0.3837 & 0.6261 \\
    & Dirichlet LM & 0.4814 & 0.4508 & 0.3831 & 0.0716 & 0.1216 & 0.1829 & 0.3655 & 0.3701 & 0.3691 & 0.3317 & 0.5817 \\
    \multirow{-5}{*}{Light stemming} & Hiemstra LM & 0.5492 & 0.5203 & 0.4542 & 0.0880 & 0.1496 & 0.2305 & 0.4360 & 0.4369 & 0.4421 & 0.3906 & 0.6301\\ \hline
    \end{tabular}%
}
\end{table}

%% file: tables/long-text-retrieval-final.tex
\begin{table}[!ht]
\centering
\caption{Effectiveness of Combined Text Preprocessing Techniques in Long-Text Retrieval. Values highlighted with a \textit{green} background indicate the best score at the respective metric and cutoff.}
\label{tab:evaluation-contents}
\resizebox{\textwidth}{!}{%
      \begin{tabular}{p{2.5cm}llrrrrrrrrrrr}
      \hline
     &  & \multicolumn{3}{c}{Precision at Cutoff} & \multicolumn{3}{c}{MAP at Cutoff} & \multicolumn{3}{c}{NDCG at Cutoff} & \multicolumn{1}{l}{} & \multicolumn{1}{l}{} \\ \cline{3-11}
    \multirow{-2}{*}{Retrieval Strategies} & \multirow{-2}{*}{Model} & \multicolumn{1}{c}{@5} & \multicolumn{1}{c}{@10} & \multicolumn{1}{c}{@20} & \multicolumn{1}{c}{@5} & \multicolumn{1}{c}{@10} & \multicolumn{1}{c}{@20} & \multicolumn{1}{c}{@5} & \multicolumn{1}{c}{@10} & \multicolumn{1}{c}{@20} & \multicolumn{1}{l}{\multirow{-2}{*}{MAP}} & \multicolumn{1}{l}{\multirow{-2}{*}{NDCG}} \\ \hline
     & BM25 & 0.4847 & 0.4525 & 0.3839 & 0.0769 & 0.1281 & 0.1931 & 0.3883 & 0.3800 & 0.3765 & 0.3429 & 0.5764 \\
     & DFR BM25 & 0.4746 & 0.4475 & 0.3839 & 0.0758 & 0.1259 & 0.1925 & 0.3826 & 0.3763 & 0.3753 & 0.3416 & 0.5754 \\
     & TF-IDF & 0.5288 & 0.4746 & 0.4110 & 0.0855 & 0.1382 & 0.2068 & 0.4275 & 0.4086 & 0.4045 & 0.3564 & 0.5927 \\
     & Dirichlet LM & 0.4576 & 0.4186 & 0.3669 & 0.0696 & 0.1111 & 0.1701 & 0.3577 & 0.3544 & 0.3545 & 0.3110 & 0.5558 \\
    \multirow{-5}{*}{Baseline} & Hiemstra LM & 0.5390 & 0.4915 & 0.4314 & 0.0856 & 0.1402 & 0.2138 & 0.4289 & 0.4158 & 0.4189 & 0.3655 & 0.5990 \\ \hline
     & BM25 & 0.5322 & 0.4966 & 0.4407 & 0.0869 & 0.1437 & 0.2265 & 0.4315 & 0.4270 & 0.4324 & 0.4042 & 0.6570 \\
     & DFR BM25 & 0.5288 & 0.4949 & 0.4407 & 0.0860 & 0.1425 & 0.2250 & 0.4259 & 0.4247 & 0.4302 & 0.4026 & 0.6547 \\
     & TF-IDF & 0.5966 & 0.5390 & 0.4686 & 0.0942 & 0.1580 & 0.2425 & 0.4812 & 0.4677 & 0.4649 & 0.4224 & 0.6783 \\
     & \cellcolor[HTML]{FFFFFF}Dirichlet LM & 0.5051 & 0.4729 & 0.4153 & 0.0813 & 0.1338 & 0.2034 & 0.3961 & 0.3967 & 0.4014 & 0.3729 & 0.6364 \\
    \multirow{-5}{*}{Remove hyphens} & \cellcolor[HTML]{FFFFFF}Hiemstra LM & 0.5797 & 0.5542 & \cellcolor[HTML]{B7E1CD}0.4907 & 0.0946 & 0.1620 & 0.2503 & 0.4688 & 0.4726 & 0.4771 & 0.4338 & 0.6827 \\ \hline
     & BM25 & 0.5322 & 0.4949 & 0.4407 & 0.0869 & 0.1434 & 0.2271 & 0.4300 & 0.4254 & 0.4328 & 0.4047 & 0.6567 \\
     & DFR BM25 & 0.5254 & 0.4932 & 0.4407 & 0.0857 & 0.1423 & 0.2256 & 0.4230 & 0.4232 & 0.4307 & 0.4031 & 0.6543 \\
     & TF-IDF & 0.5966 & 0.5390 & 0.4678 & 0.0944 & 0.1581 & 0.2432 & 0.4788 & 0.4664 & 0.4650 & 0.4232 & 0.6790 \\
     & \cellcolor[HTML]{FFFFFF}Dirichlet LM & 0.5017 & 0.4729 & 0.4136 & 0.0809 & 0.1348 & 0.2043 & 0.3953 & 0.3983 & 0.4011 & 0.3737 & 0.6369 \\
    \multirow{-5}{*}{\parbox{2.3cm}{Remove apostrophes and hyphens}} & \cellcolor[HTML]{FFFFFF}Hiemstra LM & 0.5831 & \cellcolor[HTML]{B7E1CD}0.5576 & \cellcolor[HTML]{B7E1CD}0.4907 & 0.0955 & 0.1636 & \cellcolor[HTML]{B7E1CD}0.2523 & 0.4721 & 0.4759 & \cellcolor[HTML]{B7E1CD}0.4790 & 0.4355 & 0.6840 \\ \hline
     & BM25 & 0.5322 & 0.4932 & 0.4280 & 0.0875 & 0.1442 & 0.2222 & 0.4349 & 0.4251 & 0.4273 & 0.4020 & 0.6585 \\
     & DFR BM25 & 0.5254 & 0.4932 & 0.4280 & 0.0861 & 0.1439 & 0.2211 & 0.4316 & 0.4272 & 0.4278 & 0.4007 & 0.6578 \\
     & TF-IDF & 0.5831 & 0.5288 & 0.4525 & 0.0948 & 0.1541 & 0.2339 & 0.4744 & 0.4588 & 0.4555 & 0.4175 & 0.6769 \\
     & \cellcolor[HTML]{FFFFFF}Dirichlet LM & 0.5017 & 0.4576 & 0.4025 & 0.0779 & 0.1261 & 0.1923 & 0.3882 & 0.3847 & 0.3875 & 0.3645 & 0.6288 \\
    \multirow{-5}{*}{\parbox{2.3cm}{Remove hyphens and accents}} & \cellcolor[HTML]{FFFFFF}Hiemstra LM & 0.5831 & 0.5508 & 0.4763 & 0.0955 & \cellcolor[HTML]{B7E1CD}0.1645 & 0.2474 & 0.4812 & 0.4748 & 0.4737 & 0.4330 & \cellcolor[HTML]{B7E1CD}0.6855 \\ \hline
     & BM25 & 0.5797 & 0.5220 & 0.4619 & 0.0926 & 0.1547 & 0.2361 & 0.4614 & 0.4492 & 0.4521 & 0.4195 & 0.6722 \\ 
     & DFR BM25 & 0.5763 & 0.5203 & 0.4568 & 0.0918 & 0.1549 & 0.2344 & 0.4585 & 0.4483 & 0.4491 & 0.4189 & 0.6718 \\
     & TF-IDF & \cellcolor[HTML]{B7E1CD}0.6102 & 0.5339 & 0.4661 & 0.0961 & 0.1574 & 0.2403 & \cellcolor[HTML]{B7E1CD}0.4849 & 0.4622 & 0.4612 & 0.4232 & 0.6777 \\
     & \cellcolor[HTML]{FFFFFF}Dirichlet LM & 0.4983 & 0.4695 & 0.4068 & 0.0742 & 0.1276 & 0.1946 & 0.3841 & 0.3860 & 0.3886 & 0.3604 & 0.6235 \\
    \multirow{-5}{*}{\parbox{2.3cm}{Remove hyphens and stopwords}} & \cellcolor[HTML]{FFFFFF}Hiemstra LM & 0.5932 & 0.5458 & 0.4797 & 0.0947 & 0.1601 & 0.2451 & 0.4780 & 0.4722 & 0.4714 & 0.4339 & 0.6840 \\ \hline
     & BM25 & 0.5729 & 0.5153 & 0.4610 & 0.0922 & 0.1540 & 0.2366 & 0.4575 & 0.4457 & 0.4514 & 0.4199 & 0.6726 \\
     & DFR BM25 & 0.5695 & 0.5136 & 0.4576 & 0.0913 & 0.1542 & 0.2352 & 0.4545 & 0.4447 & 0.4492 & 0.4193 & 0.6720 \\
     & TF-IDF & \cellcolor[HTML]{B7E1CD}0.6102 & 0.5322 & 0.4661 & 0.0960 & 0.1570 & 0.2410 & 0.4838 & 0.4618 & 0.4608 & 0.4236 & 0.6779 \\
     & \cellcolor[HTML]{FFFFFF}Dirichlet LM & 0.4881 & 0.4678 & 0.4042 & 0.0731 & 0.1270 & 0.1940 & 0.3786 & 0.3848 & 0.3858 & 0.3611 & 0.6232 \\
    \multirow{-5}{*}{\parbox{2.3cm}{Remove hyphens, apostrophes, and stopwords}} & \cellcolor[HTML]{FFFFFF}Hiemstra LM & 0.6000 & 0.5492 & 0.4797 & 0.0959 & 0.1628 & 0.2471 & 0.4826 & \cellcolor[HTML]{B7E1CD}0.4760 & 0.4727 & \cellcolor[HTML]{B7E1CD}0.4358 & 0.6853 \\ \hline
     & BM25 & 0.5322 & 0.5017 & 0.4322 & 0.0849 & 0.1459 & 0.2236 & 0.4156 & 0.4176 & 0.4202 & 0.3981 & 0.6450 \\
     & DFR BM25 & 0.5254 & 0.4966 & 0.4297 & 0.0844 & 0.1447 & 0.2214 & 0.4110 & 0.4136 & 0.4173 & 0.3968 & 0.6435 \\
     & TF-IDF & 0.6000 & 0.5441 & 0.4636 & 0.0942 & 0.1583 & 0.2364 & 0.4681 & 0.4589 & 0.4561 & 0.4160 & 0.6700 \\
     & \cellcolor[HTML]{FFFFFF}Dirichlet LM & 0.4915 & 0.4712 & 0.4076 & 0.0788 & 0.1330 & 0.1989 & 0.3795 & 0.3872 & 0.3911 & 0.3663 & 0.6274 \\
    \multirow{-5}{*}{\parbox{2.3cm}{Remove hyphens, apostrophes, and moderate stemmer}} & \cellcolor[HTML]{FFFFFF}Hiemstra LM & 0.5864 & 0.5559 & 0.4805 & \cellcolor[HTML]{B7E1CD}0.0965 & 0.1631 & 0.2475 & 0.4695 & 0.4734 & 0.4702 & 0.4301 & 0.6780\\ \hline

    \multicolumn{2}{l}{\parbox{4.5cm}{Average performance gains of the best model compared to the baseline}} & \textit{15.39\%} & \textit{13.45\%} & \textit{13.75\%} & \textit{12.73\%} & \textit{17.33\%} & \textit{18.01\%} & \textit{13.43\%} & \textit{14.48\%} & \textit{14.35\%} &\textit{19.23\%} & \textit{14.44\%} \\ \hline
    \multicolumn{2}{l}{\parbox{4.5cm}{Average performance gains of the best model compared to others within the same retrieval strategy}} & \textbf{9.61\%} & \textbf{11.77\%} & \textbf{11.47\%} & \textbf{13.23\%} & \textbf{16.40\%} & \textbf{12.54\%} & \textbf{9.64\%} & \textbf{10.15\%} & \textbf{11.08\%} &\textbf{7.82\%} & \textbf{4.65\%} \\ \hline
    \end{tabular}%
}
\end{table}

%% file: tables/tetun-stopword-list.tex
\begin{table}[!ht]
\centering
\caption{The~\textit{Labadain-Stopwords}.}
\label{tab:list-of-stopwords}
\resizebox{\textwidth}{!}{%
    \begin{tabular}{llp{2cm}llp{2.3cm}llp{2.3cm}llp{1.8cm}} \\ \hline
    No. & Tetun & English & No. & Tetun & English & No. & Tetun & English & No. & Tetun & English \\ \hline
    1 & an & self & 41 & Hamutuk & with, together & 81 & laek & less & 121 & nunka & never \\
    2 & aleinde & besides & 42 & hanesan & such as & 82 & lai & for a while & 122 & o & you \\
    3 & ami & we & 43 & hela & remain & 83 & laiha & without & 123 & oin & next/sort/front \\
    4 & ami-nia & our & 44 & hikas & again & 84 & lalais & quickly & 124 & oin-oin & various \\
    5 & antes & before, previously & 45 & hira & how much & 85 & laran & inside & 125 & oinsá & how \\
    6 & atu & so that, to & 46 & hirak & those & 86 & leten & on, in & 126 & oioin & diverse \\
    7 & atubele & in order to & 47 & hirak-ne’e & these & 87 & liu & exceed, more than & 127 & oituan & few \\
    8 & ba & to, for & 48 & ho & with & 88 & liubá & ago & 128 & okos & below \\
    9 & baibain & usuallly, commonly & 49 & hodi & so that & 89 & liuhosi & through & 129 & ona & already \\
    10 & bainhira & when, while & 50 & hosi & from & 90 & liuhusi & through & 130 & ou & or \\
    11 & balu & some & 51 & hotu & too, also & 91 & liuliu & especially, particularly & 131 & para & to, in order to \\
    12 & barak & many & 52 & hotu-hotu & all & 92 & liután & further, more & 132 & portantu & so, therefore \\
    13 & bazeia & based (on), according (to) & 53 & husi & from & 93 & loloos & exactly, correctly & 133 & rasik & self, own \\
    14 & beibeik & often, always & 54 & i & and & 94 & loos & very, correct & 134 & resin & over, excess \\
    15 & bele & be able to, could, may & 55 & ida & a, an & 95 & lubuk & a lot of, many & 135 & ruma & some, any \\
    16 & besik & near, nearby, almost & 56 & ida-idak & each & 96 & mai & to, toward & 136 & sai & become, out \\
    17 & buat & thing & 57 & ida-ne’e & this & 97 & maibé & but & 137 & saida & what \\
    18 & dala & time(s) & 58 & ida-ne’ebé & which one & 98 & mais & however & 138 & se & if, whether \\
    19 & dalaruma & sometimes & 59 & iha & be, exist & 99 & maizumenus & more or less, approximately & 139 & sé & who \\
    20 & daudauk & currently & 60 & imi & you (plural) & 100 & mak & to be & 140 & sei & will, still \\
    21 & daudaun & currently & 61 & inklui & include & 101 & maka & to be & 141 & seidauk & not yet \\
    22 & de’it & only, just & 62 & ita & we & 102 & malu & each other & 142 & sein & without \\
    23 & depois & after that, then, later & 63 & ita-boot & you & 103 & mas & but & 143 & seluk & other \\
    24 & dezde & since, from & 64 & ita-nia & yours & 104 & maski & despite, although & 144 & sempre & always \\
    25 & didi’ak & carefully, thoroughly & 65 & ka & or & 105 & menus & less & 145 & sira & they \\
    26 & duke & than & 66 & kada & each, every & 106 & mezmu & despite, although & 146 & sira-ne’e & these \\
    27 & duni & indeed & 67 & karik & maybe & 107 & molok & before & 147 & sira-ne’ebé & those who \\
    28 & durante & during & 68 & katak & that & 108 & mós & also & 148 & sira-nia & their \\
    29 & eh & or & 69 & kedas & beforehand, immediately & 109 & nafatin & still, remain & 149 & sira-nian & theirs \\
    30 & enkuantu & while & 70 & komesa & from, begin & 110 & ne’e & this & 150 & só & only, unless \\
    31 & entaun & so, then & 71 & kona-ba & about & 111 & ne’ebá & that & 151 & tan & more \\
    32 & entre & between, among & 72 & kotuk & behind, last & 112 & ne’ebé & where & 152 & tanba & because \\
    33 & entretantu & meanwhile & 73 & kraik & below & 113 & nia & he, she & 153 & tantu & so \\
    34 & fali & again & 74 & kuandu & whenever, while & 114 & nian & of & 154 & tebes & very, so \\
    35 & filafali & again & 75 & kuaze & almost & 115 & ninia & his, her & 155 & tenke & must \\
    36 & foin & only just & 76 & la & not & 116 & ninian & his, hers & 156 & tiha & already \\
    37 & ha’u & I & 77 & la’ós & not & 117 & no & and & 157 & to’o & until \\
    38 & ha’u-nia & my & 78 & labele & unable, don't & 118 & nomós & also & 158 & tomak & entire \\
    39 & hafoin & after & 79 & ladún & not very, not so & 119 & nu’udar & as & 159 & tuir & according to \\
    40 & hahú & from, begin & 80 & lae & no & 120 & nune’e & like this, in this way & 160 & uitoan & few,  a little\\ \hline
    \end{tabular}%
    }
\end{table}

%% file: tables/linguistic-regions.tex
\begin{table}[!ht]
\centering
  \caption{Linguistic Regions Used in the \textit{Labadain-Stemmer}.}
  \label{tab:linguistic-regions}
  \begin{tabularx}{\textwidth}{lX}
    \toprule
    Region & Definition\\
    \midrule
    R1 & The region starting after the first non-vowel that follows a vowel, or, if no such non-vowel exists, it is the null region at the end of the word.\\
    \rowcolor{gray!10}R2 & The region starting after the first non-vowel that follows a vowel within R1, or, if no such non-vowel exists, it is the null region at the end of the word.\\
    RV & If the second letter is a consonant, RV starts after the next vowel. If the first two letters are vowels, it begins after the following consonant. In the case of a consonant-vowel combination, RV starts after the third letter. If none of these conditions are met, RV starts at the end of the word.\\
  \bottomrule
\end{tabularx}
\end{table}

%% file: tables/pt-derived-suffixes.tex
\begin{table}[!ht]
\centering
    \caption{List of Portuguese-Derived Suffixes.}
    \label{tab:suffixes-description}
    \begin{tabularx}{\textwidth}{Xll}
        \toprule
        Suffix & Variable & Description\\
        \midrule
        \rowcolor{orange!10}eza, ezas, iku, ika, ikus, ikas, izmu, izmus, ável, ível, ista, istas, ozu, oza, ozus, ozas, amentu, amentus, imentu, imentus, adora, adór, asaun, adoras, adores, asoens, ante, antes, ánsia, atória, atóriu, atórias, atórius, amentál & $general\_suf$ & A list contains general suffixes\\
        
        lojia, lojias & $lojia\_suf$ & A list contains \textit{loj} and \textit{lojia} suffixes\\
        
        \rowcolor{orange!10}usaun, usoens & $usaun\_suf$ & A list contains \textit{usaun} and \textit{usoens} suffixes\\
        
        énsia, énsias & $ensia\_suf$ & A list contains \textit{énsia} and \textit{énsias} suffixes\\
        
        \rowcolor{orange!10}amente & $amente\_suf$ & A string with \textit{amente} value\\

        iv (appears before the \textit{amente} suffix) & $iv\_suf$ & A string with \textit{iv} value\\
        
        \rowcolor{orange!10}at (takes precedence over the \textit{iv}, \textit{iva},~\textit{ivu},~\textit{ivas}, or \textit{ivus} suffixes) & $at\_suf$ & A string with \textit{at} value\\

        oz, ik, ad (presents before the \textit{amente} suffix) & $ozikad\_suf$ & A list contains \textit{oz},~\textit{ik}, and \textit{ad} suffixes\\
        
        \rowcolor{orange!10}mente & $mente\_suf$ & A string with \textit{mente} value\\
        
        ante, avel, ivel (appears before the \textit{mente} suffix) & $ante\_suf$ & A list contains \textit{ante},~\textit{avel}, and \textit{ivel} suffixes\\
        
        \rowcolor{orange!10}idade, idades & $idade\_suf$ & A list contains \textit{idade} and \textit{idades} suffixes\\
        
        abil, is, iv (takes precedence over the \textit{idade} or \textit{idades} suffixes)  &  $abil\_suf$ & A list contains \textit{abil},~\textit{is}, and \textit{iv} suffixes\\
        
        \rowcolor{orange!10}iva, ivu, ivas, ivus & $iva\_suf$ & A list contains \textit{iva},~\textit{ivu},~\textit{ivas} and \textit{ivus} suffixes\\
        
         ada, adu, adas, adus, ida, idu, idas, idus, ária, áriu, árias, árius & $verb\_suf$ & A list contains \textit{verb} suffixes\\
        
        \rowcolor{orange!10}a, e, i, u, us, as & $residual\_suf$ & A list contains \textit{residual} suffixes\\ 
        \bottomrule
    \end{tabularx}
\end{table}

%% file: algorithms/light_stemmer.tex
\newcounter{algcounter} 
\newcommand{\algcaption}[1]{%
  \refstepcounter{algcounter} 
  \par\noindent
  \textbf{Algorithm \thealgcounter:} #1\par
  \label{alg:tetun-light-stemmer}
}

\algcaption{Tetun Light Stemmer Algorithm}
\begin{algorithmic}[1]
    \Require $R1$, $R2$, $RV$, $word\_list$
    \Require $general\_suf \gets$ list of \textbf{general} suffixes
    \Require $lojia\_suf \gets$ list of \textbf{lojia} suffixes
    \Require $usaun\_suf \gets$ list of \textbf{usaun} suffixes
    \Require $ensia\_suf \gets$ list of \textbf{ensia} suffixes
    \Require $amente\_suf \gets$ \textbf{amente} suffix
    \Require $iv\_suf \gets$ \textbf{iv} suffix
    \Require $at\_suf \gets$ \textbf{at} suffix
    \Require $ozikad\_suf \gets$ list of \textbf{ozikad} suffixes 
    \Require $mente\_suf \gets$ \textbf{mente} suffix
    \Require $ante\_suf \gets$ list of \textbf{ante} suffixes 
    \Require $idade\_suf \gets$ list of \textbf{idade} suffixes
    \Require $abil\_suf \gets$ list of \textbf{abil} suffixes 
    \Require $iva\_suf \gets$ list of \textbf{iva} suffixes
    \Require $verb\_suf \gets$ list of \textbf{verb} suffixes
    \Require $residual\_suf \gets$ list of \textbf{residual} suffixes

    \ForAll{$word$ \textbf{in} $word\_list$}
        \Comment{\textbf{Step 1:} Word length validation}
        \If{$\text{length}(word) < 4$}
            \State \textbf{Return} $word$
        \Else
            \Comment{\textbf{Step 2:} Standard suffix removal}
            \If{$word$ \textbf{ends with} any $suffix$ \textbf{in} $general\_suf$ sorted by length descending}
                 \If{\textbf{Position of} $suffix$ \textbf{in} $word$ \textbf{is in} $R2$}
                     \State \textbf{Delete} $suffix$ \textbf{from} $word$
                     \State $stem \gets$ $word$ \textbf{without} $suffix$
                     \State \textbf{Return} $stem$
                 \EndIf
            \ElsIf{$word$ \textbf{ends with} any $suffix$ \textbf{in} $lojia\_suf$}
                 \If{\textbf{Position of} $suffix$ \textbf{in} $word$ \textbf{is in} $R2$}
                     \State \textbf{Replace} $suffix$ \textbf{in} $word$ \textbf{with} $loj$
                     \State $stem \gets$ $word$ \textbf{without} $suffix$ \textbf{concatenates with} $loj$
                     \State \textbf{Return} $stem$
                 \EndIf
            \ElsIf{$word$ \textbf{ends with} any $suffix$ \textbf{in} $usaun\_suf$}
                 \If{\textbf{Position of} $suffix$ \textbf{in} $word$ \textbf{is in} $R2$}
                     \State \textbf{Replace} $suffix$ \textbf{in} $word$ \textbf{with} $u$
                     \State $stem \gets$ $word$ \textbf{without} $suffix$ \textbf{concatenates with} $u$
                     \State \textbf{Return} $stem$
                 \EndIf
            \ElsIf{$word$ \textbf{ends with} any $suffix$ \textbf{in} $ensia\_suf$}
                 \If{\textbf{Position of} $suffix$ \textbf{in} $word$ \textbf{is in} $R2$}
                     \State \textbf{Replace} $suffix$ \textbf{in} $word$ \textbf{with} $ente$
                     \State $stem \gets$ $word$ \textbf{without} $suffix$ \textbf{concatenates with} $ente$
                     \State \textbf{Return} $stem$
                 \EndIf
            \ElsIf{$word$ \textbf{ends with} $suffix$ \textbf{equals} $amente\_suf$}
                 \If{\textbf{Position of} $suffix$ \textbf{in} $word$ \textbf{is in} $R1$}
                    \State \textbf{Delete} $suffix$ \textbf{in} $word$
                    \State $stem\_amente \gets$ $word$ \textbf{without} $suffix$
                    \If{$stem\_amente$ \textbf{preceded by} $iv\_suf$ \textbf{and the position of} $iv\_suf$ \textbf{is in} $R2$}
                         \State \textbf{Delete} $iv\_suf$ \textbf{in} $stem\_amente$
                         \State $stem\_iv \gets$ $stem\_amente$ \textbf{without} $iv\_suf$
                        \If{$stem\_iv$ \textbf{further proceeded by} $at\_suf$ \textbf{and the position of} $at\_suf$ \textbf{is in} $R2$}
                             \State \textbf{Delete} $at\_suf$ \textbf{in} $stem\_iv$
                             \State $stem\_at \gets$ $stem\_iv$ \textbf{without} $at\_suf$
                             \State \textbf{Return} $stem\_at$  
                        \EndIf
                            \State \textbf{Return} $stem\_iv$
                    \ElsIf{$stem\_amente$ \textbf{preceded by} any $suffix\_oid$ \textbf{in} $ozikad\_suf$}
                        \If{\textbf{Position of} $suffix\_oid$ \textbf{in} $stem\_amente$ \textbf{is in} $R2$}
                            \State \textbf{Delete} $suffix\_oid$ \textbf{in} $stem\_amente$
                            \State $stem\_ozikad \gets$ $stem\_amente$ \textbf{without} $suffix\_oid$
                            \State \textbf{Return} $stem\_ozikad$
                        \EndIf
                    \EndIf
                    \State \textbf{Return} $stem\_amente$  
                \EndIf
            \ElsIf{$word$ \textbf{ends with} $suffix$ \textbf{equals} $mente\_suf$}
                 \If{\textbf{Position of} $suffix$ \textbf{in} $word$ \textbf{is in} $R2$}
                    \State \textbf{Delete} $suffix$ \textbf{in} $word$
                    \State $stem\_mente \gets$ $word$ \textbf{without} $suffix$
                    \If{$stem\_mente$ \textbf{preceded by} any $suffix\_ant$ in $ante\_suf$}
                        \If{\textbf{Position of} $suffix\_ant$ \textbf{in} $stem\_mente$ \textbf{is in} $R2$}
                            \State \textbf{Delete} $suffix\_ant$ \textbf{in} $stem\_mente$
                            \State $stem\_ante \gets$ $stem\_mente$ \textbf{without} $suffix\_ant$
                            \State \textbf{Return} $stem\_ante$
                        \EndIf
                    \EndIf
                    \State \textbf{Return} $stem\_mente$ 
                \EndIf
            \ElsIf{$word$ \textbf{ends with} $suffix$ \textbf{equals} $idade\_suf$}
                \If{\textbf{Position of} $suffix$ \textbf{in} $word$ \textbf{is in} $R2$}
                    \State \textbf{Delete} $suffix$ \textbf{in} $word$
                    \State $stem\_idade \gets$ $word$ \textbf{without} $suffix$
                    \If{$stem\_idade$ \textbf{preceded by} any $suffix\_abl$ in $abil\_suf$}
                        \If{\textbf{Position of} $suffix\_abl$ \textbf{in} $stem\_idade$ \textbf{is in} $R2$}
                            \State \textbf{Delete} $suffix\_abl$ \textbf{in} $stem\_idade$
                            \State $stem\_abil \gets$ $stem\_idade$ \textbf{without} $suffix\_abl$
                            \State \textbf{Return} $stem\_abil$
                        \EndIf
                    \EndIf
                    \State \textbf{Return} $stem\_idade$ 
                \EndIf
            \ElsIf{$word$ \textbf{ends with} $suffix$ \textbf{equals} $iva\_suf$}
                \If{\textbf{Position of} $suffix$ \textbf{in} $word$ \textbf{is in} $R2$}
                    \State \textbf{Delete} $suffix$ \textbf{in} $word$
                    \State $stem\_iva \gets$ $word$ \textbf{without} $suffix$
                    \If{$stem\_iva$ \textbf{preceded by} $at\_suf$ \textbf{and the position of} $at\_suf$ \textbf{is in} $R2$}
                        \State \textbf{Delete} $at\_suf$ \textbf{in} $stem\_iva$
                        \State $stem\_at \gets$ $stem\_iva$ \textbf{without} $at\_suf$
                        \State \textbf{Return} $stem\_at$
                    \EndIf
                    \State \textbf{Return} $stem\_iva$
                \EndIf
            \Comment{\textbf{Step 3:} Verb suffix removal}
            \ElsIf{$word$ \textbf{ends with} $suffix$ \textbf{equals} $verb\_suf$}
                 \If{\textbf{Position of} $suffix$ \textbf{in} $word$ \textbf{is in} $RV$}
                    \State \textbf{Delete} $suffix$ \textbf{in} $word$
                    \State $stem\_verb \gets$ $word$ \textbf{without} $suffix$
                    \State \textbf{Return} $stem\_verb$
                \EndIf
            \Comment{\textbf{Step 4:} Residual suffix removal}
            \ElsIf{$word$ \textbf{ends with} $suffix$ \textbf{equals} $residual\_suf$}
                 \If{\textbf{Position of} $suffix$ \textbf{in} $word$ \textbf{is in} $RV$}
                    \State \textbf{Delete} $suffix$ \textbf{in} $word$
                    \State $stem\_residual \gets$ $word$ \textbf{without} $suffix$
                    \State \textbf{Return} $stem\_residual$
                \EndIf
            \Else
                \State \textbf{Return} $word$
            \EndIf
        \EndIf
    \EndFor
\end{algorithmic}

%% file: main.bbl

\begin{thebibliography}{84}


\ifx \showCODEN    \undefined \def \showCODEN     #1{\unskip}     \fi
\ifx \showDOI      \undefined \def \showDOI       #1{#1}\fi
\ifx \showISBNx    \undefined \def \showISBNx     #1{\unskip}     \fi
\ifx \showISBNxiii \undefined \def \showISBNxiii  #1{\unskip}     \fi
\ifx \showISSN     \undefined \def \showISSN      #1{\unskip}     \fi
\ifx \showLCCN     \undefined \def \showLCCN      #1{\unskip}     \fi
\ifx \shownote     \undefined \def \shownote      #1{#1}          \fi
\ifx \showarticletitle \undefined \def \showarticletitle #1{#1}   \fi
\ifx \showURL      \undefined \def \showURL       {\relax}        \fi
\providecommand\bibfield[2]{#2}
\providecommand\bibinfo[2]{#2}
\providecommand\natexlab[1]{#1}
\providecommand\showeprint[2][]{arXiv:#2}

\bibitem[Adriani et~al\mbox{.}(2007)]%
        {adriani-et-al-2007}
\bibfield{author}{\bibinfo{person}{Mirna Adriani}, \bibinfo{person}{Jelita Asian}, \bibinfo{person}{Bobby A.~A. Nazief}, \bibinfo{person}{Seyed M.~M. Tahaghoghi}, {and} \bibinfo{person}{Hugh~E. Williams}.} \bibinfo{year}{2007}\natexlab{}.
\newblock \showarticletitle{Stemming Indonesian: {A} confix-stripping approach}.
\newblock \bibinfo{journal}{\emph{{ACM} Transactions on Asian Language Information Processing}} \bibinfo{volume}{6}, \bibinfo{number}{4} (\bibinfo{year}{2007}), \bibinfo{pages}{1--33}.
\newblock
\urldef\tempurl%
\url{https://doi.org/10.1145/1316457.1316459}
\showDOI{\tempurl}


\bibitem[AleAhmad et~al\mbox{.}(2009)]%
        {aleahmad-et-al-2009}
\bibfield{author}{\bibinfo{person}{Abolfazl AleAhmad}, \bibinfo{person}{Hadi Amiri}, \bibinfo{person}{Ehsan Darrudi}, \bibinfo{person}{Masoud Rahgozar}, {and} \bibinfo{person}{Farhad Oroumchian}.} \bibinfo{year}{2009}\natexlab{}.
\newblock \showarticletitle{Hamshahri: {A} standard Persian text collection}.
\newblock \bibinfo{journal}{\emph{Knowledge-Based Systems}} \bibinfo{volume}{22}, \bibinfo{number}{5} (\bibinfo{year}{2009}), \bibinfo{pages}{382--387}.
\newblock
\urldef\tempurl%
\url{https://doi.org/10.1016/J.KNOSYS.2009.05.002}
\showDOI{\tempurl}


\bibitem[Ali et~al\mbox{.}(2024)]%
        {ali-etal-2024-network}
\bibfield{author}{\bibinfo{person}{Felermino D. M.~A. Ali}, \bibinfo{person}{Gabriel de Jesus}, \bibinfo{person}{Henrique~Lopes Cardoso}, \bibinfo{person}{S{\'e}rgio Nunes}, {and} \bibinfo{person}{Rui Sousa-Silva}.} \bibinfo{year}{2024}\natexlab{}.
\newblock \showarticletitle{Network-based Approach for Stopwords Detection}. In \bibinfo{booktitle}{\emph{Proceedings of the 16th International Conference on Computational Processing of Portuguese - Vol. 2}}, \bibfield{editor}{\bibinfo{person}{Pablo Gamallo}, \bibinfo{person}{Daniela Claro}, \bibinfo{person}{Ant{\'o}nio Teixeira}, \bibinfo{person}{Livy Real}, \bibinfo{person}{Marcos Garcia}, \bibinfo{person}{Hugo~Gon{\c{c}}alo Oliveira}, {and} \bibinfo{person}{Raquel Amaro}} (Eds.). \bibinfo{publisher}{Association for Computational Lingustics}, \bibinfo{address}{Santiago de Compostela, Galicia/Spain}, \bibinfo{pages}{55--63}.
\newblock
\urldef\tempurl%
\url{https://aclanthology.org/2024.propor-2.9}
\showURL{%
\tempurl}


\bibitem[Ardiyanti et~al\mbox{.}(2018)]%
        {ardiyanti-et-al-2018}
\bibfield{author}{\bibinfo{person}{S.~Arie Ardiyanti}, \bibinfo{person}{Dwi~Hendratmo Widyantoro}, \bibinfo{person}{Ayu Purwarianti}, {and} \bibinfo{person}{Yayat Sudaryat}.} \bibinfo{year}{2018}\natexlab{}.
\newblock \showarticletitle{The Rule-Based Sundanese Stemmer}.
\newblock \bibinfo{journal}{\emph{{ACM} Transactions on Asian Language Information Processing}} \bibinfo{volume}{17}, \bibinfo{number}{4} (\bibinfo{year}{2018}), \bibinfo{pages}{27:1--27:28}.
\newblock
\urldef\tempurl%
\url{https://doi.org/10.1145/3195634}
\showDOI{\tempurl}


\bibitem[Aronoff(1983)]%
        {aronoff-1983}
\bibfield{author}{\bibinfo{person}{Mark Aronoff}.} \bibinfo{year}{1983}\natexlab{}.
\newblock \showarticletitle{A Decade of Morphology and Word Formation}.
\newblock \bibinfo{journal}{\emph{Annual Review of Anthropology}}  \bibinfo{volume}{12} (\bibinfo{year}{1983}), \bibinfo{pages}{355--375}.
\newblock
\urldef\tempurl%
\url{https://www.jstor.org/stable/2155652}
\showURL{%
\tempurl}


\bibitem[Baeza{-}Yates and Ribeiro{-}Neto(2011)]%
        {baeza-yates-netos-2011}
\bibfield{author}{\bibinfo{person}{Ricardo Baeza{-}Yates} {and} \bibinfo{person}{Berthier~A. Ribeiro{-}Neto}.} \bibinfo{year}{2011}\natexlab{}.
\newblock \bibinfo{booktitle}{\emph{Modern Information Retrieval - the concepts and technology behind search, Second edition}}.
\newblock \bibinfo{publisher}{Pearson Education Ltd.}, \bibinfo{address}{Harlow, England}.
\newblock
\showISBNx{978-0-321-41691-9}
\urldef\tempurl%
\url{http://www.mir2ed.org/}
\showURL{%
\tempurl}


\bibitem[Beitzel et~al\mbox{.}(2004)]%
        {beitzel-et-al-2004}
\bibfield{author}{\bibinfo{person}{Steven~M. Beitzel}, \bibinfo{person}{Eric~C. Jensen}, \bibinfo{person}{Abdur Chowdhury}, \bibinfo{person}{David~A. Grossman}, {and} \bibinfo{person}{Ophir Frieder}.} \bibinfo{year}{2004}\natexlab{}.
\newblock \showarticletitle{Hourly analysis of a very large topically categorized web query log}. In \bibinfo{booktitle}{\emph{{SIGIR} 2004: Proceedings of the 27th Annual International {ACM} {SIGIR} Conference on Research and Development in Information Retrieval, Sheffield, UK, July 25-29, 2004}}, \bibfield{editor}{\bibinfo{person}{Mark Sanderson}, \bibinfo{person}{Kalervo J{\"{a}}rvelin}, \bibinfo{person}{James Allan}, {and} \bibinfo{person}{Peter Bruza}} (Eds.). \bibinfo{publisher}{{ACM}}, \bibinfo{pages}{321--328}.
\newblock
\urldef\tempurl%
\url{https://doi.org/10.1145/1008992.1009048}
\showDOI{\tempurl}


\bibitem[Braschler and Ripplinger(2004)]%
        {braschler-ripplinger-2004}
\bibfield{author}{\bibinfo{person}{Martin Braschler} {and} \bibinfo{person}{B{\"{a}}rbel Ripplinger}.} \bibinfo{year}{2004}\natexlab{}.
\newblock \showarticletitle{How Effective is Stemming and Decompounding for German Text Retrieval?}
\newblock \bibinfo{journal}{\emph{Information Retrieval}} \bibinfo{volume}{7}, \bibinfo{number}{3-4} (\bibinfo{year}{2004}), \bibinfo{pages}{291--316}.
\newblock
\urldef\tempurl%
\url{https://doi.org/10.1023/B:INRT.0000011208.60754.A1}
\showDOI{\tempurl}


\bibitem[Chavula and Suleman(2021)]%
        {chavula-suleman-2021}
\bibfield{author}{\bibinfo{person}{Catherine Chavula} {and} \bibinfo{person}{Hussein Suleman}.} \bibinfo{year}{2021}\natexlab{}.
\newblock \showarticletitle{Ranking by Language Similarity for Resource Scarce Southern Bantu Languages}. In \bibinfo{booktitle}{\emph{{ICTIR} '21: The 2021 {ACM} {SIGIR} International Conference on the Theory of Information Retrieval, Virtual Event, Canada, July 11, 2021}}, \bibfield{editor}{\bibinfo{person}{Faegheh Hasibi}, \bibinfo{person}{Yi~Fang}, {and} \bibinfo{person}{Akiko Aizawa}} (Eds.). \bibinfo{publisher}{{ACM}}, \bibinfo{pages}{137--147}.
\newblock
\urldef\tempurl%
\url{https://doi.org/10.1145/3471158.3472251}
\showDOI{\tempurl}


\bibitem[Cleverdon(1967)]%
        {cleverdon-1976}
\bibfield{author}{\bibinfo{person}{Cyril Cleverdon}.} \bibinfo{year}{1967}\natexlab{}.
\newblock \showarticletitle{The Cranfield tests on index language devices}. In \bibinfo{booktitle}{\emph{Aslib proceedings}}, Vol.~\bibinfo{volume}{19 (6)}. \bibinfo{publisher}{MCB UP Ltd}, \bibinfo{address}{Leeds, England}, \bibinfo{pages}{173--194}.
\newblock
\urldef\tempurl%
\url{https://doi.org/10.1108/eb050097}
\showDOI{\tempurl}


\bibitem[Cohen(1960)]%
        {cohen_1960}
\bibfield{author}{\bibinfo{person}{Jacob Cohen}.} \bibinfo{year}{1960}\natexlab{}.
\newblock \showarticletitle{A Coefficient of Agreement for Nominal Scales}.
\newblock \bibinfo{journal}{\emph{Educational and Psychological Measurement}}  \bibinfo{volume}{20} (\bibinfo{year}{1960}), \bibinfo{pages}{37--46}.
\newblock
\urldef\tempurl%
\url{https://doi.org/10.1177/001316446002000104}
\showDOI{\tempurl}


\bibitem[Correia et~al\mbox{.}(2005)]%
        {dictionary-tetun-inl-2005}
\bibfield{author}{\bibinfo{person}{Adérito José~Guterres Correia}, \bibinfo{person}{Geoffrey~Stephen Hull}, \bibinfo{person}{Geoge~William Saunders}, {and} \bibinfo{person}{Domingos dos~Santos Rosa~da Costa~Tilman, Mário Adriano~Soares}.} \bibinfo{year}{2005}\natexlab{}.
\newblock \bibinfo{booktitle}{\emph{Disionáriu Nasionál ba Tetun Ofisiál}}.
\newblock \bibinfo{publisher}{Instituto Nacional de Linguística, Universidade Nacional Timor Lorosa'e}, \bibinfo{address}{Avenida Cidade de Lisboa, Dili, Timor-Leste}.
\newblock
\showISBNx{1-74138-160-6}


\bibitem[Croft et~al\mbox{.}(2009)]%
        {croft-et-al-2009}
\bibfield{author}{\bibinfo{person}{W.~Bruce Croft}, \bibinfo{person}{Donald Metzler}, {and} \bibinfo{person}{Trevor Strohman}.} \bibinfo{year}{2009}\natexlab{}.
\newblock \bibinfo{booktitle}{\emph{Search Engines - Information Retrieval in Practice}}.
\newblock \bibinfo{publisher}{Pearson Education}, \bibinfo{address}{New York City, {USA}}.
\newblock
\showISBNx{978-0-13-136489-9}
\urldef\tempurl%
\url{http://www.search-engines-book.com/}
\showURL{%
\tempurl}


\bibitem[de~Jesus(2022)]%
        {de-jesus-2022-ir}
\bibfield{author}{\bibinfo{person}{Gabriel de Jesus}.} \bibinfo{year}{2022}\natexlab{}.
\newblock \bibinfo{title}{Text Information Retrieval in Tetun: A Preliminary Study}.
\newblock
\newblock
\showeprint[arxiv]{2406.07331}~[cs.IR]
\urldef\tempurl%
\url{https://arxiv.org/abs/2406.07331}
\showURL{%
\tempurl}
\newblock
\shownote{This work was published on the 10th edition of the PhD Symposium on FDIA, July 20, 2022, Lisbon, Portugal}.


\bibitem[de~Jesus(2023)]%
        {de-jesus-2023}
\bibfield{author}{\bibinfo{person}{Gabriel de Jesus}.} \bibinfo{year}{2023}\natexlab{}.
\newblock \showarticletitle{Text Information Retrieval in Tetun}. In \bibinfo{booktitle}{\emph{Advances in Information Retrieval - 45th European Conference on Information Retrieval, {ECIR} 2023, Dublin, Ireland, April 2-6, 2023, Proceedings, Part {III}}} \emph{(\bibinfo{series}{Lecture Notes in Computer Science}, Vol.~\bibinfo{volume}{13982})}, \bibfield{editor}{\bibinfo{person}{Jaap Kamps}, \bibinfo{person}{Lorraine Goeuriot}, \bibinfo{person}{Fabio Crestani}, \bibinfo{person}{Maria Maistro}, \bibinfo{person}{Hideo Joho}, \bibinfo{person}{Brian Davis}, \bibinfo{person}{Cathal Gurrin}, \bibinfo{person}{Udo Kruschwitz}, {and} \bibinfo{person}{Annalina Caputo}} (Eds.). \bibinfo{publisher}{Springer}, \bibinfo{address}{Switzerland}, \bibinfo{pages}{429--435}.
\newblock
\urldef\tempurl%
\url{https://doi.org/10.1007/978-3-031-28241-6\_48}
\showDOI{\tempurl}


\bibitem[de~Jesus and Nunes(2024a)]%
        {de-jesus-nunes-2024-labadain}
\bibfield{author}{\bibinfo{person}{Gabriel de Jesus} {and} \bibinfo{person}{S{\'e}rgio Nunes}.} \bibinfo{year}{2024}\natexlab{a}.
\newblock \showarticletitle{Labadain-30k+: A Monolingual {T}etun Document-Level Audited Dataset}. In \bibinfo{booktitle}{\emph{Proceedings of the 3rd Annual Meeting of the Special Interest Group on Under-resourced Languages @ LREC-COLING 2024}}, \bibfield{editor}{\bibinfo{person}{Maite Melero}, \bibinfo{person}{Sakriani Sakti}, {and} \bibinfo{person}{Claudia Soria}} (Eds.). \bibinfo{publisher}{ELRA and ICCL}, \bibinfo{address}{Torino, Italia}, \bibinfo{pages}{177--188}.
\newblock
\urldef\tempurl%
\url{https://aclanthology.org/2024.sigul-1.22}
\showURL{%
\tempurl}


\bibitem[de~Jesus and Nunes(2024b)]%
        {labadain30k-dataset}
\bibfield{author}{\bibinfo{person}{Gabriel de Jesus} {and} \bibinfo{person}{Sérgio Nunes}.} \bibinfo{year}{2024}\natexlab{b}.
\newblock \bibinfo{title}{Labadain-30k+: A Monolingual {T}etun Document-Level Audited Dataset [Data set]. {INESC TEC}}.
\newblock \bibinfo{howpublished}{\url{https://doi.org/10.25747/YDWR-N696}}.
\newblock


\bibitem[de~Jesus and Nunes(2024c)]%
        {dejesus-nunes-tetun-lid-pypi}
\bibfield{author}{\bibinfo{person}{Gabriel de Jesus} {and} \bibinfo{person}{S{\'e}rgio Nunes}.} \bibinfo{year}{2024}\natexlab{c}.
\newblock \bibinfo{title}{Tetun Language Identification Model: A Python Package for {T}etun {LID}}.
\newblock \bibinfo{howpublished}{PyPI Package}.
\newblock
\urldef\tempurl%
\url{https://pypi.org/project/tetun-lid/}
\showURL{%
\tempurl}


\bibitem[de~Jesus and Nunes(2024d)]%
        {dejesus-nunes-tetun-tokenizer-pypi}
\bibfield{author}{\bibinfo{person}{Gabriel de Jesus} {and} \bibinfo{person}{S{\'e}rgio Nunes}.} \bibinfo{year}{2024}\natexlab{d}.
\newblock \bibinfo{title}{Tetun Tokenizer: A Python Package for Tokenizing {T}etun Text}.
\newblock \bibinfo{howpublished}{PyPI Package}.
\newblock
\urldef\tempurl%
\url{https://pypi.org/project/tetun-tokenizer/}
\showURL{%
\tempurl}


\bibitem[de~Jesus and Nunes(2025a)]%
        {de-jesus-nunes-2025-avaliador}
\bibfield{author}{\bibinfo{person}{Gabriel de Jesus} {and} \bibinfo{person}{S{\'e}rgio Nunes}.} \bibinfo{year}{2025}\natexlab{a}.
\newblock \bibinfo{title}{{Labadain-Avaliad{\'o}r: A Test Collection for Tetun Ad-hoc Text Retrieval Task} [Dataset]}.
\newblock
\newblock
\urldef\tempurl%
\url{https://doi.org/10.25747/2k6s-e518}
\showDOI{\tempurl}


\bibitem[de~Jesus and Nunes(2025b)]%
        {labadain-stemmer-algorithm}
\bibfield{author}{\bibinfo{person}{Gabriel de Jesus} {and} \bibinfo{person}{Sérgio Nunes}.} \bibinfo{year}{2025}\natexlab{b}.
\newblock \bibinfo{title}{Labadain-{S}temmer: A stemming algorithm designed for the {T}etun language}.
\newblock \bibinfo{howpublished}{GitHub Repository}.
\newblock
\urldef\tempurl%
\url{https://github.com/gabriel-de-jesus/labadain-stemmer}
\showURL{%
\tempurl}


\bibitem[de~Jesus and Nunes(2025c)]%
        {de-jesus-nunes-2025-stopwords}
\bibfield{author}{\bibinfo{person}{Gabriel de Jesus} {and} \bibinfo{person}{S{\'e}rgio Nunes}.} \bibinfo{year}{2025}\natexlab{c}.
\newblock \bibinfo{title}{{Labadain-Stopwords: A Curated List of 160 Tetun Stopwords} [Dataset]}.
\newblock
\newblock
\urldef\tempurl%
\url{https://doi.org/10.25747/pg2v-kx70}
\showDOI{\tempurl}


\bibitem[de~Jesus and Nunes(2024e)]%
        {de-jesus-nunes-2024-labadain-crawler}
\bibfield{author}{\bibinfo{person}{Gabriel de Jesus} {and} \bibinfo{person}{S{\'e}rgio~Sobral Nunes}.} \bibinfo{year}{2024}\natexlab{e}.
\newblock \showarticletitle{Data Collection Pipeline for Low-Resource Languages: A Case Study on Constructing a {T}etun Text Corpus}. In \bibinfo{booktitle}{\emph{Proceedings of the 2024 Joint International Conference on Computational Linguistics, Language Resources and Evaluation (LREC-COLING 2024)}}, \bibfield{editor}{\bibinfo{person}{Nicoletta Calzolari}, \bibinfo{person}{Min-Yen Kan}, \bibinfo{person}{Veronique Hoste}, \bibinfo{person}{Alessandro Lenci}, \bibinfo{person}{Sakriani Sakti}, {and} \bibinfo{person}{Nianwen Xue}} (Eds.). \bibinfo{publisher}{ELRA and ICCL}, \bibinfo{address}{Torino, Italia}, \bibinfo{pages}{4368--4380}.
\newblock
\urldef\tempurl%
\url{https://aclanthology.org/2024.lrec-main.390}
\showURL{%
\tempurl}


\bibitem[DL~01/2004(2004)]%
        {standard-tetun-inl-2004}
\bibfield{author}{\bibinfo{person}{Democratic Republic of {T}imor-{L}este DL~01/2004, Government Decree-Law No. 1/2004 of 14~April}.} \bibinfo{year}{2004}\natexlab{}.
\newblock \bibinfo{title}{The Standard Orthography of the Tetun Language}.
\newblock
\newblock
\urldef\tempurl%
\url{http://mj.gov.tl/jornal/lawsTL/RDTL-Law/RDTL-Gov-Decrees/Gov-Decree-2004-01.pdf}
\showURL{%
\tempurl}
\newblock
\shownote{Accessed on September 21, 2023.}.


\bibitem[Dolamic and Savoy(2009)]%
        {dolamic-savoy-2009}
\bibfield{author}{\bibinfo{person}{Ljiljana Dolamic} {and} \bibinfo{person}{Jacques Savoy}.} \bibinfo{year}{2009}\natexlab{}.
\newblock \showarticletitle{Indexing and stemming approaches for the Czech language}.
\newblock \bibinfo{journal}{\emph{Information Processing \& Management}} \bibinfo{volume}{45}, \bibinfo{number}{6} (\bibinfo{year}{2009}), \bibinfo{pages}{714--720}.
\newblock
\urldef\tempurl%
\url{https://doi.org/10.1016/J.IPM.2009.06.001}
\showDOI{\tempurl}


\bibitem[Dolamic and Savoy(2010)]%
        {dolamic-savoy-2010}
\bibfield{author}{\bibinfo{person}{Ljiljana Dolamic} {and} \bibinfo{person}{Jacques Savoy}.} \bibinfo{year}{2010}\natexlab{}.
\newblock \showarticletitle{Comparative Study of Indexing and Search Strategies for the Hindi, Marathi, and Bengali Languages}.
\newblock \bibinfo{journal}{\emph{{ACM} Transactions on Asian Language Information Processing}} \bibinfo{volume}{9}, \bibinfo{number}{3} (\bibinfo{year}{2010}), \bibinfo{pages}{11:1--11:24}.
\newblock
\urldef\tempurl%
\url{https://doi.org/10.1145/1838745.1838748}
\showDOI{\tempurl}


\bibitem[Ferilli(2021)]%
        {ferilli-2021}
\bibfield{author}{\bibinfo{person}{Stefano Ferilli}.} \bibinfo{year}{2021}\natexlab{}.
\newblock \showarticletitle{Automatic Multilingual Stopwords Identification from Very Small Corpora}.
\newblock \bibinfo{journal}{\emph{Electronics}} \bibinfo{volume}{10}, \bibinfo{number}{17} (\bibinfo{year}{2021}).
\newblock
\showISSN{2079-9292}
\urldef\tempurl%
\url{https://doi.org/10.3390/electronics10172169}
\showDOI{\tempurl}


\bibitem[Flores and Moreira(2016)]%
        {flores-moreira-2016}
\bibfield{author}{\bibinfo{person}{Felipe~N. Flores} {and} \bibinfo{person}{Viviane~Pereira Moreira}.} \bibinfo{year}{2016}\natexlab{}.
\newblock \showarticletitle{Assessing the impact of Stemming Accuracy on Information Retrieval - {A} multilingual perspective}.
\newblock \bibinfo{journal}{\emph{Information Processing \& Management}} \bibinfo{volume}{52}, \bibinfo{number}{5} (\bibinfo{year}{2016}), \bibinfo{pages}{840--854}.
\newblock
\urldef\tempurl%
\url{https://doi.org/10.1016/J.IPM.2016.03.004}
\showDOI{\tempurl}


\bibitem[Fox(1990)]%
        {fox-1990}
\bibfield{author}{\bibinfo{person}{Christopher~J. Fox}.} \bibinfo{year}{1990}\natexlab{}.
\newblock \showarticletitle{A Stop List for General Text}.
\newblock \bibinfo{journal}{\emph{{SIGIR} Forum}} \bibinfo{volume}{24}, \bibinfo{number}{1-2} (\bibinfo{year}{1990}), \bibinfo{pages}{19--35}.
\newblock
\urldef\tempurl%
\url{https://doi.org/10.1145/378881.378888}
\showDOI{\tempurl}


\bibitem[Frakes and Fox(2003)]%
        {frakes-and-fox-2003}
\bibfield{author}{\bibinfo{person}{William~B. Frakes} {and} \bibinfo{person}{Christopher~J. Fox}.} \bibinfo{year}{2003}\natexlab{}.
\newblock \showarticletitle{Strength and similarity of affix removal stemming algorithms}.
\newblock \bibinfo{journal}{\emph{SIGIR Forum}} \bibinfo{volume}{37}, \bibinfo{number}{1} (\bibinfo{date}{apr} \bibinfo{year}{2003}), \bibinfo{pages}{26--30}.
\newblock
\showISSN{0163-5840}
\urldef\tempurl%
\url{https://doi.org/10.1145/945546.945548}
\showDOI{\tempurl}


\bibitem[Ghosh and Bhattacharya(2017)]%
        {ghosh-bhattacharya-2017}
\bibfield{author}{\bibinfo{person}{Kripabandhu Ghosh} {and} \bibinfo{person}{Arnab Bhattacharya}.} \bibinfo{year}{2017}\natexlab{}.
\newblock \showarticletitle{Stopword Removal: Why Bother? {A} Case Study on Verbose Queries}. In \bibinfo{booktitle}{\emph{Proceedings of the 10th Annual {ACM} India Compute Conference, Compute 2017, Bhopal, India, November 16-18, 2017}}, \bibfield{editor}{\bibinfo{person}{Partha~Pratim Chakraborty}, \bibinfo{person}{Manish Gupta}, \bibinfo{person}{Lipika Dey}, {and} \bibinfo{person}{Shourya Roy}} (Eds.). \bibinfo{publisher}{{ACM}}, \bibinfo{pages}{99--102}.
\newblock
\urldef\tempurl%
\url{https://doi.org/10.1145/3140107.3140125}
\showDOI{\tempurl}


\bibitem[Greks\'{a}kov\'{a}(2018)]%
        {greksakova-2018}
\bibfield{author}{\bibinfo{person}{Zuzana Greks\'{a}kov\'{a}}.} \bibinfo{year}{2018}\natexlab{}.
\newblock \emph{\bibinfo{title}{Tetun in Timor-Leste: The role of language contact in its development}}.
\newblock \bibinfo{thesistype}{Ph.\,D. Dissertation}. \bibinfo{school}{Universidade de Coimbra, Portugal}.
\newblock
\urldef\tempurl%
\url{http://hdl.handle.net/10316/80665}
\showURL{%
\tempurl}


\bibitem[Hajek and van Klinken(2019)]%
        {hajek-klinken-2019}
\bibfield{author}{\bibinfo{person}{John Hajek} {and} \bibinfo{person}{Catharina~Williams van Klinken}.} \bibinfo{year}{2019}\natexlab{}.
\newblock \showarticletitle{Language Contact and Gender in Tetun Dili: What Happens When Austronesian Meets Romance?}
\newblock \bibinfo{journal}{\emph{Oceanic Linguistics}}  \bibinfo{volume}{58} (\bibinfo{date}{06} \bibinfo{year}{2019}), \bibinfo{pages}{59--91}.
\newblock
\urldef\tempurl%
\url{https://doi.org/10.1353/ol.2019.0003}
\showDOI{\tempurl}


\bibitem[Harman(1992)]%
        {harman-et-al-1992}
\bibfield{editor}{\bibinfo{person}{Donna~K. Harman}} (Ed.). \bibinfo{year}{1992}\natexlab{}.
\newblock \bibinfo{booktitle}{\emph{Proceedings of The First Text REtrieval Conference, {TREC} 1992, Gaithersburg, Maryland, USA, November 4-6, 1992}}. \bibinfo{series}{{NIST} Special Publication}, Vol.~\bibinfo{volume}{500-207}. \bibinfo{publisher}{National Institute of Standards and Technology {(NIST)}}.
\newblock
\urldef\tempurl%
\url{http://trec.nist.gov/pubs/trec1/t1\_proceedings.html}
\showURL{%
\tempurl}


\bibitem[Hawking(2000)]%
        {hawking-2000}
\bibfield{author}{\bibinfo{person}{David Hawking}.} \bibinfo{year}{2000}\natexlab{}.
\newblock \showarticletitle{Overview of the {TREC-9} Web Track}. In \bibinfo{booktitle}{\emph{Proceedings of The Ninth Text REtrieval Conference, {TREC} 2000, Gaithersburg, Maryland, USA, November 13-16, 2000}} \emph{(\bibinfo{series}{{NIST} Special Publication}, Vol.~\bibinfo{volume}{500-249})}, \bibfield{editor}{\bibinfo{person}{Ellen~M. Voorhees} {and} \bibinfo{person}{Donna~K. Harman}} (Eds.). \bibinfo{publisher}{National Institute of Standards and Technology {(NIST)}}.
\newblock
\urldef\tempurl%
\url{http://trec.nist.gov/pubs/trec9/papers/web9.pdf}
\showURL{%
\tempurl}


\bibitem[Hiemstra(2001)]%
        {hiemstra-2001}
\bibfield{author}{\bibinfo{person}{Djoerd Hiemstra}.} \bibinfo{year}{2001}\natexlab{}.
\newblock \emph{\bibinfo{title}{Using Language Models for Information Retrieval}}.
\newblock \bibinfo{thesistype}{Ph.\,D. Dissertation}. \bibinfo{school}{University of Twente, Enschede, Netherlands}.
\newblock
\urldef\tempurl%
\url{https://ris.utwente.nl/ws/files/6042641/}
\showURL{%
\tempurl}


\bibitem[Hiemstra and Kraaij(1999)]%
        {hiemstra-kraaij-1999}
\bibfield{author}{\bibinfo{person}{Djoerd Hiemstra} {and} \bibinfo{person}{Wessel Kraaij}.} \bibinfo{year}{1999}\natexlab{}.
\newblock \showarticletitle{Twenty-One at TREC-7: ad-hoc and cross-language track}. In \bibinfo{booktitle}{\emph{Proceedings of the seventh Text Retrieval Conference ({TREC})}} \emph{(\bibinfo{series}{{NIST} Special Publications})}, \bibfield{editor}{\bibinfo{person}{E.M. Voorhees} {and} \bibinfo{person}{D.K. Harman}} (Eds.). \bibinfo{publisher}{National Institute of Standards and Technology}, \bibinfo{address}{United States}, \bibinfo{pages}{227--238}.
\newblock
\urldef\tempurl%
\url{https://trec.nist.gov/pubs/trec7/t7\_proceedings.html}
\showURL{%
\tempurl}


\bibitem[Hollink et~al\mbox{.}(2004)]%
        {hollink-et-al-2004}
\bibfield{author}{\bibinfo{person}{Vera Hollink}, \bibinfo{person}{Jaap Kamps}, \bibinfo{person}{Christof Monz}, {and} \bibinfo{person}{Maarten de Rijke}.} \bibinfo{year}{2004}\natexlab{}.
\newblock \showarticletitle{Monolingual Document Retrieval for European Languages}.
\newblock \bibinfo{journal}{\emph{Information Retrieval}} \bibinfo{volume}{7}, \bibinfo{number}{1-2} (\bibinfo{year}{2004}), \bibinfo{pages}{33--52}.
\newblock
\urldef\tempurl%
\url{https://doi.org/10.1023/B:INRT.0000009439.19151.4C}
\showDOI{\tempurl}


\bibitem[Hull and Correia(2005)]%
        {hull-correia-2005}
\bibfield{author}{\bibinfo{person}{Geoffrey~Stephen Hull} {and} \bibinfo{person}{Adérito José~Guterres Correia}.} \bibinfo{year}{2005}\natexlab{}.
\newblock \bibinfo{booktitle}{\emph{Kursu Gramátika Tetun ba Profesór, Tradutór, Jornalista, no Estudante-Universidade Sira}}.
\newblock \bibinfo{publisher}{Instituto Nacional de Linguística ({INL})}.
\newblock
\newblock
\shownote{ISBN: 1-7413-8137-1}.


\bibitem[INETL(2022)]%
        {census-timor-leste-2022}
\bibfield{author}{\bibinfo{person}{Instituto Nacional de Estatística Timor-Leste INETL}.} \bibinfo{year}{2022}\natexlab{}.
\newblock \bibinfo{title}{Timor-Leste Population and Housing Census}.
\newblock
\newblock
\urldef\tempurl%
\url{https://inetl-ip.gov.tl/2023/10/04/2022-census-wall-chart/}
\showURL{%
\tempurl}
\newblock
\shownote{Accessed on February 19, 2024}.


\bibitem[Jones and Galliers(1996)]%
        {jones-galliers-1996}
\bibfield{editor}{\bibinfo{person}{Karen~Sparck Jones} {and} \bibinfo{person}{Julia~Rose Galliers}} (Eds.). \bibinfo{year}{1996}\natexlab{}.
\newblock \bibinfo{booktitle}{\emph{Evaluating Natural Language Processing Systems, An Analysis and Review}}. \bibinfo{series}{Lecture Notes in Computer Science}, Vol.~\bibinfo{volume}{1083}.
\newblock \bibinfo{publisher}{Springer}.
\newblock
\showISBNx{3-540-61309-9}
\urldef\tempurl%
\url{https://doi.org/10.1007/BFB0027470}
\showDOI{\tempurl}


\bibitem[Kek{\"{a}}l{\"{a}}inen(2005)]%
        {kekalainen-2005}
\bibfield{author}{\bibinfo{person}{Jaana Kek{\"{a}}l{\"{a}}inen}.} \bibinfo{year}{2005}\natexlab{}.
\newblock \showarticletitle{Binary and graded relevance in {IR} evaluations--Comparison of the effects on ranking of {IR} systems}.
\newblock \bibinfo{journal}{\emph{Information Processing \& Management}} \bibinfo{volume}{41}, \bibinfo{number}{5} (\bibinfo{year}{2005}), \bibinfo{pages}{1019--1033}.
\newblock
\urldef\tempurl%
\url{https://doi.org/10.1016/J.IPM.2005.01.004}
\showDOI{\tempurl}


\bibitem[Kudugunta et~al\mbox{.}(2023)]%
        {kudugunta-et-al-2023}
\bibfield{author}{\bibinfo{person}{Sneha Kudugunta}, \bibinfo{person}{Isaac Caswell}, \bibinfo{person}{Biao Zhang}, \bibinfo{person}{Xavier Garcia}, \bibinfo{person}{Derrick Xin}, \bibinfo{person}{Aditya Kusupati}, \bibinfo{person}{Romi Stella}, \bibinfo{person}{Ankur Bapna}, {and} \bibinfo{person}{Orhan Firat}.} \bibinfo{year}{2023}\natexlab{}.
\newblock \showarticletitle{{MADLAD-400:} {A} Multilingual And Document-Level Large Audited Dataset}. In \bibinfo{booktitle}{\emph{Advances in Neural Information Processing Systems 36: Annual Conference on Neural Information Processing Systems 2023, NeurIPS 2023, New Orleans, LA, USA, December 10 -- 16, 2023}}, \bibfield{editor}{\bibinfo{person}{Alice Oh}, \bibinfo{person}{Tristan Naumann}, \bibinfo{person}{Amir Globerson}, \bibinfo{person}{Kate Saenko}, \bibinfo{person}{Moritz Hardt}, {and} \bibinfo{person}{Sergey Levine}} (Eds.).
\newblock
\urldef\tempurl%
\url{http://papers.nips.cc/paper\_files/paper/2023/hash/d49042a5d49818711c401d34172f9900-Abstract-Datasets\_and\_Benchmarks.html}
\showURL{%
\tempurl}


\bibitem[Landis and Koch(1977)]%
        {landis-koch-1977}
\bibfield{author}{\bibinfo{person}{J~Richard Landis} {and} \bibinfo{person}{Gary~G Koch}.} \bibinfo{year}{1977}\natexlab{}.
\newblock \showarticletitle{The measurement of observer agreement for categorical data.}
\newblock \bibinfo{journal}{\emph{Biometrics}}  \bibinfo{volume}{33 1} (\bibinfo{year}{1977}), \bibinfo{pages}{159--74}.
\newblock
\urldef\tempurl%
\url{https://api.semanticscholar.org/CorpusID:11077516}
\showURL{%
\tempurl}
\newblock
\shownote{The reference contains interpretation of k-value of inter-annotators.The interpreptation is only for two annotators and two class. It is used in interpreting Fleiss' Kappa.}.


\bibitem[Lo et~al\mbox{.}(2005)]%
        {lo-et-al-2005}
\bibfield{author}{\bibinfo{person}{Rachel~Tsz{-}Wai Lo}, \bibinfo{person}{Ben He}, {and} \bibinfo{person}{Iadh Ounis}.} \bibinfo{year}{2005}\natexlab{}.
\newblock \showarticletitle{Automatically Building a Stopword List for an Information Retrieval System}.
\newblock \bibinfo{journal}{\emph{Journal of Digital Information Management}} \bibinfo{volume}{3}, \bibinfo{number}{1} (\bibinfo{year}{2005}), \bibinfo{pages}{3--8}.
\newblock
\urldef\tempurl%
\url{http://www.dirf.org/jdim/abstractv3i1.htm\#01}
\showURL{%
\tempurl}


\bibitem[Lovins(1968)]%
        {lovins-1968}
\bibfield{author}{\bibinfo{person}{Julie~Beth Lovins}.} \bibinfo{year}{1968}\natexlab{}.
\newblock \showarticletitle{Development of a stemming algorithm}.
\newblock \bibinfo{journal}{\emph{Mechanical Translation and Computational Linguistics}} \bibinfo{volume}{11}, \bibinfo{number}{1-2} (\bibinfo{year}{1968}), \bibinfo{pages}{22--31}.
\newblock
\urldef\tempurl%
\url{http://www.mt-archive.info/MT-1968-Lovins.pdf}
\showURL{%
\tempurl}


\bibitem[Luhn(1957)]%
        {luhn-1957}
\bibfield{author}{\bibinfo{person}{Hans~Peter Luhn}.} \bibinfo{year}{1957}\natexlab{}.
\newblock \showarticletitle{A Statistical Approach to Mechanized Encoding and Searching of Literary Information}.
\newblock \bibinfo{journal}{\emph{{IBM} Journal of Research and Development}} \bibinfo{volume}{1}, \bibinfo{number}{4} (\bibinfo{year}{1957}), \bibinfo{pages}{309--317}.
\newblock
\urldef\tempurl%
\url{https://doi.org/10.1147/RD.14.0309}
\showDOI{\tempurl}


\bibitem[MacAvaney et~al\mbox{.}(2021)]%
        {macavaney-et-al-2021}
\bibfield{author}{\bibinfo{person}{Sean MacAvaney}, \bibinfo{person}{Andrew Yates}, \bibinfo{person}{Sergey Feldman}, \bibinfo{person}{Doug Downey}, \bibinfo{person}{Arman Cohan}, {and} \bibinfo{person}{Nazli Goharian}.} \bibinfo{year}{2021}\natexlab{}.
\newblock \showarticletitle{Simplified Data Wrangling with ir{\_}datasets}. In \bibinfo{booktitle}{\emph{{SIGIR} '21: The 44th International {ACM} {SIGIR} Conference on Research and Development in Information Retrieval, Virtual Event, Canada, July 11-15, 2021}}, \bibfield{editor}{\bibinfo{person}{Fernando Diaz}, \bibinfo{person}{Chirag Shah}, \bibinfo{person}{Torsten Suel}, \bibinfo{person}{Pablo Castells}, \bibinfo{person}{Rosie Jones}, {and} \bibinfo{person}{Tetsuya Sakai}} (Eds.). \bibinfo{publisher}{{ACM}}, \bibinfo{pages}{2429--2436}.
\newblock
\urldef\tempurl%
\url{https://doi.org/10.1145/3404835.3463254}
\showDOI{\tempurl}


\bibitem[Macdonald and Tonellotto(2020)]%
        {macdonald-tonellotto-2020}
\bibfield{author}{\bibinfo{person}{Craig Macdonald} {and} \bibinfo{person}{Nicola Tonellotto}.} \bibinfo{year}{2020}\natexlab{}.
\newblock \showarticletitle{Declarative Experimentation in Information Retrieval using PyTerrier}. In \bibinfo{booktitle}{\emph{{ICTIR} '20: The 2020 {ACM} {SIGIR} International Conference on the Theory of Information Retrieval, Virtual Event, Norway, September 14-17, 2020}}, \bibfield{editor}{\bibinfo{person}{Krisztian Balog}, \bibinfo{person}{Vinay Setty}, \bibinfo{person}{Christina Lioma}, \bibinfo{person}{Yiqun Liu}, \bibinfo{person}{Min Zhang}, {and} \bibinfo{person}{Klaus Berberich}} (Eds.). \bibinfo{publisher}{{ACM}}, \bibinfo{pages}{161--168}.
\newblock
\urldef\tempurl%
\url{https://doi.org/10.1145/3409256.3409829}
\showDOI{\tempurl}


\bibitem[MacKay and Peto(1995)]%
        {mackay-peto-1995}
\bibfield{author}{\bibinfo{person}{David J.~C. MacKay} {and} \bibinfo{person}{Linda C.~Bauman Peto}.} \bibinfo{year}{1995}\natexlab{}.
\newblock \showarticletitle{A hierarchical Dirichlet language model}.
\newblock \bibinfo{journal}{\emph{Nat. Lang. Eng.}} \bibinfo{volume}{1}, \bibinfo{number}{3} (\bibinfo{year}{1995}), \bibinfo{pages}{289--308}.
\newblock
\urldef\tempurl%
\url{https://doi.org/10.1017/S1351324900000218}
\showDOI{\tempurl}


\bibitem[Manning et~al\mbox{.}(2009)]%
        {manning-et-al-2009}
\bibfield{author}{\bibinfo{person}{Christopher~D. Manning}, \bibinfo{person}{Christopher~D. Manning}, {and} \bibinfo{person}{Christopher~D. Manning}.} \bibinfo{year}{2009}\natexlab{}.
\newblock \bibinfo{booktitle}{\emph{An Introduction to Information Retrieval}}.
\newblock \bibinfo{publisher}{Cambridge University Press}, \bibinfo{address}{Cambridge, England}.
\newblock
\urldef\tempurl%
\url{https://nlp.stanford.edu/IR-book/pdf/irbookonlinereading.pdf}
\showURL{%
\tempurl}


\bibitem[Moll\'{a} and Hutchinson(2003)]%
        {molla-hutchinson-2003}
\bibfield{author}{\bibinfo{person}{Diego Moll\'{a}} {and} \bibinfo{person}{Ben Hutchinson}.} \bibinfo{year}{2003}\natexlab{}.
\newblock \showarticletitle{Intrinsic versus extrinsic evaluations of parsing systems} \emph{(\bibinfo{series}{Evalinitiatives '03})}. \bibinfo{publisher}{Association for Computational Linguistics}, \bibinfo{address}{USA}, \bibinfo{pages}{43–50}.
\newblock


\bibitem[Niyongabo et~al\mbox{.}(2020)]%
        {niyongabo-et-al-2020}
\bibfield{author}{\bibinfo{person}{Rubungo~Andre Niyongabo}, \bibinfo{person}{Hong Qu}, \bibinfo{person}{Julia Kreutzer}, {and} \bibinfo{person}{Li Huang}.} \bibinfo{year}{2020}\natexlab{}.
\newblock \showarticletitle{{KINNEWS} and {KIRNEWS:} Benchmarking Cross-Lingual Text Classification for Kinyarwanda and Kirundi}. In \bibinfo{booktitle}{\emph{Proceedings of the 28th International Conference on Computational Linguistics, {COLING} 2020, Barcelona, Spain (Online), December 8--13, 2020}}, \bibfield{editor}{\bibinfo{person}{Donia Scott}, \bibinfo{person}{N{\'{u}}ria Bel}, {and} \bibinfo{person}{Chengqing Zong}} (Eds.). \bibinfo{publisher}{International Committee on Computational Linguistics}, \bibinfo{pages}{5507--5521}.
\newblock
\urldef\tempurl%
\url{https://doi.org/10.18653/V1/2020.COLING-MAIN.480}
\showDOI{\tempurl}


\bibitem[of~Linguistics~(INL)(2004)]%
        {inl-2004-tetum-orthography}
\bibfield{author}{\bibinfo{person}{National~Institute of Linguistics~(INL)}.} \bibinfo{year}{2004}\natexlab{}.
\newblock \bibinfo{booktitle}{\emph{The Standard Orthography of the Tetum Language: 115 Years in the Making}}.
\newblock \bibinfo{publisher}{Directorate of the National Institute of Linguistics (INL)}, \bibinfo{address}{Dili, Timor-Leste}.
\newblock
\urldef\tempurl%
\url{https://archive.org/details/the-standard-orthography-of-the-tetum-language}
\showURL{%
\tempurl}


\bibitem[Ounis et~al\mbox{.}(2005)]%
        {ounis-et-al-2005}
\bibfield{author}{\bibinfo{person}{Iadh Ounis}, \bibinfo{person}{Gianni Amati}, \bibinfo{person}{Vassilis Plachouras}, \bibinfo{person}{Ben He}, \bibinfo{person}{Craig Macdonald}, {and} \bibinfo{person}{Douglas Johnson}.} \bibinfo{year}{2005}\natexlab{}.
\newblock \showarticletitle{Terrier Information Retrieval Platform}. In \bibinfo{booktitle}{\emph{Advances in Information Retrieval, 27th European Conference on {IR} Research, {ECIR} 2005, Santiago de Compostela, Spain, March 21-23, 2005, Proceedings}} \emph{(\bibinfo{series}{Lecture Notes in Computer Science}, Vol.~\bibinfo{volume}{3408})}, \bibfield{editor}{\bibinfo{person}{David~E. Losada} {and} \bibinfo{person}{Juan~M. Fern{\'{a}}ndez{-}Luna}} (Eds.). \bibinfo{publisher}{Springer}, \bibinfo{pages}{517--519}.
\newblock
\urldef\tempurl%
\url{https://doi.org/10.1007/978-3-540-31865-1\_37}
\showDOI{\tempurl}


\bibitem[Paice(1994)]%
        {paice-1994}
\bibfield{author}{\bibinfo{person}{Chris~D. Paice}.} \bibinfo{year}{1994}\natexlab{}.
\newblock \showarticletitle{An Evaluation Method for Stemming Algorithms}. In \bibinfo{booktitle}{\emph{Proceedings of the 17th Annual International {ACM-SIGIR} Conference on Research and Development in Information Retrieval. Dublin, Ireland, 3-6 July 1994 (Special Issue of the {SIGIR} Forum)}}, \bibfield{editor}{\bibinfo{person}{W.~Bruce Croft} {and} \bibinfo{person}{C.~J. van Rijsbergen}} (Eds.). \bibinfo{publisher}{ACM/Springer}, \bibinfo{pages}{42--50}.
\newblock
\urldef\tempurl%
\url{https://doi.org/10.1007/978-1-4471-2099-5\_5}
\showDOI{\tempurl}


\bibitem[Plachouras et~al\mbox{.}(2004)]%
        {plachouras-et-al-2004}
\bibfield{author}{\bibinfo{person}{Vassilis Plachouras}, \bibinfo{person}{Ben He}, {and} \bibinfo{person}{Iadh Ounis}.} \bibinfo{year}{2004}\natexlab{}.
\newblock \showarticletitle{University of Glasgow at {TREC} 2004: Experiments in Web, Robust, and Terabyte Tracks with Terrier}. In \bibinfo{booktitle}{\emph{Proceedings of the Thirteenth Text REtrieval Conference, {TREC} 2004, Gaithersburg, Maryland, USA, November 16-19, 2004}} \emph{(\bibinfo{series}{{NIST} Special Publication}, Vol.~\bibinfo{volume}{500-261})}, \bibfield{editor}{\bibinfo{person}{Ellen~M. Voorhees} {and} \bibinfo{person}{Lori~P. Buckland}} (Eds.). \bibinfo{publisher}{National Institute of Standards and Technology {(NIST)}}.
\newblock
\urldef\tempurl%
\url{http://trec.nist.gov/pubs/trec13/papers/uglasgow.web.robust.tera.pdf}
\showURL{%
\tempurl}


\bibitem[Porter(1980)]%
        {porter-1980}
\bibfield{author}{\bibinfo{person}{Martin~F. Porter}.} \bibinfo{year}{1980}\natexlab{}.
\newblock \showarticletitle{An algorithm for suffix stripping}.
\newblock \bibinfo{journal}{\emph{Program: electronic library and information systems}} \bibinfo{volume}{14}, \bibinfo{number}{3} (\bibinfo{year}{1980}), \bibinfo{pages}{130--137}.
\newblock
\urldef\tempurl%
\url{https://doi.org/10.1108/EB046814}
\showDOI{\tempurl}


\bibitem[Radlinski et~al\mbox{.}(2008)]%
        {radlinski-et-al-2008}
\bibfield{author}{\bibinfo{person}{Filip Radlinski}, \bibinfo{person}{Madhu Kurup}, {and} \bibinfo{person}{Thorsten Joachims}.} \bibinfo{year}{2008}\natexlab{}.
\newblock \showarticletitle{How does clickthrough data reflect retrieval quality?}. In \bibinfo{booktitle}{\emph{Proceedings of the 17th {ACM} Conference on Information and Knowledge Management, {CIKM} 2008, Napa Valley, California, USA, October 26-30, 2008}}, \bibfield{editor}{\bibinfo{person}{James~G. Shanahan}, \bibinfo{person}{Sihem Amer{-}Yahia}, \bibinfo{person}{Ioana Manolescu}, \bibinfo{person}{Yi~Zhang}, \bibinfo{person}{David~A. Evans}, \bibinfo{person}{Aleksander Kolcz}, \bibinfo{person}{Key{-}Sun Choi}, {and} \bibinfo{person}{Abdur Chowdhury}} (Eds.). \bibinfo{publisher}{{ACM}}, \bibinfo{pages}{43--52}.
\newblock
\urldef\tempurl%
\url{https://doi.org/10.1145/1458082.1458092}
\showDOI{\tempurl}


\bibitem[Robertson and Zaragoza(2009)]%
        {robertson-zaragoza-2009}
\bibfield{author}{\bibinfo{person}{Stephen~E. Robertson} {and} \bibinfo{person}{Hugo Zaragoza}.} \bibinfo{year}{2009}\natexlab{}.
\newblock \showarticletitle{The Probabilistic Relevance Framework: {BM25} and Beyond}.
\newblock \bibinfo{journal}{\emph{Foundations and Trends in Information Retrieval}} \bibinfo{volume}{3}, \bibinfo{number}{4} (\bibinfo{year}{2009}), \bibinfo{pages}{333--389}.
\newblock
\urldef\tempurl%
\url{https://doi.org/10.1561/1500000019}
\showDOI{\tempurl}


\bibitem[Rohatgi et~al\mbox{.}(2021)]%
        {rohatgi-et-al-2021}
\bibfield{author}{\bibinfo{person}{Shaurya Rohatgi}, \bibinfo{person}{C.~Lee Giles}, {and} \bibinfo{person}{Jian Wu}.} \bibinfo{year}{2021}\natexlab{}.
\newblock \showarticletitle{What Were People Searching For? {A} Query Log Analysis of An Academic Search Engine}. In \bibinfo{booktitle}{\emph{{ACM/IEEE} Joint Conference on Digital Libraries, {JCDL} 2021, Champaign, IL, USA, September 27-30, 2021}}, \bibfield{editor}{\bibinfo{person}{J.~Stephen Downie}, \bibinfo{person}{Dana McKay}, \bibinfo{person}{Hussein Suleman}, \bibinfo{person}{David~M. Nichols}, {and} \bibinfo{person}{Faryaneh Poursardar}} (Eds.). \bibinfo{publisher}{{IEEE}}, \bibinfo{pages}{342--343}.
\newblock
\urldef\tempurl%
\url{https://doi.org/10.1109/JCDL52503.2021.00062}
\showDOI{\tempurl}


\bibitem[Sahu et~al\mbox{.}(2023)]%
        {sahu-et-al-2023}
\bibfield{author}{\bibinfo{person}{Siba~Sankar Sahu}, \bibinfo{person}{Debrup Dutta}, \bibinfo{person}{Sukomal Pal}, {and} \bibinfo{person}{Imran Rasheed}.} \bibinfo{year}{2023}\natexlab{}.
\newblock \showarticletitle{Effect of Stopwords and Stemming Techniques in Urdu {IR}}.
\newblock \bibinfo{journal}{\emph{{SN} Computer Science}} \bibinfo{volume}{4}, \bibinfo{number}{5} (\bibinfo{year}{2023}), \bibinfo{pages}{547}.
\newblock
\urldef\tempurl%
\url{https://doi.org/10.1007/S42979-023-01953-4}
\showDOI{\tempurl}


\bibitem[Sahu and Pal(2023)]%
        {sahu-pal-2023}
\bibfield{author}{\bibinfo{person}{Siba~Sankar Sahu} {and} \bibinfo{person}{Sukomal Pal}.} \bibinfo{year}{2023}\natexlab{}.
\newblock \showarticletitle{Building a text retrieval system for the Sanskrit language: Exploring indexing, stemming, and searching issues}.
\newblock \bibinfo{journal}{\emph{Computer Speech and Language}}  \bibinfo{volume}{81} (\bibinfo{year}{2023}), \bibinfo{pages}{101518}.
\newblock
\urldef\tempurl%
\url{https://doi.org/10.1016/J.CSL.2023.101518}
\showDOI{\tempurl}


\bibitem[Salton et~al\mbox{.}(1975)]%
        {salton-et-al-1975}
\bibfield{author}{\bibinfo{person}{Gerard Salton}, \bibinfo{person}{Anita Wong}, {and} \bibinfo{person}{Chung{-}Shu Yang}.} \bibinfo{year}{1975}\natexlab{}.
\newblock \showarticletitle{A Vector Space Model for Automatic Indexing}.
\newblock \bibinfo{journal}{\emph{Commun. ACM}} \bibinfo{volume}{18}, \bibinfo{number}{11} (\bibinfo{year}{1975}), \bibinfo{pages}{613--620}.
\newblock
\urldef\tempurl%
\url{https://doi.org/10.1145/361219.361220}
\showDOI{\tempurl}


\bibitem[Sanderson(2010)]%
        {sanderson-2010}
\bibfield{author}{\bibinfo{person}{Mark Sanderson}.} \bibinfo{year}{2010}\natexlab{}.
\newblock \showarticletitle{Test Collection Based Evaluation of Information Retrieval Systems}.
\newblock \bibinfo{journal}{\emph{Foundations and Trends in Information Retrieval}} \bibinfo{volume}{4}, \bibinfo{number}{4} (\bibinfo{year}{2010}), \bibinfo{pages}{247--375}.
\newblock
\urldef\tempurl%
\url{https://doi.org/10.1561/1500000009}
\showDOI{\tempurl}


\bibitem[Savoy(1999)]%
        {savoy-1999}
\bibfield{author}{\bibinfo{person}{Jacques Savoy}.} \bibinfo{year}{1999}\natexlab{}.
\newblock \showarticletitle{A Stemming Procedure and Stopword List for General French Corpora}.
\newblock \bibinfo{journal}{\emph{Journal of the American Society for Information Science}} \bibinfo{volume}{50}, \bibinfo{number}{10} (\bibinfo{year}{1999}), \bibinfo{pages}{944--952}.
\newblock
\urldef\tempurl%
\url{https://doi.org/10.5555/318976.318984}
\showDOI{\tempurl}


\bibitem[Snow et~al\mbox{.}(2008)]%
        {snow-et-al-2008}
\bibfield{author}{\bibinfo{person}{Rion Snow}, \bibinfo{person}{Brendan O'Connor}, \bibinfo{person}{Daniel Jurafsky}, {and} \bibinfo{person}{Andrew~Y. Ng}.} \bibinfo{year}{2008}\natexlab{}.
\newblock \showarticletitle{Cheap and Fast - But is it Good? Evaluating Non-Expert Annotations for Natural Language Tasks}. In \bibinfo{booktitle}{\emph{2008 Conference on Empirical Methods in Natural Language Processing, {EMNLP} 2008, Proceedings of the Conference, 25-27 October 2008, Honolulu, Hawaii, USA, A meeting of SIGDAT, a Special Interest Group of the {ACL}}}. \bibinfo{publisher}{{ACL}}, \bibinfo{pages}{254--263}.
\newblock
\urldef\tempurl%
\url{https://aclanthology.org/D08-1027/}
\showURL{%
\tempurl}


\bibitem[Snowball(2002)]%
        {snowball-2002}
\bibfield{author}{\bibinfo{person}{Snowball}.} \bibinfo{year}{2002}\natexlab{}.
\newblock \bibinfo{title}{Stemming algorithms for use in Information Retrieval}.
\newblock
\newblock
\urldef\tempurl%
\url{https://snowballstem.org}
\showURL{%
\tempurl}


\bibitem[Snowball(2005a)]%
        {snowball-pt-2005}
\bibfield{author}{\bibinfo{person}{Snowball}.} \bibinfo{year}{2005}\natexlab{a}.
\newblock \bibinfo{title}{Portuguese stemming algorithm}.
\newblock
\newblock
\urldef\tempurl%
\url{https://snowballstem.org/algorithms/portuguese/stemmer.html}
\showURL{%
\tempurl}


\bibitem[Snowball(2005b)]%
        {snowball-es-2005}
\bibfield{author}{\bibinfo{person}{Snowball}.} \bibinfo{year}{2005}\natexlab{b}.
\newblock \bibinfo{title}{Spanish stemming algorithm}.
\newblock
\newblock
\urldef\tempurl%
\url{https://snowballstem.org/algorithms/spanish/stemmer.html}
\showURL{%
\tempurl}


\bibitem[Sormunen(2002)]%
        {sormunen-2002}
\bibfield{author}{\bibinfo{person}{Eero Sormunen}.} \bibinfo{year}{2002}\natexlab{}.
\newblock \showarticletitle{Liberal relevance criteria of {TREC} -: counting on negligible documents?}. In \bibinfo{booktitle}{\emph{{SIGIR} 2002: Proceedings of the 25th Annual International {ACM} {SIGIR} Conference on Research and Development in Information Retrieval, August 11-15, 2002, Tampere, Finland}}, \bibfield{editor}{\bibinfo{person}{Kalervo J{\"{a}}rvelin}, \bibinfo{person}{Micheline Beaulieu}, \bibinfo{person}{Ricardo~A. Baeza{-}Yates}, {and} \bibinfo{person}{Sung{-}Hyon Myaeng}} (Eds.). \bibinfo{publisher}{{ACM}}, \bibinfo{pages}{324--330}.
\newblock
\urldef\tempurl%
\url{https://doi.org/10.1145/564376.564433}
\showDOI{\tempurl}


\bibitem[Sparck~Jones(1972)]%
        {sparck-jones-1972}
\bibfield{author}{\bibinfo{person}{Karen Sparck~Jones}.} \bibinfo{year}{1972}\natexlab{}.
\newblock \showarticletitle{A Statistical Interpretation of Term Specificity and Its Application in Retrieval}.
\newblock \bibinfo{journal}{\emph{Journal of Documentation}} \bibinfo{volume}{28}, \bibinfo{number}{1} (\bibinfo{year}{1972}), \bibinfo{pages}{11--21}.
\newblock
\urldef\tempurl%
\url{https://doi.org/10.1108/eb026526}
\showDOI{\tempurl}


\bibitem[Sp{\"a}rck-Jones and van Rijsbergen(1975)]%
        {sparck-jones-rijsbergen-1975}
\bibfield{author}{\bibinfo{person}{Karen Sp{\"a}rck-Jones} {and} \bibinfo{person}{Cornelis~Joost van Rijsbergen}.} \bibinfo{year}{1975}\natexlab{}.
\newblock \bibinfo{booktitle}{\emph{Report on the Need for and the Provision of an ‘Ideal’ Information Retrieval Test Collection}}.
\newblock \bibinfo{type}{{T}echnical {R}eport} British Library Research and Development Report No. 5266. \bibinfo{institution}{Computer Laboratory, University of Cambridge}, \bibinfo{address}{Cambridge, United Kingdom}. \bibinfo{pages}{43} pages.
\newblock
\urldef\tempurl%
\url{https://sigir.org/files/museum/pub-14/pub_14.pdf}
\showURL{%
\tempurl}


\bibitem[Urbanczyk(2017)]%
        {suzanne-2017}
\bibfield{author}{\bibinfo{person}{Suzanne Urbanczyk}.} \bibinfo{year}{2017}\natexlab{}.
\newblock \bibinfo{title}{Phonological and Morphological Aspects of Reduplication}.
\newblock
\newblock
\urldef\tempurl%
\url{https://doi.org/10.1093/acrefore/9780199384655.013.80}
\showURL{%
\tempurl}


\bibitem[van Klinken and Hajek(2018)]%
        {klinken-hajek-2018}
\bibfield{author}{\bibinfo{person}{Catharina~Williams van Klinken} {and} \bibinfo{person}{John Hajek}.} \bibinfo{year}{2018}\natexlab{}.
\newblock \showarticletitle{Language contact and functional expansion in Tetun Dili: The evolution of a new press register}.
\newblock \bibinfo{journal}{\emph{Multilingua}}  \bibinfo{volume}{37} (\bibinfo{year}{2018}), \bibinfo{pages}{613--647}.
\newblock


\bibitem[van Klinken et~al\mbox{.}(2002)]%
        {klinken-et-al-2002}
\bibfield{author}{\bibinfo{person}{Catharina~Williams van Klinken}, \bibinfo{person}{John Hajek}, {and} \bibinfo{person}{Rachel Nordlinger}.} \bibinfo{year}{2002}\natexlab{}.
\newblock \bibinfo{booktitle}{\emph{Tetun Dili: a grammar of an East Timorese language}}.
\newblock \bibinfo{publisher}{Pacific Linguistics}, \bibinfo{address}{Canberra, Australia}.
\newblock
\urldef\tempurl%
\url{https://doi.org/10.15144/PL-528}
\showDOI{\tempurl}
\newblock
\shownote{ISBN: 85883-509-6}.


\bibitem[Vasconcelos et~al\mbox{.}(2011)]%
        {vasconcelos-et-al-2011}
\bibfield{author}{\bibinfo{person}{Pedro Carlos Bacelar~de Vasconcelos}, \bibinfo{person}{Andreia Sofia~Pinto Oliveira}, \bibinfo{person}{Ricardo Sousa~da Cunha}, \bibinfo{person}{Andreia Rute da~Silva Baptista}, \bibinfo{person}{Alexandre Corte-Real~de Ara{\'u}jo}, \bibinfo{person}{Benedita McCrorie~Gra{\c{c}}a Moura}, \bibinfo{person}{Bernardo Almeida}, \bibinfo{person}{Cl{\'a}udio Ximenes}, \bibinfo{person}{Fernando~Conde Monteiro}, \bibinfo{person}{Henrique Curado}, {et~al\mbox{.}}} \bibinfo{year}{2011}\natexlab{}.
\newblock \bibinfo{title}{Constitui{\c{c}}{\~a}o Anotada da Rep{\'u}blica Democr{\'a}tica de Timor-Leste}.
\newblock
\newblock
\urldef\tempurl%
\url{http://hdl.handle.net/10400.22/4008}
\showURL{%
\tempurl}


\bibitem[Voorhees(2004)]%
        {voorhees-et-al-trec-2004}
\bibfield{author}{\bibinfo{person}{Ellen~M. Voorhees}.} \bibinfo{year}{2004}\natexlab{}.
\newblock \showarticletitle{Overview of {TREC} 2004}. In \bibinfo{booktitle}{\emph{Proceedings of the Thirteenth Text REtrieval Conference, {TREC} 2004, Gaithersburg, Maryland, USA, November 16-19, 2004}} \emph{(\bibinfo{series}{{NIST} Special Publication}, Vol.~\bibinfo{volume}{500-261})}, \bibfield{editor}{\bibinfo{person}{Ellen~M. Voorhees} {and} \bibinfo{person}{Lori~P. Buckland}} (Eds.). \bibinfo{publisher}{National Institute of Standards and Technology {(NIST)}}.
\newblock
\urldef\tempurl%
\url{http://trec.nist.gov/pubs/trec13/papers/OVERVIEW13.pdf}
\showURL{%
\tempurl}


\bibitem[Voorhees(2006)]%
        {voorhees-2006}
\bibfield{author}{\bibinfo{person}{Ellen~M. Voorhees}.} \bibinfo{year}{2006}\natexlab{}.
\newblock \showarticletitle{Overview of the {TREC} 2006}. In \bibinfo{booktitle}{\emph{Proceedings of the Fifteenth Text REtrieval Conference, {TREC} 2006, Gaithersburg, Maryland, USA, November 14-17, 2006}} \emph{(\bibinfo{series}{{NIST} Special Publication}, Vol.~\bibinfo{volume}{500-272})}, \bibfield{editor}{\bibinfo{person}{Ellen~M. Voorhees} {and} \bibinfo{person}{Lori~P. Buckland}} (Eds.). \bibinfo{publisher}{National Institute of Standards and Technology {(NIST)}}.
\newblock
\urldef\tempurl%
\url{http://trec.nist.gov/pubs/trec15/papers/OVERVIEW.pdf}
\showURL{%
\tempurl}


\bibitem[Voorhees and Harman(1998)]%
        {voorhees-harman-1998-trec}
\bibfield{author}{\bibinfo{person}{Ellen~M. Voorhees} {and} \bibinfo{person}{Donna Harman}.} \bibinfo{year}{1998}\natexlab{}.
\newblock \showarticletitle{The {T}ext {RE}trieval {C}onferences ({TREC}s)}. In \bibinfo{booktitle}{\emph{TIPSTER TEXT PROGRAM PHASE III: Proceedings of a Workshop held at Baltimore, {M}aryland, October 13-15, 1998}}. \bibinfo{publisher}{Association for Computational Linguistics}, \bibinfo{address}{Baltimore, Maryland, USA}, \bibinfo{pages}{241--273}.
\newblock
\urldef\tempurl%
\url{https://doi.org/10.3115/1119089.1119127}
\showDOI{\tempurl}


\bibitem[Wenzek et~al\mbox{.}(2020)]%
        {wenzek-et-al-2020}
\bibfield{author}{\bibinfo{person}{Guillaume Wenzek}, \bibinfo{person}{Marie{-}Anne Lachaux}, \bibinfo{person}{Alexis Conneau}, \bibinfo{person}{Vishrav Chaudhary}, \bibinfo{person}{Francisco Guzm{\'{a}}n}, \bibinfo{person}{Armand Joulin}, {and} \bibinfo{person}{Edouard Grave}.} \bibinfo{year}{2020}\natexlab{}.
\newblock \showarticletitle{CCNet: Extracting High Quality Monolingual Datasets from Web Crawl Data}. In \bibinfo{booktitle}{\emph{Proceedings of The 12th Language Resources and Evaluation Conference, {LREC} 2020, Marseille, France, May 11-16, 2020}}, \bibfield{editor}{\bibinfo{person}{Nicoletta Calzolari}, \bibinfo{person}{Fr{\'{e}}d{\'{e}}ric B{\'{e}}chet}, \bibinfo{person}{Philippe Blache}, \bibinfo{person}{Khalid Choukri}, \bibinfo{person}{Christopher Cieri}, \bibinfo{person}{Thierry Declerck}, \bibinfo{person}{Sara Goggi}, \bibinfo{person}{Hitoshi Isahara}, \bibinfo{person}{Bente Maegaard}, \bibinfo{person}{Joseph Mariani}, \bibinfo{person}{H{\'{e}}l{\`{e}}ne Mazo}, \bibinfo{person}{Asunci{\'{o}}n Moreno}, \bibinfo{person}{Jan Odijk}, {and} \bibinfo{person}{Stelios Piperidis}} (Eds.). \bibinfo{publisher}{European Language Resources Association}, \bibinfo{pages}{4003--4012}.
\newblock
\urldef\tempurl%
\url{https://aclanthology.org/2020.lrec-1.494/}
\showURL{%
\tempurl}


\bibitem[Wolf et~al\mbox{.}(2009)]%
        {wolf-et-al-2009}
\bibfield{author}{\bibinfo{person}{Elisabeth Wolf}, \bibinfo{person}{Delphine Bernhard}, {and} \bibinfo{person}{Iryna Gurevych}.} \bibinfo{year}{2009}\natexlab{}.
\newblock \showarticletitle{Combining Probabilistic and Translation-Based Models for Information Retrieval Based on Word Sense Annotations}. In \bibinfo{booktitle}{\emph{Multilingual Information Access Evaluation I. Text Retrieval Experiments, 10th Workshop of the Cross-Language Evaluation Forum, {CLEF} 2009, Corfu, Greece, September 30 - October 2, 2009, Revised Selected Papers}} \emph{(\bibinfo{series}{Lecture Notes in Computer Science}, Vol.~\bibinfo{volume}{6241})}, \bibfield{editor}{\bibinfo{person}{Carol Peters}, \bibinfo{person}{Giorgio Maria~Di Nunzio}, \bibinfo{person}{Mikko Kurimo}, \bibinfo{person}{Thomas Mandl}, \bibinfo{person}{Djamel Mostefa}, \bibinfo{person}{Anselmo Pe{\~{n}}as}, {and} \bibinfo{person}{Giovanna Roda}} (Eds.). \bibinfo{publisher}{Springer}, \bibinfo{pages}{120--127}.
\newblock
\urldef\tempurl%
\url{https://doi.org/10.1007/978-3-642-15754-7\_14}
\showDOI{\tempurl}


\bibitem[Wu and Fang(2013)]%
        {wu-fang-2013}
\bibfield{author}{\bibinfo{person}{Hao Wu} {and} \bibinfo{person}{Hui Fang}.} \bibinfo{year}{2013}\natexlab{}.
\newblock \showarticletitle{Tie Breaker: {A} Novel Way of Combining Retrieval Signals}. In \bibinfo{booktitle}{\emph{International Conference on the Theory of Information Retrieval, {ICTIR} '13, Copenhagen, Denmark, September 29 - October 02, 2013}}, \bibfield{editor}{\bibinfo{person}{Oren Kurland}, \bibinfo{person}{Donald Metzler}, \bibinfo{person}{Christina Lioma}, \bibinfo{person}{Birger Larsen}, {and} \bibinfo{person}{Peter Ingwersen}} (Eds.). \bibinfo{publisher}{{ACM}}, \bibinfo{pages}{16}.
\newblock
\urldef\tempurl%
\url{https://doi.org/10.1145/2499178.2499192}
\showDOI{\tempurl}


\bibitem[Zhai and Lafferty(2001)]%
        {zhai-lafferty-2001}
\bibfield{author}{\bibinfo{person}{ChengXiang Zhai} {and} \bibinfo{person}{John~D. Lafferty}.} \bibinfo{year}{2001}\natexlab{}.
\newblock \showarticletitle{A Study of Smoothing Methods for Language Models Applied to Ad Hoc Information Retrieval}. In \bibinfo{booktitle}{\emph{{SIGIR} 2001: Proceedings of the 24th Annual International {ACM} {SIGIR} Conference on Research and Development in Information Retrieval, September 9-13, 2001, New Orleans, Louisiana, {USA}}}, \bibfield{editor}{\bibinfo{person}{W.~Bruce Croft}, \bibinfo{person}{David~J. Harper}, \bibinfo{person}{Donald~H. Kraft}, {and} \bibinfo{person}{Justin Zobel}} (Eds.). \bibinfo{publisher}{{ACM}}, \bibinfo{pages}{334--342}.
\newblock
\urldef\tempurl%
\url{https://doi.org/10.1145/383952.384019}
\showDOI{\tempurl}


\end{thebibliography}
